\documentclass[rpp]{iopart}

\usepackage{amsthm,amssymb}
\usepackage{marginnote}

\usepackage{scrextend}
\deffootnote[1em]{1.5em}{1em}{%
\textsuperscript{\thefootnotemark}%
}

\expandafter\let\csname equation*\endcsname\relax 
\expandafter\let\csname endequation*\endcsname\relax 
\usepackage{amsmath}
\usepackage{soul,xcolor}

\usepackage{iopams}

\usepackage{graphicx}
\usepackage{color}
\usepackage{bm}
\usepackage{comment}
\usepackage{hyperref}
\usepackage{soul}
\usepackage{subfig}
\usepackage{url}

\newcommand{\beq}{\begin{equation}}
\newcommand{\eeq}{\end{equation}}
\newcommand{\res}{{\cal R}}
\renewcommand{\P}{{\cal P}}
\newcommand{\R}{{\rm R}}
\renewcommand{\S}{{\rm S}}
\renewcommand{\e}{\epsilon}

\begin{document}

\setstcolor{red}

\title{Self-force and radiation reaction in general relativity}

\author{Leor Barack and Adam Pound}
\address{Mathematical Sciences, University of Southampton, Southampton SO17 1BJ, United Kingdom}

\date{\today}

\begin{abstract}
The detection of gravitational waves from binary black-hole mergers by the LIGO-Virgo Collaboration marks the dawn of an era when general-relativistic dynamics in its most extreme manifestation is directly accessible to observation. In the future, planned (space-based) observatories operating in the millihertz band will detect the intricate gravitational-wave signals from the inspiral of compact objects into massive black holes residing in galactic centers. Such inspiral events are extremely effective probes of black-hole geometries, offering unparalleled precision tests of General Relativity in its most extreme regime. This prospect has in the past two decades motivated a programme to obtain an accurate theoretical model of the strong-field radiative dynamics in a two-body system with a small mass ratio. The problem naturally lends itself to a perturbative treatment based on a systematic expansion of the field equations in the small mass ratio. At leading order one has a pointlike particle moving in a geodesic orbit around the large black hole. At subsequent orders, interaction of the particle with its own gravitational perturbation gives rise to an effective ``self-force'', which drives the radiative evolution of the orbit, and whose effects can be accounted for order by order in the mass ratio. 

This review surveys the theory of gravitational self-force in curved spacetime and its application to the astrophysical inspiral problem. We first lay the relevant formal foundation, describing the rigorous derivation of the equation of self-forced motion using matched asymptotic expansions and other ideas. We then review the progress that has been achieved in numerically calculating the self-force and its physical effects in astrophysically realistic inspiral scenarios. We highlight the way in which, nowadays, self-force calculations make a fruitful contact with other approaches to the two-body problem and help inform an accurate universal model of binary black hole inspirals, valid across all mass ratios. We conclude with a summary of the state of the art, open problems and prospects. 

Our review is aimed at non-specialist readers and is for the most part self-contained and non-technical; only elementary-level acquaintance with General Relativity is assumed. Where useful, we draw on analogies with familiar concepts from Newtonian gravity or classical electrodynamics.
\end{abstract}



\tableofcontents

\section{Introduction}

Black holes are ``the simplest macroscopic objects in the universe'',  goes the famous quote from S.\ Chandrasekhar \cite{Chandrasekhar:1985kt}.  Indeed, we expect astrophysical black holes, when in isolation, to be described in {\em exact} form in terms of the 2-parameter family of Kerr solutions to the field equations of General Relativity (GR). But put two such black holes in orbit around each other, and they form a strikingly complex dynamical system. No closed-form solutions are known, and even numerical solutions have been forbiddingly hard to obtain until well into the 21st century. The complexity of the gravitational two-body problem in GR stands in stark contrast to its elementary nature in the context of point-particle Newtonian gravity, where all possible orbital configurations are simple conical sections. Point-particle idealizations are problematic in GR, but even in scenarios where they can make sense in some effective way, the orbital dynamics remains very complicated. In classical GR, a gravitationally bound system of two masses (subject only to gravitational forces) admits no stationary configurations: gravitational waves constantly carry orbital energy away from the system, and back-reaction from that radiation gradually drives the two objects closer together. Given enough time, the two bodies eventually merge. If the two bodies are initially Kerr black holes, a single, larger Kerr black hole eventually forms. The detailed description of this inspiral and merger process has been a major computational challenge and a key theme in gravitational research for almost 50 years.  A central motivation has been the desire to predict the exact pattern of gravitational waves through which such systems can manifest themselves observationally. 

The landmark observation of merging black holes by the LIGO-Virgo Collaboration in 2015 \cite{Abbott:2016blz} has conclusively established the existence in nature of black holes, inspiralling black-hole binaries, and gravitational waves, all consistent with GR. Once mathematical curiosities, black holes and gravitational waves are now firmly in the realm of observational astronomy. For the first time, we are given direct observational access to a natural process where general-relativistic dynamics plays out at its most extreme.  The continued advance of {\em gravitational-wave astronomy} will bring unprecedented opportunities to probe relativistic theory in its most dynamical regime. The theoretical modelling of gravitational-wave sources is an integral part of that programme, as the realization of the exciting science prospects relies crucially on the availability of accurate source models. 

The LIGO-Virgo discoveries can serve as a case in point: without an accurate model of the inspiral and merger it would have not been possible to extract the physical parameters of GW150914 (the first event detected and the brightest so far) at the precision with which they were reported \cite{Abbott:2016blz}, and some of the other events would have likely been missed altogether \cite{Abbott:2016nmj}. Analysis by the LIGO-Virgo Collaboration \cite{Abbott:2016wiq} concluded that the quality of science extractable from future observations may well be limited not by experimental precision but by the accuracy of available theoretical models. (This would be the case if detected mergers were to involve more rapidly spinning black holes than in the mergers already observed, or larger mass disparities between the two merging black holes.) The desire to
maximize the science return from gravitational-wave experiments thus continues to drive the theory programme to improve waveform models across the full parameter space relevant to observation. 



Of particular interest is the inspiral scenario where one of the black holes is much lighter than the other: the so-called ``extreme-mass-ratio inspiral'', or EMRI. Nature abounds with EMRIs. They come in the form of stellar-mass black holes (or neutron stars) that are captured into inspiral orbits around massive black holes---the kind of behemoth black holes, of masses of order $10^6$ to $10^9$ solar masses, that reside in the cores of many galaxies, including our own Milky Way. Astrophysical EMRIs emit gravitational waves in millihertz frequencies, which cannot be detected by existing detectors (seismic gravity-gradient perturbations restrict the operation of ground-based detectors to frequencies well above 1 Hz). But they will be prime targets for the planned space-based detector LISA (the Laser Interferometer Space Antenna \cite{LISAESA,LISANASA}), whose peak sensitivity will be in the millihertz band.  EMRIs are extraordinary natural laboratories for strong-gravity physics. In EMRIs, back-reaction from emitted gravitational waves modifies the orbit on a timescale much larger than the orbital period, so the inspiral is slow and gradual---``adiabatic''. In a typical LISA EMRI, the captured object spends the last few years of inspiral in a tight orbit around the massive black hole (with an orbital revolution period of order an hour), moving at a significant fraction of the speed of light and emitting some $10^5$--$10^6$ detectable gravitational-wave cycles. The intricate gravitational-wave signal cleanly encodes within it an extremely detailed map of the spacetime geometry around the black hole: nature's ``ideal experiment'' in strong gravity! Studies have shown how information from EMRI signals could be used to measure the central object's mass and spin with exquisite accuracies, confirm whether it is a Kerr black hole as GR predicts, and eliminate or tightly constrain a host of proposed alternatives to GR. LISA could observe up to hundreds of such EMRIs per year \cite{Babak:2017tow}.

However, even the most luminous EMRI signals are expected to be weaker than the instrumental noise of LISA in its current design.  They could be extracted only by filtering the signal against accurate theoretical waveform templates. A detailed model of the radiative evolution and emitted radiation in EMRI systems is thus a prerequisite to being able to tap into the rich science encapsulated in their signals. Gravitational-wave detectors measure {\it phase} modulations, so the model has to capture correctly the coherent phase evolution of the emitted EMRI signal over the entire inspiral; and it must ideally do so over the entire relevant parameter space of astrophysical EMRIs, allowing for arbitrary orbital configurations, masses, and any other relevant physical attributes. The exciting prospects of observing EMRIs with LISA have, over the past 20 years or so, driven a concerted effort by theorists to develop a faithful model of EMRIs within GR.

The EMRI domain of the relativistic two-body problem presents unique challenges. The disparate lengthscales and long inspiral time make it hard to tackle using direct Numerical Relativity (NR) methods of the kind that inform the LIGO-Virgo searches (i.e., ones based on numerically solving the full nonlinear set of Einstein's equations).  These methods are extremely computationally expensive already in the comparable-masses regime relevant to LIGO-Virgo searches (indeed, these searches rely on phenomenological approximants that interpolate the very sparse database of NR waveforms available), and they become completely intractable for EMRIs.
Also inadequate are methods based on weak-gravity or slow-motion approximations, such as the post-Newtonian (PN) approach, in which GR is treated perturbatively as an expansion about Newtonian gravity. PN methods have been greatly successful in predicting and explaining a variety of observed GR phenomena, from Mercury's perihelion advance to the orbital decay in the famous Hulse--Taylor binary pulsar \cite{HuTa.75}. They also play an important part in modelling the early stage of inspiral for LIGO-Virgo merger searches. But PN methods are wholly inappropriate in the EMRI case, where the entire interesting part of the inspiral occurs low inside the deep gravitational well of the central black hole, where gravity is extreme. One must not rely on any weak-field approximation when modelling EMRIs.

Fortunately, the EMRI problem is naturally amenable to a different type of perturbative treatment: one based on an expansion in the small mass ratio (the small body's mass $m$ divided by the massive black hole's mass $M$), which, for EMRIs detectable by LISA, is as small as $10^{-5}$--$10^{-7}$. At ``zeroth order'', the small object is a pointlike test particle, whose motion is unaffected by its own gravitational field or internal structure. The particle traces a {\it geodesic} orbit---the curved-space equivalent of a straight line---in the Kerr spacetime associated with the large black hole. Geodesic orbits around a Kerr black hole can be highly complex (in particular, they are generically {\it ergodic}---i.e., space-filling; see Sec.\ \ref{sec:orbits} and Fig.\ \ref{fig:orbit} below), but they are well understood and can be described, essentially, in closed form. Such geodesic orbits approximate true EMRI orbits well only over periods much shorter than the timescale of gravitational radiation-reaction. Radiation-reaction effects come into play at the next order of the expansion: one now treats the gravitational field of the small object as a small perturbation of the background Kerr geometry, satisfying a linearized version of Einstein's field equations, and one then considers the back-reaction from that perturbation on the particle's orbit. Still viewed as a trajectory in the fixed Kerr background, the particle's orbit now experiences a small acceleration with respect to the original geodesic. This acceleration is interpreted as being caused by an effective ``gravitational self-force'' (GSF) attributed to the particle's interaction with its own gravitational field. In this picture, it is the GSF which drives the slow radiative inspiral, and whose work upon the particle converts orbital energy into gravitational-wave energy.  Higher orders in the perturbative expansion account for nonlinear interactions of the gravitational field with the particle and with itself, as well as for dynamical effects coming from the small object's internal structure. (The leading-order effect of the particle's spin enters the equation of motion already at linear order.) 

In this manner, one obtains successive approximations for the EMRI orbit and emitted radiation through a systematic expansion in the mass ratio. At each order, one derives an effective equation of motion that describes the EMRI orbit as a trajectory on the fixed Kerr geometry, coupled to a set of field equations that govern the gravitational field, including observable radiation. Analysis suggests \cite{Hinderer:2008dm} that to construct a sufficiently accurate EMRI model for future LISA searches, one must derive and solve the equations of motion through second order in the perturbative expansion (i.e., accounting for self-acceleration terms that scale with the square of $m/M$). This is a formidable theoretical and computational challenge. 
To tackle it with sufficient rigour one must take a step back and revisit a fundamental question about the nature of motion in GR: how does a ``small'' object move in a curved spacetime, which it itself influences?

Historically, many attempts to address this question have been in the PN context of weak fields, going at least as far back as the seminal work of Einstein, Infeld, and Hoffmann~\cite{Einstein-Infeld-Hoffmann:38}. In the fully relativistic arena, significant effort has gone into simply establishing that the equivalence principle (or geodesic principle) is actually a derivable result of the Einstein field equations, not an independent postulate: in the limit of zero mass and size, the Einstein equations alone dictate that all objects, no matter their internal composition, move on geodesics of the external spacetime. Ref.~\cite{Infeld-Schild:49} provides an early review of work along these lines, and Ref.~\cite{Geroch-Weatherall:17} a very recent one. 

A major step beyond this was taken in the 1970s by Dixon~\cite{Dixon:74}, who showed precisely how the geodesic approximation is altered by the finite size of a body (building on earlier work by Mathisson~\cite{Mathisson:37} and Papapetrou~\cite{Papapetrou:52}, among others). He derived an exact equation of motion for an arbitrary material body, in which the body's multipole moments couple to the curvature of the external spacetime to induce corrections to geodesic motion. However, his method fails if the body's own gravity is strong, and he hence restricted his result to {\em test} bodies, which are of finite size but do not affect the spacetime geometry. He reasoned that extending his method to gravitating bodies would require identifying and somehow subtracting the body's own (highly nonlinear) ``self-field'', which dominates the metric in the body's neighbourhood but, in analogy with Newtonian gravity, might be expected to have no direct effect on its motion. 

In the 1980s, another major step was taken by Thorne and Hartle~\cite{Thorne-Hartle:85}, this time building on work by D'Eath~\cite{DEath:75} and Kates~\cite{Kates:80}, among others. Using a perturbative expansion in the limit of small mass and size, they established a sort of {\em generalized} equivalence principle: at least at low orders in perturbation theory, a gravitating object, be it a material body or a black hole, moves as a test body, governed by Dixon's laws, in what locally appears to be the external spacetime. If finite-size effects are neglected, the object moves on a geodesic of that external metric. However, unlike the metric in the usual equivalence principle, this metric is only {\it effectively} external. It is influenced by the object itself, in a way that Thorne and Hartle did not determine. Hence, like Dixon, they left open the difficult task of finding an appropriate division of the physical metric into a self-field and an ``external" remainder. 

A decade later, motivated by the emerging need for EMRI models, Mino, Sasaki, and Tanaka~\cite{Mino-Sasaki-Tanaka:97} and Quinn and Wald~\cite{Quinn-Wald:97} overcame that hurdle, deriving an equation of motion, now known as the MiSaTaQuWa equation, that included the effect of the object's own field at first order in perturbation theory. Detweiler and Whiting~\cite{Detweiler:01, Detweiler-Whiting:03} then showed that the MiSaTaQuWa equation is equivalent to the geodesic equation in a certain linearized vacuum metric, thereby identifying (through first perturbative order) the effective external metric required to complete Thorne and Hartle's results.

This article reviews the significant progress that has been made on the EMRI problem over the two decades since the derivation of the MiSaTaQuWa equation. Our intention is to provide readers with a non-specialist introduction to the subject (the first of this kind, we believe), covering both foundational and computational aspects. We begin with an elementary-level introduction to self-force theory in curved spacetime: 
Section \ref{sec:EM self-force} reviews the foundations of electromagnetic self-force theory in flat and curved spacetimes, and Sec.\ \ref{sec:grav self-force} then covers the essentials of gravitational self-force theory. In Sec.\ \ref{sec:computation} we survey the computational techniques that have been developed in order to enable the application of GSF theory to the astrophysical EMRI problem. Then, in Sec.\ \ref{sec:evolution}, we describe a perturbative approach to the problem of the long-term radiative evolution of the orbit (given the GSF), based on a systematic two-timescale expansion. At this point we turn to discuss actual calculations of the GSF and its effects in EMRI systems. Section \ref{sec:dissipative} covers calculations of dissipative effects and the long-term orbital evolution in EMRIs, while in Sec.\ \ref{sec:conservative} we focus on non-dissipative physical effects, such as the GSF-induced modification of the rates of periastron advance and spin precession. Then, in Sec.\ \ref{sec:synergy},
we review work comparing the predictions of GSF calculations with those of full NR simulations and of PN calculations. Such synergistic studies can inform the development of a universal model of binary inspirals, valid across all mass ratios; we discuss this idea, and the ``effective one body'' (EOB) framework that enables it. 
We conclude, in Sec.\ \ref{sec:frontiers}, with a summery of progress and a discussion of open problems and prospects. 

There already exist several, more expert-oriented review texts on EMRI physics and the self-force. The most comprehensive review of self-force foundations, including self-contained derivations, is the Living Review article by Poisson, Pound and Vega \cite{Poisson:2011nh} (last updated in 2011). There is a more recent, pedagogical review of foundations by A. Pound \cite{Pound:2015tma}, which covers also more recent work of importance. A. Harte's \cite{Harte:2014wya} reviews the non-perturbative approach to the problem of motion in GR, carrying on Dixon's programme. Computational methods for EMRIs have been reviewed by Barack in \cite{Barack:2009ux} and, more recently, by Wardell in \cite{Wardell:2015kea}. For reviews of EMRI science with LISA, see, for example, \cite{AmaroSeoane:2011ei,Amaro-Seoane:2014ela,Babak:2017tow}.



\section{Electromagnetic self-force in flat and curved spacetimes}\label{sec:EM self-force}

In order to understand GSF physics, it is worthwhile to examine the ways in which it differs from Newtonian physics---and the perhaps surprising ways in which it remains the same. Consider the Newtonian analog of an EMRI: the Kepler problem, specifically the case in which a smaller body, such as a planet, orbits a much larger one, such as the Sun. We learn early in our physics education that if the smaller body is perfectly spherical and of uniform density, then it can be treated as a point mass $m$. It creates a gravitational potential\footnote{ To keep expressions simple, throughout this review we use geometric units in which $G=c=1$.} $\Phi(\vec x)=-\frac{m}{|\vec x-\vec z(t)|}$, which diverges at its location, $\vec x=\vec z(t)$. But we also learn that it does not ``feel'' the field created by this potential; instead, it only responds to the {\em external} field of the larger body, obeying the equation of motion
\beq\label{Kepler}
m\frac{d^2 \vec z}{dt^2} =  -m\vec\nabla \Phi_{\rm ext}(\vec z),
\eeq
where $\Phi_{\rm ext}(\vec x) = -\frac{M}{|\vec x-\vec Z(t)|}$ and $\vec Z(t)$ is the trajectory of the larger body (of mass $M$). Finally, when we solve the equation of motion~\eqref{Kepler}, we learn that the (bound) solutions are eternal, periodic ellipses.

To a large extent, GSF physics describes the breakdown of each of these results, and in this review we will detail how that breakdown occurs. However, we will also frequently emphasise the countervailing fact: that so long as its various elements are appropriately generalized, much of the Newtonian picture remains remarkably valid.

\subsection{Nonrelativistic self-force}
Perhaps some years after learning about the Kepler problem, we first encounter a self-force, through which an object does ``feel its own field'', in the case of a nonrelativistic accelerating charge in flat spacetime. At first, the facts in electromagnetism appear very much the same as in Newtonian gravity: if a point charge $q$ is static, it produces a Coulomb potential $\phi = \frac{q}{|\vec x-\vec z|}$, just like the gravitational potential; and if we can treat it as a test charge, then it feels only the external fields, just like the point mass does, obeying the Lorentz-force law
\beq
m\frac{d^2\vec z}{dt^2} = \vec F_{\rm ext} := q (\vec E_{\rm ext} + \vec v\times \vec B_{\rm ext}),
\eeq
where $ \vec v=d\vec z/dt$. However, this is no longer the case if we stop treating it as a test charge. If we actually take into account the change in the field due to the particle's acceleration, we find the motion obeys the Abraham-Lorentz equation:\footnote{ The force $ \frac{2}{3}\frac{q^2}{m} \frac{d\vec F_{\rm ext}}{dt}$ is often written as $\frac{2}{3}q^2 \frac{d^3\vec z}{dt^3}$, making the equation of motion third order in time and leading to physically pathological solutions. We instead write it in the ``order-reduced'' form~\cite{Landau-Lifshitz} (sometimes called the Landau-Lifshitz equation), in which $\frac{d^3\vec z}{dt^3}$ is replaced by its leading approximant $\frac{d}{dt}(\vec F_{\rm ext}/m)$. Solutions to the order-reduced equation approximate solutions to the third-order equation in a meaningful sense, but they are physically well behaved~\cite{Flanagan-Wald:96,Spohn:00,Rohrlich:08}. Moreover, the order-reduced form is in fact the correct one from a more fundamental perspective. The third-order form is derived from treating the charge as an exact point particle. The order-reduced form, on the other hand, automatically follows from considering an asymptotically small but extended charge distribution~\cite{Gralla-Harte-Wald:09}. As we discuss below, here we take the position that in a classical theory, point particles are only ever approximations to extended objects, and so we favour the order-reduced form as a matter of principle.} 
\beq\label{AL equation}
m\frac{d^2\vec z}{dt^2} = \vec F_{\rm ext} + \frac{2}{3}\frac{q^2}{m} \frac{d\vec F_{\rm ext}}{dt}.
\eeq
The second term is a self-force. More specifically, it is a radiation-reaction force. Unlike the Coulomb potential of the static charge, the Li\'enard-Wiechert potentials of the accelerating charge contain an unbound piece, which carries energy-momentum out to infinity in the form of radiation. That emission causes a recoil, pushing the particle in the opposite direction. Because of this effect, Eq.~\eqref{AL equation} differs from Newtonian gravity not only in the presence of a self-force, but in the fact that the self-force is dissipative. For that reason, the classical hydrogen atom, consisting of an electron in orbit around a proton, is famously unstable: due to its emission of radiation, the orbiting electron would lose energy, causing it to spiral inward until the atom collapsed. Hence, it is impossible to construct an electromagnetic analog of the Kepler problem.

This connection between local and global effects is an important theme in self-force theory. Locally, the particle's energy can only be changed by the self-force. But the self-force removes an amount of energy from the particle precisely equal to that carried off to infinity in the form of radiation. In fact, the most straightforward way of arriving at Eq.~\eqref{AL equation} is by finding the simplest force that ensures this energy balance. 

A more rigorous way of deriving Eq.~\eqref{AL equation}, and one which introduces a second central theme, is by considering a small charge distribution and then taking the limit as it shrinks to zero size in a self-similar way, such that its charge, mass, and size all go to zero~\cite{Gralla-Harte-Wald:09}. The notion of a point charge, and in particular a test charge, arises as the leading nontrivial approximation in this limit; the Abraham-Lorentz force (along with, in general, some finite-size effects) appears at the first subleading order. This type of limiting procedure plays a crucial role in self-force theory.

\subsection{Relativistic self-force in flat spacetime}
Despite our description of it, the Abraham-Lorentz equation in the form~\eqref{AL equation} does not evince a particularly direct relationship between the self-force and the particle's field. To obtain a more direct, physically compelling picture, and to begin to recover some of the Newtonian description, it is convenient to consider the relativistic generalization of Eq.~\eqref{AL equation}, known as the Abraham-Lorentz-Dirac equation:
\beq\label{ALD equation}
m\frac{D^2 z^\mu}{d\tau^2} = F^\mu_{\rm ext}+ \frac{2}{3}\frac{q^2}{m}(\delta^\mu_\nu + u^\mu u_\nu) \frac{DF^{\nu}_{\rm ext}}{d\tau}. 
\eeq
Here $F_\mu^{\rm ext}:=qF^{\rm ext}_{\mu\nu}u^\nu$ is the covariant form of the Lorentz force, $\tau$ is the particle's proper time, $u^\mu:=dz^\mu/d\tau$ its four-velocity, and $D/d\tau:=u^\mu\nabla_\mu$ the covariant derivative along its worldline.

Like the Abraham-Lorentz force, the self-force in Eq.~\eqref{ALD equation} is dissipative. And like Eq.~\eqref{AL equation}, Eq.~\eqref{ALD equation} can be surmised from a simple conservation law, as Dirac did~\cite{Dirac:38}---specifically, from the conservation of stress-energy inside a small tube around the particle's worldline. Alternatively, it can be rigorously derived from conservation of stress-energy of a small, extended charge distribution, using the limiting procedure mentioned above~\cite{Gralla-Harte-Wald:09}; a second-order extension of the equation has also been recently derived using the same method~\cite{Moxon:2017ozd}. This, besides being more physically compelling, bypasses the need for an infinite mass renormalization, which is essential in Dirac's derivation. However, we are primarily interested in the form the equation takes, not in how it is derived. Specifically, we wish to establish how the radiation-reaction force in it relates to the particle's own field. 

\begin{figure}[t]
\begin{center}
\includegraphics[width=.2325\textwidth]{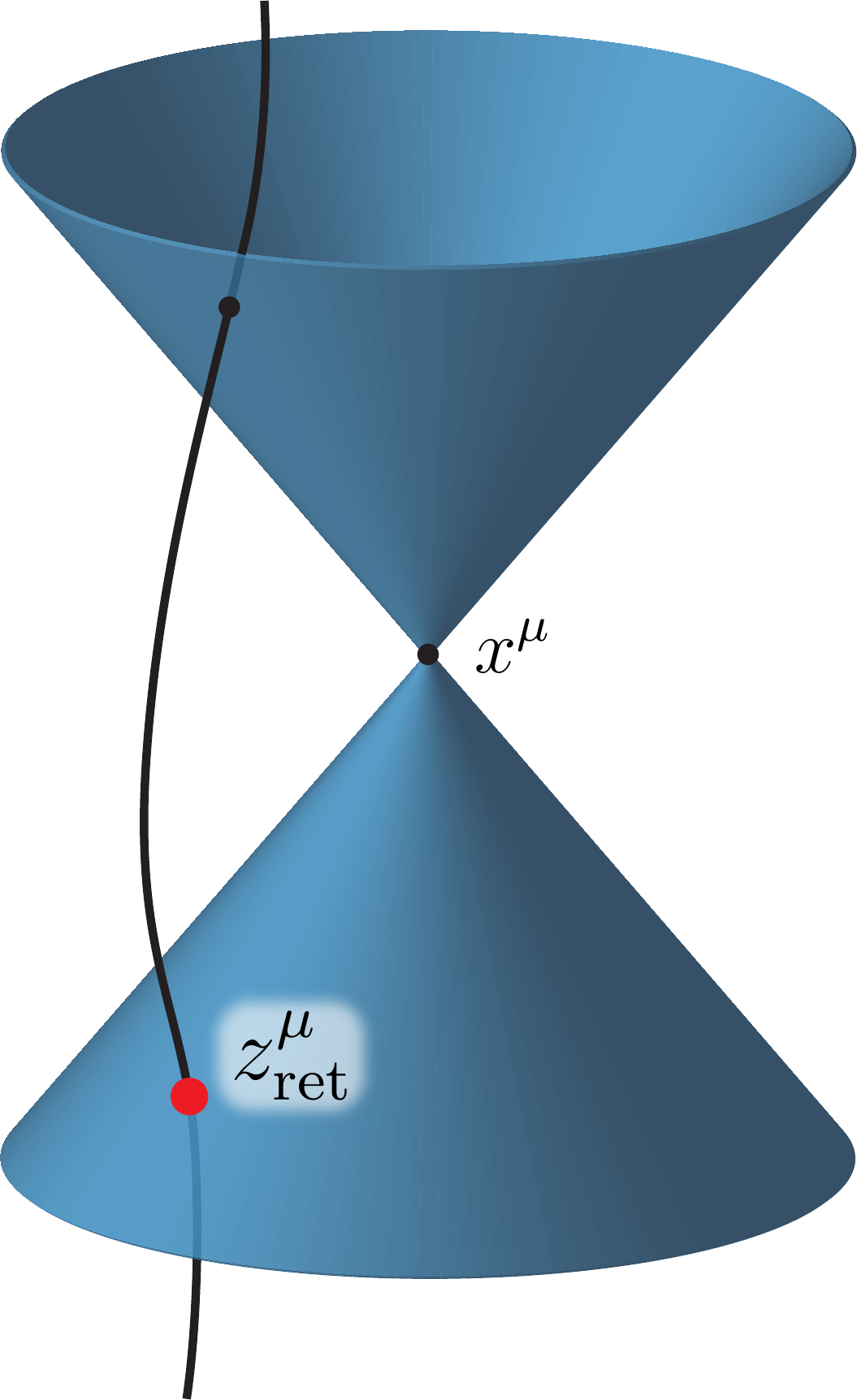}\hspace{25pt}
\includegraphics[width=.2325\textwidth]{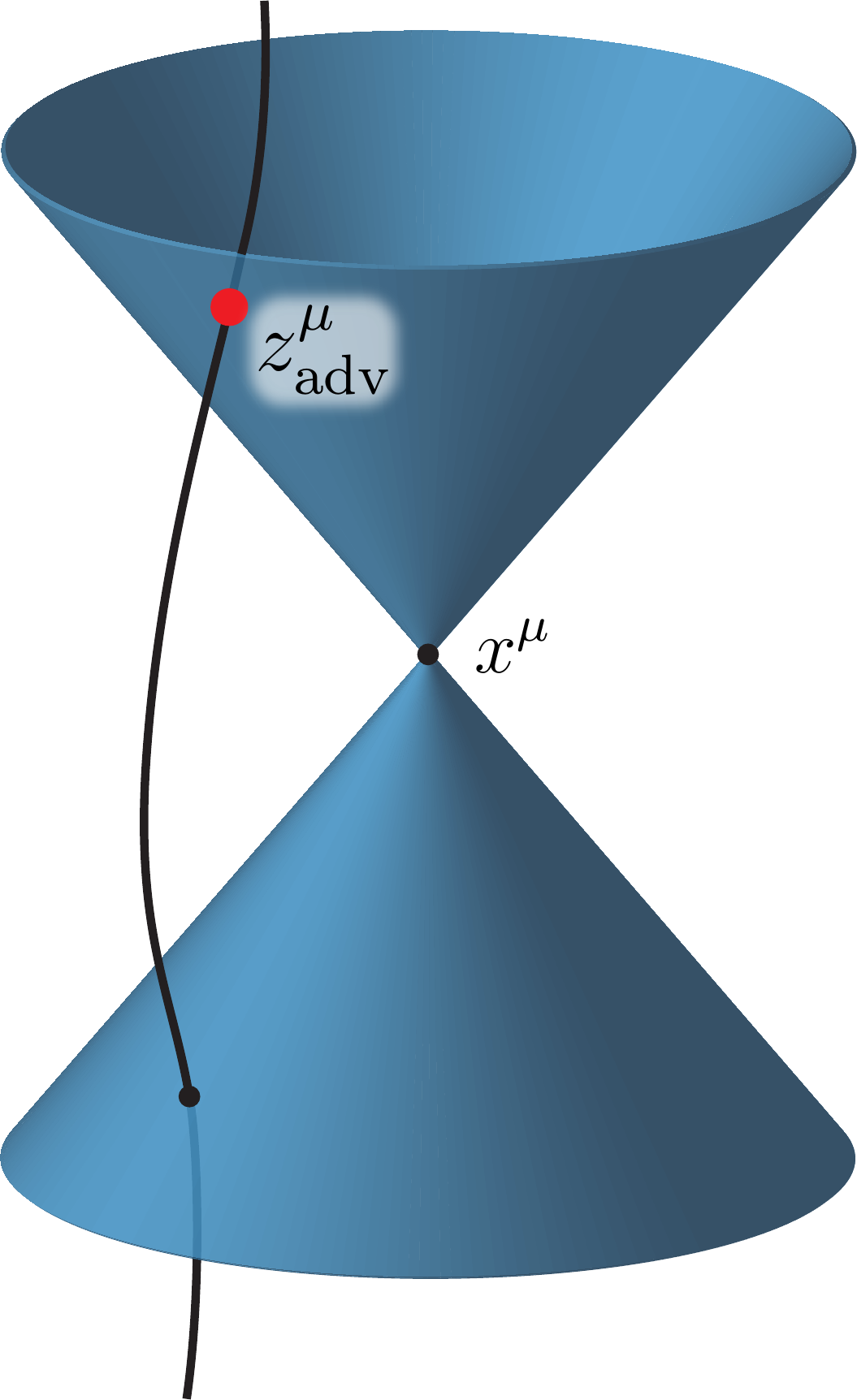}\hspace{25pt}
\includegraphics[width=.2325\textwidth]{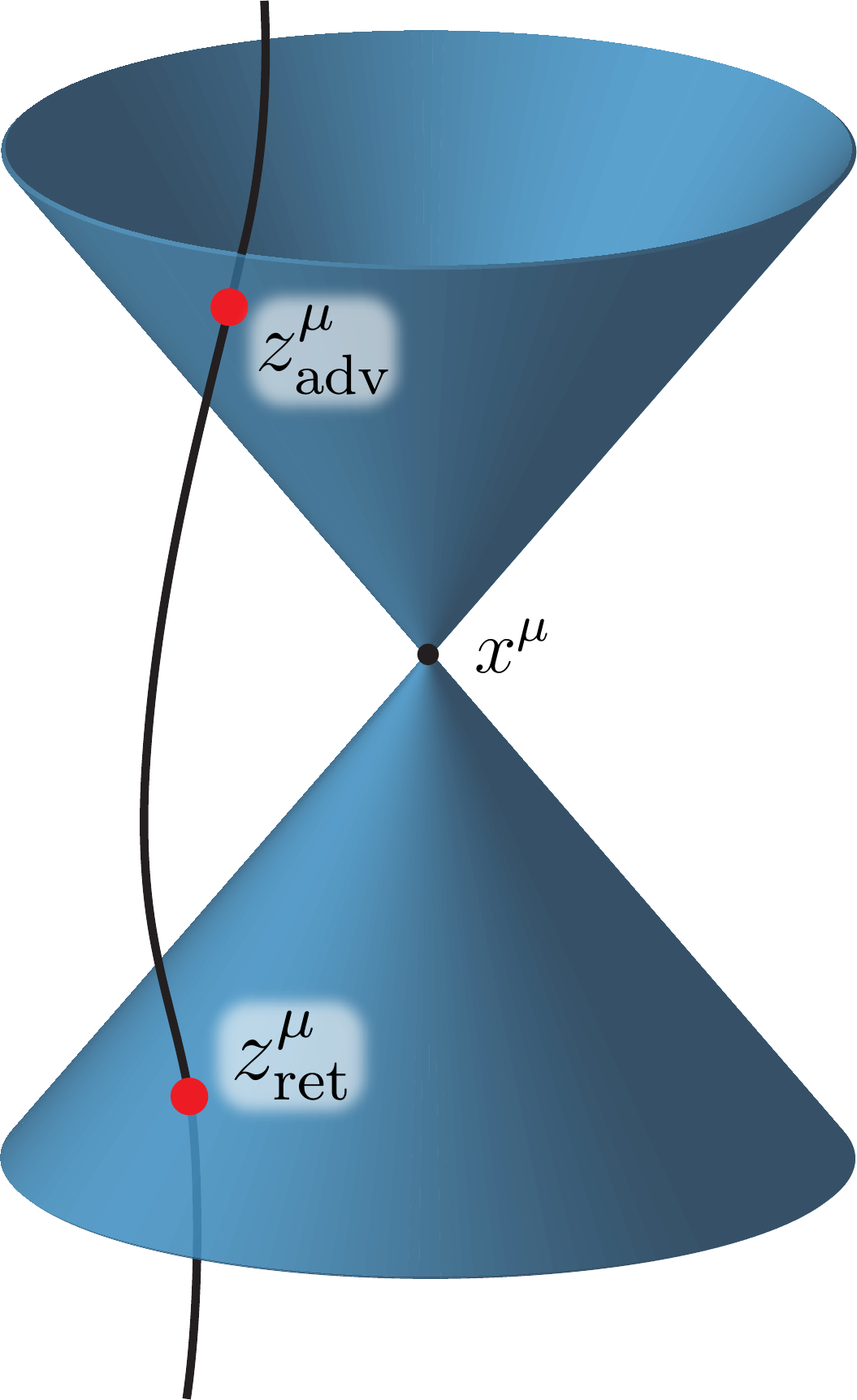}
\caption{Relevant points for the retarded field $A^\mu_+$ (left), advanced field $A^\mu_-$ (middle), and singular and regular fields $A^\mu_\S$ and $A^\mu_\R$ (right) in flat spacetime. $A^\mu_+$ at the point $x^\mu$ depends on the state of the particle at the retarded point $z^\mu_{\rm ret}=z^\mu(\tau_{\rm ret})$, where the particle's worldline intersects $x^\mu$'s past light cone. $A^\mu_-$ at $x^\mu$ depends on the state of the particle at the advanced point $z^\mu_{\rm adv}=z^\mu(\tau_{\rm adv})$, where the particle's worldline intersects $x^\mu$'s future light cone. $A^\mu_\S$ and $A^\mu_\R$ each depend on the state of the particle at both $z^\mu_{\rm ret}$ and $z^\mu_{\rm adv}$.}
\label{flat cones} 
\end{center}
\end{figure}

We first examine the form of the field. If the Lorenz gauge condition  $\nabla_\mu A^\mu=0$ is satisfied, then the potential $A^\mu$ sourced by the particle satisfies the wave equation
\beq\label{Maxwell}
\Box A^\mu = -4\pi j^\mu,
\eeq
where $\Box:=\eta^{\mu\nu}\nabla_\mu \nabla_\nu $ is the flat-space d'Alembertian, $\eta_{\mu\nu}$ is the metric of flat spacetime, and
\beq
j^\mu = q u^\mu\frac{\delta^3(\vec x-\vec z)}{u^t}
\eeq
is the particle's charge-current density. The standard retarded and advanced solutions to this equation are
\beq\label{ret/adv fields}
A_\pm^\mu(x) = \int G_\pm^{\mu}{}_{\mu'}(x,x') j^{\mu'}(x') d^4x',
\eeq
where, in Cartesian coordinates $(t,\vec x)$,
\beq\label{ret-adv Minkowski Greens functions}
G_\pm^{\mu}{}_{\mu'} = \delta^{\mu}_{\mu'}\frac{\delta(t-t' \mp |\vec x-\vec x'|)}{|\vec x-\vec x'|}
\eeq
are the retarded (upper sign) and advanced (lower sign) Green's functions for $\Box$, and $d^4x'=dt'd^3x'$ is the spacetime volume element. Primed indices correspond to tensors at $x'^\mu=(t',\vec x')$. Due to the delta function in Eq.~\eqref{ret-adv Minkowski Greens functions}, the retarded and advanced solutions $A_\pm^\mu$ are entirely determined by the state of the particle at the retarded and advanced time, respectively; see Fig.~\ref{flat cones}.

Of course we are primarily interested in the physical, retarded solution. It contains both time-symmetric and time-antisymmetric pieces, which we can obtain by splitting the retarded Green's function into corresponding pieces:
\beq
G_+^{\mu}{}_{\mu'} = G_\S^{\mu}{}_{\mu'}+G_\R^{\mu}{}_{\mu'},
\eeq
where
\beq
G_\S^{\mu}{}_{\mu'} = \frac{1}{2}(G_+^{\mu}{}_{\mu'} + G_-^{\mu}{}_{\mu'}) 
\eeq
and
\beq
G_\R^{\mu}{}_{\mu'} = \frac{1}{2}(G_+^{\mu}{}_{\mu'} - G_-^{\mu}{}_{\mu'}).
\eeq
$G^\S_{\mu\mu'}$ is a symmetric Green's function, satisfying $G^\S_{\mu\mu'}(x,x') = G^\S_{\mu'\mu}(x',x)$ and $\Box G_\S^\mu{}_{\mu'} = -4\pi \delta_{\mu'}^\mu\delta(x-x')$. $G^\R_{\mu\mu'}$, on the other hand, is an antisymmetric homogeneous solution, satisfying $G^\R_{\mu\mu'}(x,x') = - G^\R_{\mu'\mu}(x',x)$ and $\Box G^\R_\mu{}^{\mu'} =0$. (Note that we use the symbols $x$ and $x^\mu$ interchangeably to label a point.) Substituting this split into Eq.~\eqref{ret/adv fields} gives us the corresponding split of the retarded field,
\beq\label{A=AS+AR flat}
A_+^\mu = A_\S^\mu + A_\R^\mu.
\eeq
The {\em singular field} $A_\S^{\mu}=\int G_\S^\mu{}_{\mu'}j^{\mu'}dV'$ is a relativistic generalization of the Coulomb field. It satisfies Eq.~\eqref{Maxwell}, it is time-symmetric, and locally, near the particle, it behaves as $A_\S^\mu \sim qu^\mu/|\vec x-\vec z|$, becoming singular at the particle's location. The {\em regular field} $A_\R^\mu=\int G_\R^\mu{}_{\mu'}j^{\mu'}dV'$, on the other hand, is an unbound field. It satisfies the homogeneous equation
\beq
\Box A_\R^\mu = 0,
\eeq
it is time-antisymmetric, and it is regular (actually, smooth) at the particle's location.

Now return to the equation of motion~\eqref{ALD equation}. By explicitly evaluating the Faraday tensor associated with the regular field, $F^\R_{\mu\nu}=A^\R_{\mu;\nu}-A^\R_{\nu;\mu}$, on the particle, and comparing the result to the right-hand side of Eq.\ \eqref{ALD equation} (before the order reduction described in footnote 3), one finds that the equation of motion can also be written as 
\beq\label{ALD FR}
m\frac{D^2 z^\mu}{d\tau^2} = F^\mu_{\rm ext} + q F_{\R}^\mu{}_\nu u^\nu. 
\eeq
 In line with its interpretation as a generalization of the Coulomb field, $A_\S^\mu$ does not appear in the equation of motion. 
But $A_\R^{\mu}$ exerts an ordinary Lorentz force on the particle. Combined with the fact that $A_\R^{\mu}$ is a homogeneous field, this suggests that from the particle's perspective, $A_\R^{\mu}$ is indistinguishable from an {\em external} field. If we define the {\em effective} external field $\tilde A_{\rm ext}^\mu = A_{\rm ext}^\mu+A_\R^\mu$ (and associated Faraday tensor $\tilde F^{\rm ext}_{\mu\nu}$), then the equation of motion is simply the Lorentz-force law 
\beq\label{ALD Fext}
m\frac{D^2 z^\mu}{d\tau^2} = q\tilde F_{\rm ext}^{\mu}{}_{\,\nu}u^\nu. 
\eeq

Equation~\eqref{ALD Fext} provides  an alternative description of the self-force, one that is not tied to dissipation or radiation-reaction, and one much closer to the Newtonian picture: whatever its field does, a particle is always governed by the Lorentz force exerted by what it perceives to be the ``external'' field. However, we stress that this is only an effective external field. Away from the particle, $A_\R^\mu$ {\em is not physical}. It depends not only on the retarded point $z^\mu(\tau_{\rm ret})$ on the particle's worldline, but also on the advanced point $z^\mu(\tau_{\rm adv})$ (see Fig.~\ref{flat cones}). Hence, it is not causal. Only in the limit to the particle, where the retarded and advanced points merge, does it become physically meaningful. 

After the inception of the EMRI modelling programme in 1996, this idea of a particle (or small object) behaving as a test particle in an effective external field was most famously advocated by Detweiler. It occupies a central place in self-force theory, and we will return to it at every stage of this review.


\subsection{Electromagnetic self-force in curved spacetime}
\begin{figure}[t]
\begin{center}
\includegraphics[width=.2325\textwidth]{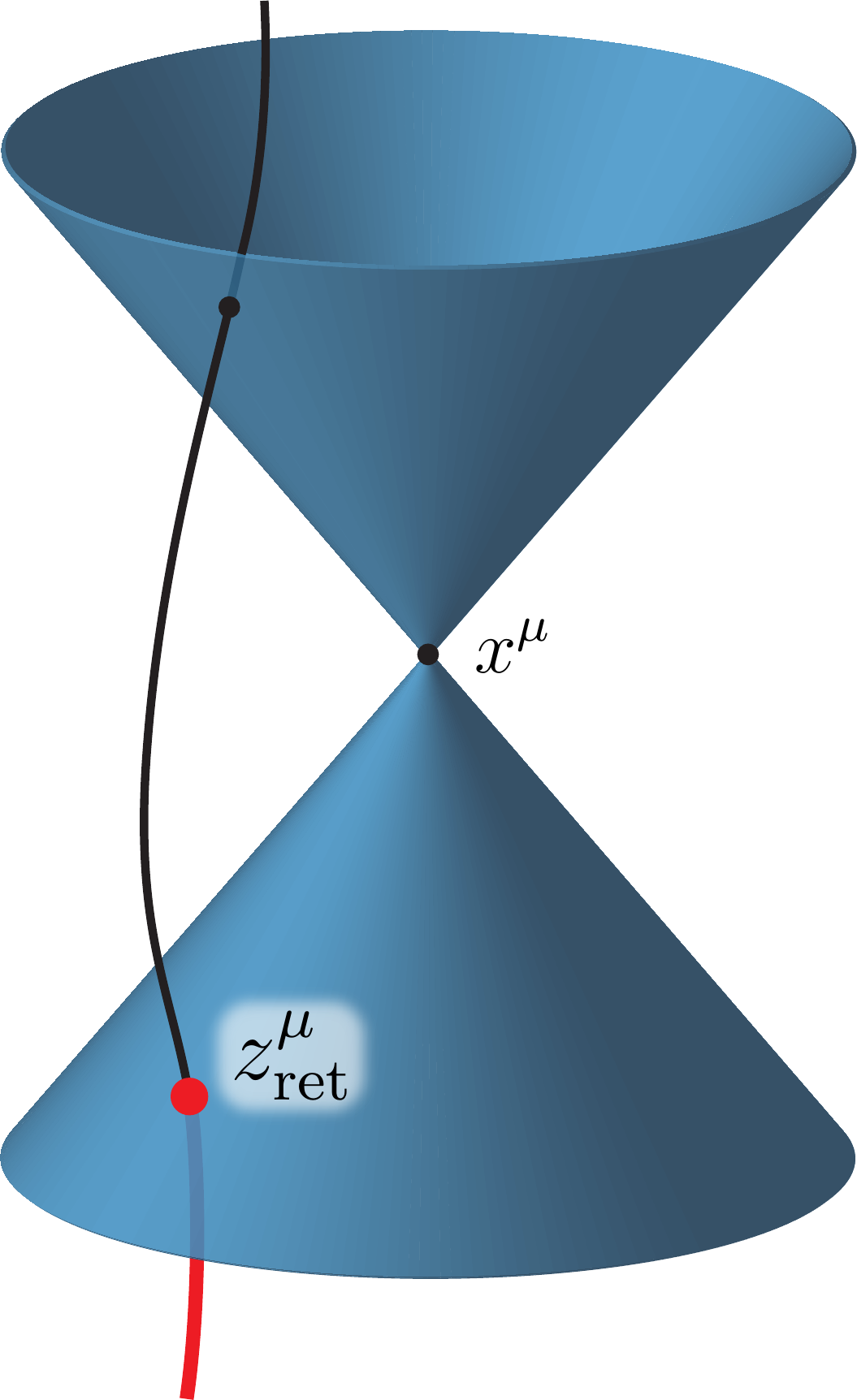}\hspace{5pt}
\includegraphics[width=.2325\textwidth]{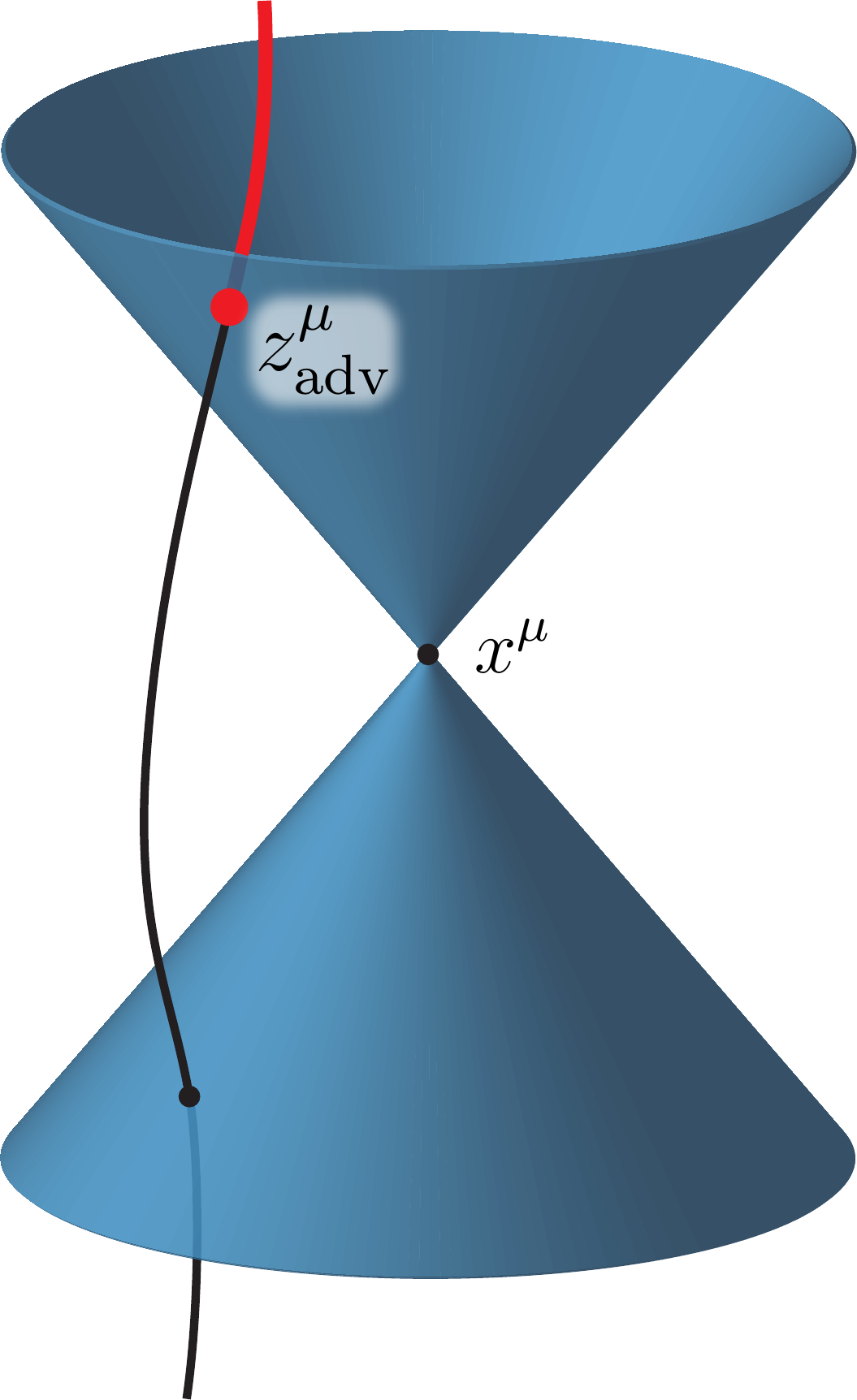}\hspace{5pt}
\includegraphics[width=.2325\textwidth]{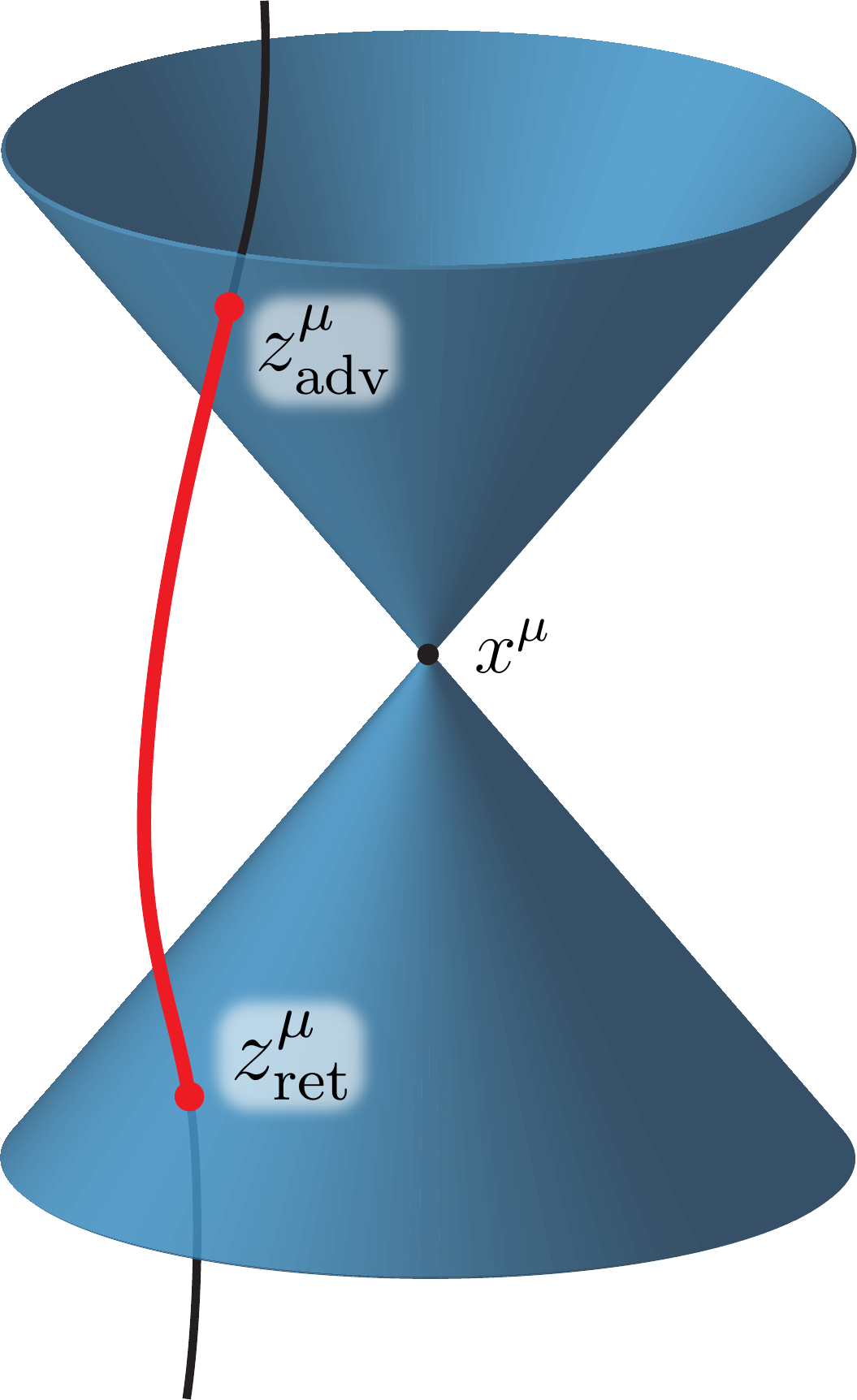}\hspace{5pt}
\includegraphics[width=.2325\textwidth]{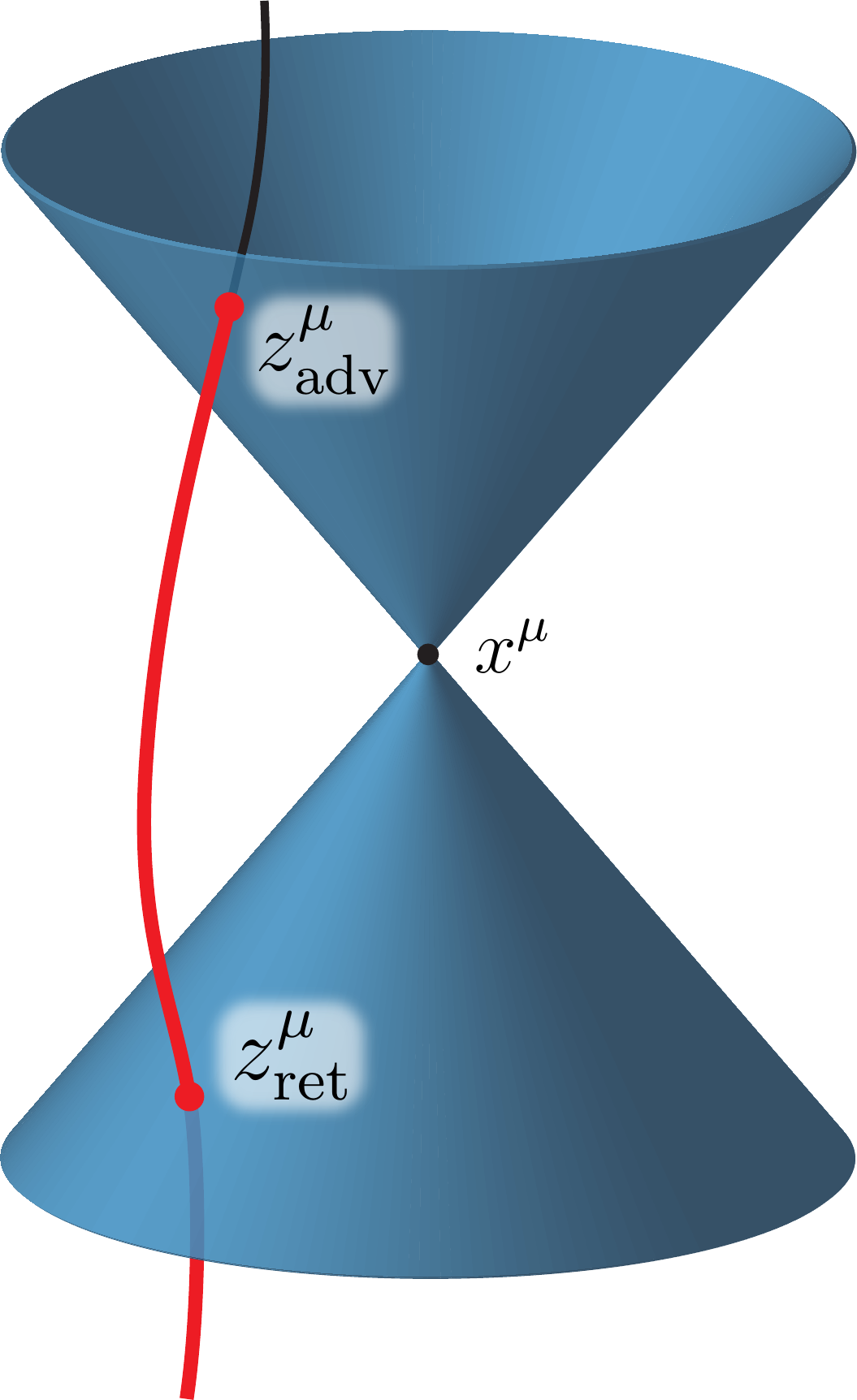}
\caption{Relevant points for (left to right) the retarded field $A^\mu_+$, advanced field $A^\mu_-$, singular field $A^\mu_\S$, and regular field $A^\mu_\R$ in curved spacetime. $A^\mu_+$ at the point $x^\mu$ depends not just on the state of the particle at the retarded point on $x^\mu$'s past light cone, but also on the particle's state at all points {\em within} the past light cone. Analogously,  $A^\mu_-$ depends on the state of the particle at all points on and within $x^\mu$'s future light cone. $A^\mu_\S$ depends on the state of the particle at all points on {\em and outside} $x^\mu$'s past and future light cones. $A^\mu_\R$ depends on the state of the particle at the advanced point $z^\mu_{\rm adv}=z^\mu(\tau_{\rm adv})$ and at all prior points $z^\mu(\tau<\tau_{\rm adv})$.}
\label{curved cones} 
\end{center}
\end{figure}

The move to curved spacetime brings a major change to the physics of the problem. In flat spacetime, waves propagate at the speed of light, along null rays: but in curved spacetime, waves scatter off the spacetime curvature, causing solutions to propagate not just {\em on} lightcones, but also {\em within} them. Because of this, the retarded potential $A^\mu_+$ depends not only on the state of the particle at the retarded point $z^\mu(\tau_{\rm ret})$, but on its state at all prior points $z^\mu(\tau<\tau_{\rm ret})$, as illustrated in Fig.~\ref{curved cones}. This causes an important change to the equation of motion~\eqref{ALD equation}, which becomes  
\beq\label{DeWitt-Brehme}
m\frac{D^2 z^\mu}{d\tau^2} = F_{\rm ext}^\mu + q^2(\delta^\mu_\nu+u^\mu u_\nu)\left(\frac{2}{3m}\frac{DF^{\nu}_{\rm ext}}{d\tau} + \frac{1}{3}R^\nu{}_\rho u^\rho\right) + 2q^2u_\nu\int^{\tau^-}_{-\infty}\!\!\!\!\! \nabla^{[\mu}G_+^{\nu]}{}_{\mu'}u^{\mu'}d\tau,
\eeq
where $R^\nu{}_\rho$ is the Ricci tensor of the spacetime, and $G_+^{\mu}{}_{\mu'}$ is the retarded Green's function for the curved-space wave equation [Eq.~\eqref{Maxwell curved}, below]. The final term in this equation is a ``tail''. It is an integral over the entire past history of the particle, up to $\tau^-=\tau-0^+$, accounting for all the waves that have scattered back to the particle after having been created by it in its past. 

Equation~\eqref{DeWitt-Brehme} was first derived by DeWitt and Brehme~\cite{DeWitt-Brehme:60} (as corrected by Hobbs~\cite{Hobbs:68}) using the same approach as Dirac, considering conservation of stress-energy within a small tube around the particle. Like in the case of flat spacetime, the most rigorous derivation follows from considering the point-particle limit of an extended charge distribution; this has been done by Harte~\cite{Harte:09,Harte:2014wya}, who derived the exact equation of motion of an arbitrary charge distribution and then took the point-particle limit. But also like in the flat-space case, for the moment we are more interested in the form of the equation than its derivation.

Despite the changes in the physics of the solution, the fundamental picture from the preceding section remains valid: the particle feels a Lorentz force due to an effective external field $\tilde A^{\rm ext}_\mu = A^{\rm ext}_\mu+A^\R_\mu$, and the equation of motion~\eqref{DeWitt-Brehme} can be rewritten in the form \eqref{ALD Fext}. 

To motivate the form of the regular field $A^\R_\mu$, we begin with the field equation that the particle's potential $A^\mu$ satisfies. In the Lorenz gauge, it reads 
\beq\label{Maxwell curved}
\Box A^\mu-R^\mu{}_\nu A^\nu = -4\pi j^\mu,
\eeq
where $\Box:=g^{\mu\nu}\nabla_\mu\nabla_\nu$, and $g_{\mu\nu}$ is the metric of the spacetime. The retarded solution is given by $A^\mu_+=\int G^\mu_+{}_{\mu'}j^{\mu'}dV'$, where $dV'=\sqrt{-g'}d^4x'$ is a covariant volume element, with $g'$ being the determinant of $g_{\mu\nu}$ at the integration point $x'^\mu$. We wish to split this solution into appropriate singular and regular pieces in analogy with Eq.~\eqref{A=AS+AR flat}. We first note that in curved spacetime, the self-force can plainly not be described as the Lorentz force exerted by the potential $\frac{1}{2}(A^\mu_+ - A^\mu_-)$: just as the retarded solution $A^\mu_+$ depends on the entire past history of the particle, the advanced solution $A^\mu_-$ depends on its entire {\em future} history, as shown in Fig.~\ref{curved cones}. This acausality persists even in the limit to the particle, unlike that of the regular field in flat space, and so it cannot give rise to the correct, physical self-force. An appropriate alternative was found by Detweiler and Whiting~\cite{Detweiler-Whiting:03}, who defined the modified two-point functions
\beq\label{GS curved spacetime}
G^\mu_\S{}_{\mu'} = \frac{1}{2}(G^\mu_+{}_{\mu'} + G^\mu_-{}_{\mu'} - H^\mu{}_{\mu'}) 
\eeq
and
\beq\label{GR curved spacetime}
G_\R^\mu{}_{\mu'} = \frac{1}{2}(G^\mu_+{}_{\mu'} - G^\mu_-{}_{\mu'} + H^\mu{}_{\mu'}). 
\eeq
Here $H^\mu{}_{\mu'}$ is a symmetric homogeneous solution, satisfying $H_{\mu\mu'}(x,x')=H_{\mu'\mu}(x',x)$ and $\Box H^\mu{}_{\mu'} - R^\mu{}_\nu H^{\nu}{}_{\mu'} = 0$. It is chosen such that $G_\R^\mu{}_{\mu'}(x,x')$ has support at all points $x'$ except those in the chronological past of $x$. This ensures that the corresponding field  has no dependence on points $z^\mu(\tau>\tau_{\rm adv})$ in the chronological future; once again, see Fig.~\ref{curved cones}.

With this choice of $H^\mu{}_{\mu'}$, the singular and regular fields $A_\S^{\mu}=\int G_\S^\mu{}_{\mu'}j^{\mu'}dV'$ and $A_\R^{\mu}=\int G_\R^\mu{}_{\mu'}j^{\mu'}dV'$ possess all of the same key properties as in flat spacetime. $A^\mu_\S$ satisfies the inhomogeneous Eq.~\eqref{Maxwell curved}, and near the particle it behaves as a Coulomb field. $A^\mu_\R$ satisfies the homogeneous wave equation and is smooth at the particle. Off the particle, it is acausal, depending on all points on the worldline prior to the advanced point $z^\mu(\tau_{\rm adv})$. But like in flat spacetime, when evaluated on the particle, it becomes causal, depending only on points in the past. Most importantly, evaluating $A^\R_{\mu}$ and its derivatives on the particle reveals that the DeWitt-Brehme equation, \eqref{DeWitt-Brehme}, is equivalent to Eq.~\eqref{ALD FR} and therefore to Eq.~\eqref{ALD Fext}. 

However, there is one significant change from flat spacetime. While the Green's function $G^\S_{\mu\mu'}$ remains symmetric in its arguments and indices, $G^\R_{\mu\mu'}$ does not remain antisymmetric, due to the presence of $H_{\mu\mu'}$ in Eq.~\eqref{GR curved spacetime}. Because of this, unlike the purely dissipative self-force in flat spacetime, the self-force in curved spacetime has a conservative piece; it is no longer simply a radiation-reaction force. But despite this change, and despite the more complicated physics of wave propagation in curved spacetime, the essential picture remains unchanged: the particle obeys the Lorentz-force law~\eqref{ALD Fext} in what it perceives to be the external field.

\section{Gravitational self-force and the generalized equivalence principle}\label{sec:grav self-force}

\subsection{Perturbation theory in GR and the failure of the point particle description}\label{perturbation theory}

In our discussion of the electromagnetic self-force, we said that the results are rigorously justified by considering the limit of an asymptotically small charge distribution, with the point particle and its field emerging from that limit. However, the resulting equations, and the discussion of the physics, could be couched almost entirely in terms of the point particle. In the case of a gravitating source in GR, this is no longer true: the point-particle approximation fails; the field of the small object cannot, in general, be expressed as that of a point particle.

This failure stems from the nonlinearity of the Einstein field equations. From a physical perspective, we know the Einstein equations imply that a sufficiently dense mass distribution will collapse to form a black hole, not a point particle. From a mathematical perspective, we know that the Einstein equations with a point-particle source do not have a well-defined solution within any suitable class of functions~\cite{Geroch-Traschen:87,Steinbauer-Vickers:08}. 

Let us examine how this failure manifests itself in our problem. We consider an object of mass $m$ moving in a spacetime with a much larger external length scale ${\cal L}\gg m$; in an EMRI, ${\cal L}$ can be the mass $M$ of the large black hole, for example (in geometrical units where mass has dimensions of length). Now we wish to take advantage of the separation of scales by expanding the exact metric of our system, ${\sf g}_{\mu\nu}$, in the limit $m /{\cal L}\to0$. The metric reads
\beq\label{g expansion}
{\sf g}_{\mu\nu} = g_{\mu\nu} + \e h^{(1)}_{\mu\nu} + \e^2 h^{(2)}_{\mu\nu} + O(\e^3),
\eeq
where we have introduced $\e$ as a formal expansion parameter to count powers of $m/{\cal L}$; it will be set to unity at the end of a calculation. The zeroth-order term in Eq.~\eqref{g expansion}, $g_{\mu\nu}$, is referred to as the background metric. In the case of an EMRI, it is the metric of the large black hole. The corrections $h^{(n)}_{\mu\nu}$ describe the gravitational perturbations created by the small object.

The metric \eqref{g expansion} must satisfy the Einstein equation $G_{\mu\nu}[{\sf g}]=8\pi  T_{\mu\nu}$, where $G_{\mu\nu}[{\sf g}]$ is the Einstein tensor of the spacetime and $T_{\mu\nu}$ is the stress-energy tensor of the system's matter content. For simplicity, suppose that the small object represents the only matter, such that $T_{\mu\nu}$ is the stress-energy tensor of the small object itself. If we substitute \eqref{g expansion} into the Einstein equation, then the left-hand side becomes 
\beq\label{G expansion}
G_{\mu\nu}[{\sf g}] = G_{\mu\nu}[g] + \e\delta G_{\mu\nu}[h^{(1)}] + \e^2\left(\delta G_{\mu\nu}[h^{(2)}]+\delta^2 G_{\mu\nu}[h^{(1)}]\right)+O(\e^3),
\eeq
where $\delta G_{\mu\nu}[h^{(n)}]$ is linear in $h^{(n)}_{\mu\nu}$ and $\delta^2 G_{\mu\nu}[h^{(1)}]$ has the schematic form $\partial h^{(1)}_{\mu\nu}\partial h^{(1)}_{\alpha\beta}+h^{(1)}_{\mu\nu}\partial^2 h^{(1)}_{\alpha\beta}$. Let us also suppose that in this limit, $T_{\mu\nu}$ is approximately that of a point particle, such that $T_{\mu\nu} = \e T^{(1)}_{\mu\nu} + \e^2 T^{(2)}_{\mu\nu} +O(\e^3)$, where $T^{(1)}_{\mu\nu}$ is the stress-energy of a point mass moving in the background $g_{\mu\nu}$. (We will momentarily delay the question of whether this makes sense in the case of a black hole, for which $T_{\mu\nu}$ identically vanishes at all points in the spacetime manifold.) 

Through first order in $\e$, no fundamental problem arises. 
Keeping only the first-order terms in the Einstein equation, we arrive at the {\em linearized} Einstein equation with a point-particle source:
\beq\label{first-order EFE}
\delta G_{\mu\nu}[h^{(1)}] = 8\pi T^{(1)}_{\mu\nu}.
\eeq
This equation is analogous to Eq.~\eqref{Maxwell curved} for the electromagnetic potential. Its solutions can be expressed in terms of Green's functions, just as in the preceding sections. Like in the electromagnetic case, the retarded field splits into Detweiler-Whiting singular and regular fields. The singular field behaves as $h^{\S(1)}_{\mu\nu}\sim \frac{m}{r}$ near the particle, where $r$ is a measure of distance to the particle's worldline. The regular field $h^{\R(1)}_{\mu\nu}$ is again a smooth vacuum solution that contains the backscattered waves that arise from propagation within, not just on, the light cones of the background spacetime.

But now suppose we read off  the {\em second}-order term in the Einstein equation. It is 
\beq\label{second-order EFE}
\delta G_{\mu\nu}[h^{(2)}] = 8\pi T^{(2)}_{\mu\nu} - \delta^2 G_{\mu\nu}[h^{(1)}].
\eeq
The second-order perturbation $h^{(2)}_{\mu\nu}$ is sourced by the quadratic combinations of $h^{(1)}_{\mu\nu}$ in $\delta^2 G_{\mu\nu}[h^{(1)}]$, which generically behave like $\sim m^2/r^4$ near the particle.\footnote{ We will, however, mention a fine-tuned way of skirting this generic behavior in Sec.~\ref{gauge}.} This singularity is too strong to be integrated, and because it is constructed from a quadratic operation on an integrable function (as opposed to a linear one), it is not even well defined as a distribution. The other source in Eq.~\eqref{second-order EFE}, $8\pi T^{(2)}_{\mu\nu}$, if it is well defined at all, must be a distribution solely supported on the particle's worldline. Hence, it cannot cure $\delta^2 G_{\mu\nu}$'s nondistributional divergence, and the field equation~\eqref{second-order EFE} is itself  ill defined. 


Because of this failure of the point-particle treatment, in gravity we must face the small object's extended size head on. However, since the object is small, we still wish to avoid directly including its potentially complicated internal dynamics in the Einstein equations. A principal goal of self-force theory is therefore to generalize the point-particle approximation: to reduce the object to a few ``bulk'' properties (such as mass and spin) supported on a worldline, without representing it as a delta function stress-energy tensor. Like the point-particle approximation in electromagnetism, this reduction is achieved by considering an extended object in the limit of zero mass and size. Of course, a critical ingredient in the reduction is an equation of motion for the representative worldline, and it is there that the GSF appears. 

Before proceeding to describe the limiting procedure and its results, we first note one of its crucial outcomes: at linear order, it establishes that the point-particle approximation {\em is} valid. That is, Eq.~\eqref{first-order EFE} is correct even though Eq.~\eqref{second-order EFE} is not. It  is also correct even if $T^{(1)}_{\mu\nu}$ is not the first-order approximation to $T_{\mu\nu}$; it remains valid even if the small object is a black hole and $T_{\mu\nu}$ identically vanishes, for example. Given these facts, much of the GSF literature takes Eq.~\eqref{first-order EFE} as its starting point, and from that point it describes the GSF in a manner precisely analogous to the electromagnetic case. However, rather than presenting that description immediately, we will instead examine how it emerges from the more fundamental picture of an asymptotically small object.

\subsection{Point particle limits and multipole moments}\label{point particles}


The key idea in generalizing the notion of a point particle is to focus not on the small object itself, but on the gravitational field in its immediate neighbourhood. Rather than thinking of the point-particle approximation as a statement about the object's stress-energy tensor, we can take it to be a statement about the object's field.

To illustrate this idea, we return to the simpler context of Newtonian gravity. Consider an isolated material body described by a mass distribution $\rho$. It sources a Newtonian potential $\Phi$ satisfying 
\beq
\partial^i\partial_i\Phi = 4\pi\rho,
\eeq
where we have adopted Cartesian coordinates $x^i = \vec x = (x,y,z)$. Written in terms of the Green's function $G(\vec x, \vec x')=-\frac{1}{|\vec x-\vec x'|}$, the potential reads
\beq\label{Newtonian potential}
\Phi(\vec x) = \int G(\vec x, \vec x')\rho(\vec x') d^3x'.
\eeq
Now, suppose we are only interested in the potential outside the body. Further suppose that the body is small, with a characteristic size $\ell\ll|\vec x|$, and choose any representative worldline $z^\mu=(t,z^i(t))$ in its interior. We can then expand $G(\vec x, \vec x')$ in the integrand of Eq.~\eqref{Newtonian potential} as
\beq
G(\vec x,\vec x') =- \frac{1}{r} - \frac{\delta x'^i n_i }{r^2} - \frac{(3\delta x'^i\delta x'^j - \delta r'^2\delta^{ij})n_in_j}{2r^3} + O(\ell^3/r^4),
\eeq
where $r:=|\vec x-\vec z|$, $\delta \vec{x}' := \vec{x}'-\vec{z}$, $\delta r':=|\delta\vec{x}'|$, and $\vec{n}:=\frac{\vec{x}-\vec{z}}{r}$. Equation~\eqref{Newtonian potential} then becomes an expansion in terms of the body's multipole moments:
\beq\label{Newtonian potential multipole}
\Phi = -\frac{m}{r} - \frac{m^i n_i}{r^2} - \frac{m^{ij} n_in_j}{2r^3} +O(m\ell^3/r^4),
\eeq
where $m:=\int \rho(x')d^3x'$ is the body's mass, $m^i:=\int \rho(x')\delta x'^i d^3x'$ its dipole moment, and $m^{ij}:=\int \rho(x')(3\delta x'^i \delta x'^j - \delta r'^2\delta^{ij}) d^3x'$ its quadrupole moment. The dipole moment $m^i$ measures the distance between $z^i$ and the body's center of mass. Hence, by choosing the representative worldline $z^\mu$ to be that of the body's center of mass, we can set $m^i=0$.

Equation~\eqref{Newtonian potential multipole} is the potential {\em outside} a small but extended body. But suppose we took it to be the potential at all points off the body's worldline [i.e., $x^i\neq z^i(t)$]. It then corresponds precisely to the potential sourced by a ``structured'' point particle with a mass distribution 
\beq\label{point particle source}
\rho = m\delta^3(x^i - z^i) - m^i \partial_i \delta^3(x^i - z^i) + \frac{1}{6}m^{ij}\partial_i\partial_j \delta^3(x^i-z^i) + \ldots 
\eeq 
One can easily check that this source, when substituted into Eq.~\eqref{Newtonian potential}, yields precisely the field~\eqref{Newtonian potential multipole}. If we include only the first term in the source~\eqref{point particle source}, it represents the standard, structureless point mass. The potential it sources in that case is identical to one sourced by a perfectly spherical body, with $z^i$ at the body's center.

The above expansion procedure gives us a precise way of thinking of point particles as approximations to extended objects. But it also demonstrates that, in a meaningful sense, the delta function source~\eqref{point particle source} is not fundamental to the approximation. Rather than thinking of the mass distribution $\rho = m\delta^3(\vec x-\vec z)$ as the defining characteristic of the approximation, we can instead think of the singularity in the potential, $\Phi\sim \frac{m}{|\vec x-\vec z|}$, as primitive; the delta distribution is simply an intermediary that encodes this behavior of the field. This way of thinking, far from being novel, is how point particles were thought of prior to Dirac's invention of his delta function, and it played a significant role in early derivations of equations of motion in GR~\cite{Einstein-Grommer:27,Einstein-Infeld-Hoffmann:38, Einstein-Infeld:49, Infeld-Schild:49}. Unlike a delta function, it survives in the nonlinear arena of Einstein's theory.

It is worth stressing that here we take the view that a point particle is purely an approximation to an extended object, with no fundamental status on its own. It emerges from an asymptotic expansion in the limit of small size, $\ell\ll r$---or equivalently, large relative distance, $r\gg\ell$. As we get closer to the object, we ``see'' more of its structure, in the form of sensitivity to its higher multipole moments. If we are very close, at distances comparable to the object's size, $r\sim\ell$, then the approximation breaks down entirely. 

\subsection{Matched asymptotic expansions and the local form of the metric}\label{local analysis}

Even in GR, any sufficiently well-behaved stress-energy tensor can be reduced to an infinite set of multipole moments, as Dixon showed~\cite{Dixon:74}. But unlike in Newtonian gravity, there is no simple way of translating these moments into an expression for the gravitational field outside the object. Fortunately, there is actually no need to do so: we can obtain the field in a small region outside the object, expressed in terms of a discrete set of multipole moments, directly from the field equations. Instead of specifying a stress-energy tensor, we need only specify the object's moments.\footnote{ Of course, if one wishes to model a particular type of body with a specific $T^{\mu\nu}$, then one must infer its moments from that $T^{\mu\nu}$. However, that becomes, in some sense, an independent problem.} We achieve this using the method of matched asymptotic expansions~\cite{Eckhaus:79,Kevorkian-Cole:96}. This method, which derives from singular perturbation theory, has become a standard means of obtaining equations of motion of small objects; see Refs.~\cite{DEath:75, DEath:96, Kates:80,Kates:80b, Thorne-Hartle:85, Futamase:87, Futamase-Itoh:07, Mino-Sasaki-Tanaka:97, Detweiler:01, Poisson:04, Rosenthal:06a, Rosenthal:06b, Gralla:2008fg, Gralla:12, Pound:10a, Pound:2012nt, Pound:2017psq} for a sample. In particular, this method was used in the first derivation of the MiSaTaQuWa equation~\cite{Mino-Sasaki-Tanaka:97}, and it has been the basis of most later foundational work. Our presentation of the method follows Refs.~\cite{Pound:10a,Pound:12b}.

Matched expansions are used when an ordinary expansion breaks down in a small region. The key idea of the method is to introduce a second limit, one which magnifies this problematic region, and to perform a second expansion in that new limit. In our case, this region is a neighbourhood of the small object itself. As we mentioned in the previous section, the point-particle approximation breaks down at a distance $r\sim \ell$ from the object. The expansion~\eqref{g expansion} of the metric fails at that same distance. This is intuitively sensible: Eq.~\eqref{g expansion} treats the small object's field as a small perturbation of the external universe, but sufficiently near the object, its gravity will dominate over that of external sources, and it cannot be treated as a small perturbation.

Let us assume the object is compact, such that $m\sim \ell$, and $\e$ now counts powers of $m/{\cal L}$ or $\ell/{\cal L}$. If we think of ${\cal L}$ as being of order 1, then the region $r\sim \ell$, which we call the ``body zone'', is equivalent to $r\sim \e$. To zoom in on this region, we introduce a scaled distance $\tilde r=r/\e$. In terms of this scaled distance, the body zone corresponds to $\tilde r\sim 1$. Hence, we can zoom in on the body zone by taking the limit $\e\to0$ at fixed $\tilde r$, and we perform our new expansion in this limit. This contrasts with the original, ordinary expansion~\eqref{g expansion}, which is performed in the  limit $\e\to0$ at fixed $r$. That limit shrinks the object to zero size while holding external lengths fixed. The limit at fixed $\tilde r$ instead keeps lengths of size $\e$ fixed---in particular, the size of the small object---and blows up external lengths out to infinity. 

We shall illustrate the method in the Newtonian case. Suppose we have a small body of mass density $\rho$ in an external gravitational field. Again assume the object is compact (but otherwise arbitrarily structured), and take ${\cal L}$ to be the typical scale over which the external field varies. In a region outside the external sources, the total field satisfies $\nabla^a\nabla_a \Phi = 4\pi\rho$. Outside all sources, including the body, it satisfies the homogeneous equation $\nabla^a\nabla_a\Phi=0$. If we expand in powers of $\e$ at fixed $r$, we have $\Phi(r,\e)=\Phi_0(r)+\e \Phi_1(r)+\e^2\Phi_2(r)+O(\e^3)$, analogous to Eq.~\eqref{g expansion}. If we expand in powers of $\e$ at fixed $\tilde r$, we have instead $\Phi(r,\e)=\hat\Phi_0(\tilde r)+\e\hat\Phi_1(\tilde r)+\e^2\hat\Phi_2(\tilde r)+O(\e^3)$. The background field  in the first expansion, $\Phi_0$, is the field due to external sources in the absence of the body; the background in the second expansion, $\hat\Phi_0(\tilde r)$, is the field of the small body itself in the absence of  external sources. Let us call the first expansion an ``outer expansion'' and denote it $E_{\rm out}\Phi = \Phi_{\rm out}(r,\e)$, and the second expansion an ``inner expansion'' and denote it $E_{\rm in}\Phi = \Phi_{\rm in}(\tilde r,\e)$, where $E_{\rm in/out}$ represent the expansion operations. 

\begin{figure}
\begin{center}
\includegraphics[width=.9\textwidth]{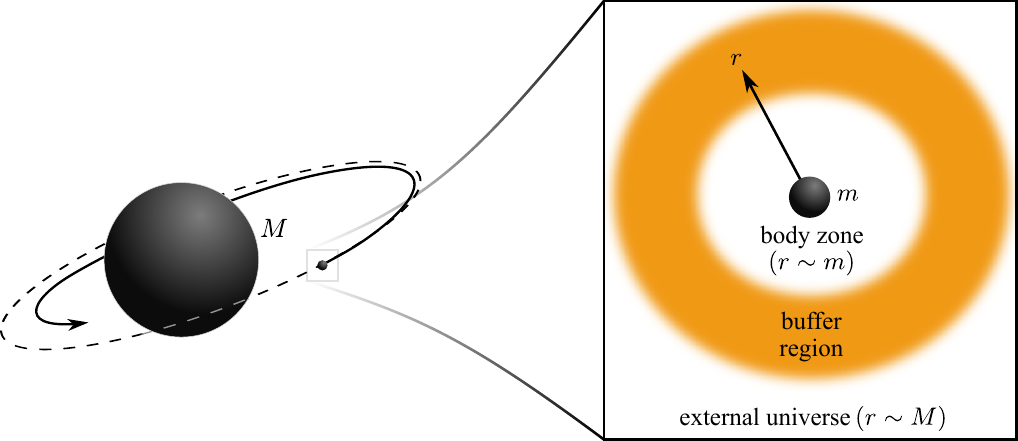}
\caption{Regions involved in matched asymptotic expansions, specialized to the case of an EMRI. The body zone corresponds to distances $r\sim m$ from the small object; the inner expansion, performed in the limit $\e\to0$ at fixed $r/\e$, is presumed to be accurate there. The external universe corresponds to distances $r\sim M$; the outer expansion, performed in the limit $\e\to0$ at fixed $r$, is presumed to be accurate there. The buffer region corresponds to $m\ll r\ll M$, lying between the other two; the double expansions in the limits $\e\to0$ and $r\to0$ are expected to be accurate there. }
\label{buffer region} 
\end{center}
\end{figure}

Since $\Phi_{\rm out}$ and $\Phi_{\rm in}$ are both expansions of the same function, they must agree with (or ``match") each other when suitably compared. To make the comparison, we perform an inner expansion of $\Phi_{\rm out}$, giving us $E_{\rm in}\Phi_{\rm out}=E_{\rm in}E_{\rm out}\Phi$, and we perform an outer expansion of $\Phi_{\rm in}$, giving us $E_{\rm out}\Phi_{\rm in}=E_{\rm out}E_{\rm in}\Phi$. When written as functions of $\e$ and $r$, each of these operations yields a double expansion in the limit $r\to0$ and $\e\to 0$, with the forms
\begin{align}
E_{\rm in}\Phi_{\rm out} &= \sum_{n=0}^\infty\sum_{p=-\infty}^\infty r^p \e^n\Phi_{n,p},\label{inner-outer}\\
E_{\rm out}\Phi_{\rm in} &= \sum_{n=0}^\infty\sum_{q=-\infty}^\infty \left(\frac{\e}{r}\right)^q \e^n \hat\Phi_{n,q},\label{outer-inner}
\end{align}
where the coefficients are independent of $r$ and $\e$. For a sufficiently well-behaved $\Phi$, we will have $E_{\rm in}E_{\rm out}\Phi=E_{\rm out}E_{\rm in}\Phi$,\footnote{ For discussions of the criteria that guarantee this condition is satisfied, see Ref.~\cite{Eckhaus:79}. Reference~\cite{Gralla:2008fg} uses a particularly clear, though unnecessarily strong set of criteria.} implying the matching condition 
\beq\label{matching condition}
\Phi_{n,p} = \hat\Phi_{n+p,-p}. 
\eeq
Equation~\eqref{inner-outer} represents the behaviour of the outer expansion in the limit $r\to0$, very near the worldline relative to the external length scale  ${\cal L}$, while Eq.~\eqref{outer-inner} represents the behaviour of the inner expansion in the limit $\tilde r\to\infty$, very far from the body relative to the internal length scale $\ell$. We can expect the resulting double expansions to be accurate when $\e/r$ and $r/{\cal L}$ are both small. As illustrated in Fig.~\ref{buffer region}, this range, $\ell\ll r\ll {\cal L}$, describes a ``buffer region'' between the body zone and the external universe. It can be thought of as a ``local far-field" region, being simultaneously in the small body's local neighbourhood ($r\ll{\cal L}$) and in its far field ($r\gg\ell$).


Without performing any calculations, we can use the matching condition~\eqref{matching condition} to constrain the forms of $\Phi_{\rm in}$ and $\Phi_{\rm out}$. Since there are no negative powers of $\e$ in the outer expansion (i.e., no $\Phi_{n,p}$ for $n<0$), Eq.~\eqref{matching condition} implies that $\hat\Phi_{n,q}$ vanishes for $q<-n$; likewise, since there are no negative powers of $\e$ in $\Phi_{\rm in}$ (i.e., no $\hat\Phi_{n,q}$ for $n<0$), Eq.~\eqref{matching condition} dictates that $\Phi_{n,p}$ vanishes for $p<-n$. Therefore we have
\begin{subequations}\label{Phihat expansions}
\begin{alignat}{5}
\hat\Phi_0 & = \hat\Phi_{0,0} &&+ \frac{\e}{r} \hat\Phi_{0,1} &&+ \frac{\e^2}{r^2} \hat\Phi_{0,2} &&+ O(\e^3/r^3),\label{Phihat0 expansion}\\
\e\hat\Phi_1 & = r \hat\Phi_{1,-1} &&+ \e \hat\Phi_{1,0} &&+ \frac{\e^2}{r} \hat\Phi_{1,1} &&+ O(\e^3/r^2),\\
\e^2\hat\Phi_2 & = r^2\hat\Phi_{2,-2} &&+ \e r \hat\Phi_{2,-1} &&+ \e^2 \hat\Phi_{2,0}&& + O(\e^3/r),
\end{alignat}
\end{subequations}
and
\begin{subequations}\label{Phi expansions}
\begin{alignat}{5}
\Phi_0 &= \Phi_{0,0} &&+ r \Phi_{0,1} &&+ r^2  \Phi_{0,2} &&+ O(r^3),\label{Phi0 expansion}\\
\e\Phi_1 &= \frac{\e}{r}\Phi_{1,-1} &&+ \e r^0 \Phi_{1,0} &&+ \e r \Phi_{1,1} &&+ O(\e r^2),\\
\e^2\Phi_2 &= \frac{\e^2}{r^2}\Phi_{2,-2} &&+ \frac{\e^2}{r} \Phi_{2,-1} &&+ \e^2r^0 \Phi_{2,0} &&+ O(\e^2 r).\label{Phi2 expansion}
\end{alignat}
\end{subequations}
The matching condition further dictates that the coefficients in the $n$th row of Eq.~\eqref{Phihat expansions} match, term by term, those in the $n$th column of Eq.~\eqref{Phi expansions}. However, we often only care about the particular case of the first row and first column. 

In our context, we are primarily interested in the outer expansion; we only use the inner expansion to inform the outer. The form~\eqref{Phi expansions} of the outer expansion near the worldline is valid regardless of what field equation $\Phi$ satisfies. We can now further constrain it using the field equation. Substituting Eq.~\eqref{Phi expansions} into $\partial^i\partial_i \Phi_n = 0$, one finds the familiar result that the terms of the form $\frac{\Phi_{n,-p}}{r^p}$ (with $p>0$) must be linear combinations of spherical harmonics $Y^{p-1,m}$, and those of the form $r^p \Phi_{n,p}$ (with $p\geq 0$) must be linear combinations of harmonics $Y^{p,m}$. We can also write this as $\Phi_{n,-p} = - m^{(n)}_{i_1\cdots i_{p-1}}n^{i_1}\cdots n^{i_{p-1}}$ and $\Phi_{n,p} = \phi^{(n)}_{i_1\cdots i_{p}}n^{i_1}\cdots n^{i_{p}}$ for some symmetric and trace-free tensors $m^{(n)}_{i_1\cdots i_{p-1}}(t)$ and $\phi^{(n)}_{i_1\cdots i_p}(t)$. 

Now let us interpret the terms in the expansion. With suggestive renamings of the tensors $m^{(n)}_{i_1\cdots i_p}$, we can conveniently sort the terms into two groups, 
\begin{subequations} \label{PhiS}
\begin{align}
\Phi^\S_{1} &= -\frac{m(t)}{r},\\
\Phi^\S_2 &= -\frac{m_i(t) n^i}{r^2} - \frac{\delta m(t)}{r},\\
 &\ \, \vdots\nonumber
\end{align}
\end{subequations}
and
\beq\label{PhiR}
\Phi^\R_{n} = \phi^{(n)}(t) +  \phi^{(n)}_i(t)x^i + \phi^{(n)}_{ij}(t)x^ix^j+\ldots,
\eeq
with the total field given by their sum, $\Phi_n=\Phi^\R_n + \Phi^\S_n$. Although derived in a very different way, these two groups represent (Newtonian versions of) Detweiler-Whiting singular and regular fields. We can think of $\Phi^\S=\sum_{n\geq1}\e^n \Phi^\S_n$ as the body's self-field. It is characterized by a set of multipole moments $m_{i_1\cdots i_p}$ and corrections $\delta^n m_{i_1\cdots i_p}$ to them, with $\Phi^{\S}_n$ containing moments up to $m_{i_1\cdots i_n}$. We can identify these moments as those of the body itself, {\em without} having to integrate over the body's interior. To understand this, note that the most singular term at a given order $\e^n$,  $\Phi_{n,-n}/r^n$, corresponds to a term in the expansion~\eqref{Phihat0 expansion}. That expansion is identical to Eq.~\eqref{Newtonian potential multipole}, the far-field expansion of the field of an isolated body, and instead of defining the body's moments as integrals over its interior, we can define them directly as the coefficients in that far-field expansion. On the other hand, the R field $\Phi^\R=\sum_n \e^n \Phi^\R_n$ defined from Eq.~\eqref{PhiR} has the form of a Taylor expansion of a smooth external field, with no direct dependence on the body's moments. Hence, we can think of $\tilde \Phi = \Phi_0 + \Phi^\R$ as an effective external field. In Sec.~\ref{gauge} we will present an example of how the fields $\Phi^\R_n$ actually arise in a Newtonian context.

From the above analysis, we see that solely from the vacuum field equation and the matching condition, we can tightly constrain the local form of the field outside the body: it is given by Eqs.~\eqref{PhiS} and \eqref{PhiR}, expressed in terms of the body's multipole moments and some smooth fields $\Phi^\R_n$. We can freely specify the multipole moments; this is equivalent to specifying the body's material composition. We can also freely specify the fields $\Phi^\R_n$; this is equivalent to specifying the external environment. However, it is more practical to leave the $\Phi^\R_n$ free at this stage. They will be determined by solving the global problem with whatever external sources are present. 

With the local form of the field known, we can now forget about the matching procedure and extend our solution down to $r=0$ as we did when discussing the point-particle approximation. This effectively replaces the physical field inside the body with a fictitious one. But crucially, it does {\em not} alter the physical field outside the body. 
At $r=0$, the S field becomes singular, while the R field remains a smooth solution to the vacuum equation $\partial^i\partial_i\Phi^\R_n=0$, in precise analogy with the Detweiler-Whiting fields.



Although the calculations become far more involved in GR, all of the essential ideas remain the same. We assume an outer expansion of the form~\eqref{g expansion} along with a complementary inner expansion at fixed $\tilde r$. The matching condition dictates that the fields $h^{(n)}_{\mu\nu}$ have forms precisely analogous to those in Eq.~\eqref{Phi expansions}. After further constraining the form of the perturbations using the vacuum Einstein equations, one finds that they can again be conveniently written in the form $h^{(n)}_{\mu\nu} = h^{\S(n)}_{\mu\nu}+ h^{\R(n)}_{\mu\nu}$, where in analogy with Eq.~\eqref{PhiS}, the S field has the schematic form~\cite{Pound:10a,Pound:12b}
\begin{align}
h^{\S(1)}_{\mu\nu} &\sim \frac{m}{r} + O(r^0),\label{h1}
\end{align}
\begin{subequations}\label{h2}
\begin{align}
h^{\S(2)}_{tt} \sim h^{\S(2)}_{ij}& \sim \frac{m^2 + m_i n^i}{r^2}+ O(1/r),\label{h2tt}\\
h^{\S(2)}_{ti} &\sim \frac{\epsilon_{ijk}s^j n^k}{r^2}+ O(1/r),\label{h2ti}
\end{align}
\end{subequations}
and so on at higher orders. [Here, in the curved-space case, $(t,x^i)$ are specialized to be comoving coordinates centred on $\vec{z}$.]
In analogy with Eq.~\eqref{PhiR}, the R field has the form
\beq\label{hR expansion}
h^{\R (n)}_{\mu\nu} = \phi^{(n)}_{\mu\nu}(t) +  \phi^{(n)}_{\mu\nu i}(t)x^i + \phi^{(n)}_{\mu\nu ij}(t)x^ix^j+\ldots
\eeq

As is clear from the presence of the object's spin $s^i$ in $h^{\S(2)}_{\mu\nu}$, the forms of these fields do differ somewhat from the Newtonian ones. In the Newtonian case, the S field was written entirely in terms of ``mass moments" $m_{i_1\cdots i_\ell}$; in GR there are an additional set of moments $s_{i_1\cdots i_\ell}$, called ``current moments", of which $s_i$ is the first. In $h^{\S(3)}_{\mu\nu}$, both the mass and current quadrupole moments,  $m_{ij}$ and $s_{ij}$, would appear; in $h^{\S(4)}_{\mu\nu}$, the octupole moments; and so forth. Similarly, the R field is more complex than its Newtonian analogue. In Newtonian gravity each of the coefficients in the expansion~\eqref{PhiR} was symmetric and trace-free; in GR the tensors $\phi^{(n)}_{\mu\nu i_1\cdots i_\ell}$ are not all symmetric and trace-free, though they are uniquely defined in terms of a certain set of symmetric-trace-free ``seed" tensors, in a procedure best explained in Refs.~\cite{Pound:12b,Pound:2015fma}. 

These differences aside, the S and R fields' properties closely parallel those of the Newtonian fields. In particular, the singular field $h^\S_{\mu\nu}=\sum_n \e^n h^{\S (n)}_{\mu\nu}$ carries local information about the object; every term in its expansion, at all orders in $r$ and $\e$, is proportional to one of the object's multipole moments (or to some nonlinear combination of them). The regular field $h^\R_{\mu\nu}=\sum_n \e^n h^{\R (n)}_{\mu\nu}$ is a smooth vacuum perturbation for all $r\geq0$, locally independent of the object; prior to imposing boundary conditions, no terms in the series~\eqref{hR expansion} depend on any of the object's moments. One can also show that, like the Detweiler-Whiting regular field in  electromagnetism, $h^\R_{\mu\nu}$ is causal on the worldline~\cite{Pound:2015fma}. We can combine it with the external background $g_{\mu\nu}$ to form a metric $\tilde g_{\mu\nu}=g_{\mu\nu}+h^\R_{\mu\nu}$ that we can interpret as the effective external geometry. This interpretation will be further bolstered when we consider the equation of motion. 

Nevertheless, though the S and R fields have desirable properties, one should keep in mind that only their sum $h^{(n)}_{\mu\nu}=h^{\S(n)}_{\mu\nu}+ h^{\R(n)}_{\mu\nu}$ represents a truly physical field. We could have split that field into alternative choices of S and R fields, or foregone the split altogether. For concreteness we have adopted the definitions in Refs.~\cite{Pound:2012nt,Pound:12b}, but there are many (in fact, infinitely many) other possible choices. Unlike in the Newtonian case, the singular fields in GR, under most sensible definitions, will contain nondivergent terms proportional to $r^n$ with $n\geq0$, and because of this, there is no obviously preferred singular-regular split. Ultimately, one can make any convenient choice and derive useful equations in terms of that choice. In Sec.~\ref{equation of motion} we will comment on some alternative choices that have been made in the literature.

Before proceeding, we recapitulate the essential point of this section: Eqs.~\eqref{h1} and \eqref{h2}, along with the Taylor series for $\tilde g_{\mu\nu}$, represent the form of the metric in the buffer region outside any compact object. This form is valid regardless of whether the small object is a material body or a black hole. If it is a material body, then $r$ is some measure of distance to a representative worldline $z^\mu$ in the body's interior, as in the Newtonian case. If the object is instead a black hole, then clearly there is no representative worldline in its physical interior. However, we can {\em associate} a worldline with it based on the behaviour of the field outside of it, with $r=0$ serving to define that worldline. This associated worldline emerges from the limiting procedure, existing not in the physical spacetime but in the background spacetime with metric $g_{\mu\nu}$ (or equivalently, in the effective spacetime with metric $\tilde g_{\mu\nu}$).

\subsection{Point particles, punctures, and effective sources}\label{subsec:punctures}

As we described in Sec.~\ref{perturbation theory}, a traditional point-particle description fails at nonlinear orders. However, after the local analysis of the last section, we are now equipped to adopt the more general viewpoint of Sec.~\ref{point particles}: rather than approximating the object as a delta function stress-energy tensor, we replace it with a local singularity in the metric. We call this singularity a {\em puncture} in the spacetime. 

To understand how we use this idea, first begin with the vacuum Einstein equations outside the object. At first and second order, they have the form of Eqs.~\eqref{first-order EFE} and \eqref{second-order EFE} with the stress-energy terms set to zero,
\begin{align}
\delta G_{\mu\nu}[h^{(1)}] &= 0,\label{vacuum EFE1}\\
\delta G_{\mu\nu}[h^{(2)}] &= -\delta^2 G_{\mu\nu}[h^{(1)}].\label{vacuum EFE2}
\end{align}
The physically correct solutions to these equations satisfy a free boundary-value problem: in a small region around $r=0$, they must satisfy Eqs.~\eqref{h1}--\eqref{h2}, which we can think of as boundary conditions; and the location of that boundary region (or equivalently, of $r=0$) is free to move in response to the solution. 

We solve this free boundary value problem by taking the following steps. We first extend the locally derived fields $h^{(n)}_{\mu\nu}=h^{\S(n)}_{\mu\nu}+h^{\R(n)}_{\mu\nu}$ down to all $r>0$, replacing the physical field in the object's interior but leaving intact the physical field in its exterior. The fields $h^{(n)}_{\mu\nu}$ then satisfy Eqs.~\eqref{vacuum EFE1}--\eqref{vacuum EFE2} for all $r>0$. But they do not satisfy these equations on the domain $r\geq0$; and for $n>1$, they cannot (in general) be made to satisfy distributional equations on that domain. To obtain equations that can be solved on that domain, we next change our field variable from $h^{(n)}_{\mu\nu}$ to, essentially, $h^{\R (n)}_{\mu\nu}$. More precisely, we introduce a ``puncture field" $h^{\P (n)}_{\mu\nu}\approx h^{\S (n)}_{\mu\nu}$ by truncating the expansions~\eqref{h1} and \eqref{h2} at some finite order, and we then solve for the ``residual field" $h^{\res(n)}_{\mu\nu}:=h^{(n)}_{\mu\nu}-h^{\P(n)}_{\mu\nu}\approx h^{\R(n)}_{\mu\nu}$. This field satisfies the equations
\begin{align}
\delta G_{\mu\nu}[h^{\res (1)}] &= -\delta G_{\mu\nu}[h^{\P(1)}],\label{effective EFE1}\\
\delta G_{\mu\nu}[h^{\res (2)}] &= -\delta^2 G_{\mu\nu}[h^{(1)}] - \delta G_{\mu\nu}[h^{\P(2)}]\label{effective EFE2}
\end{align}
for all $r>0$. If $h^{\P (n)}_{\mu\nu}$ approximates $h^{\S (n)}_{\mu\nu}$ sufficiently well, then the sources in these equations (referred to as {\em effective} sources) are integrable, and we can define the equations for all $r\geq0$; note that here the sources are defined at $r=0$ nondistributionally, by taking the limit from $r>0$. We can also make $h^{\P(n)}_{\mu\nu}$ go to zero outside some region containing $r=0$, such that outside that region, $h^{\res (n)}_{\mu\nu}$ reduces to the physical field $h^{(n)}_{\mu\nu}$. 

With this procedure, we recover the physical field $h^{(n)}_{\mu\nu}=h^{\P(n)}_{\mu\nu}+h^{\res(n)}_{\mu\nu}$ outside the object, along with the effective field $h^{\res (n)}_{\mu\nu}$ in a region containing $r=0$. We can also say that we are recovering a local approximation to $h^{\R (n)}_{\mu\nu}$: if we make $h^{\P (n)}_{\mu\nu}$ approximate $h^{\S (n)}_{\mu\nu}$ to sufficiently high order in $r$, then we can make the Taylor expansions of $h^{\res(n)}_{\mu\nu}$ and $h^{\R(n)}_{\mu\nu}$ identical to any finite order in $r$. This allows us to {\em exactly} obtain $h^{\R(n)}_{\mu\nu}$ at $r=0$, along with any number of its derivatives there.

However, recall that in Sec.~\ref{perturbation theory}, we mentioned that at first order, the point-particle approximation {\em is} valid. To see how the approximation is justified, note that $\delta G_{\mu\nu}[h^{(1)}]$ is well defined as a distribution on $r\geq0$, since it is a linear operator acting on an integrable function. We can hence {\em define} the stress-energy $T^{(1)}_{\mu\nu}:=\frac{1}{8\pi}\delta G_{\mu\nu}[h^{(1)}]$, and a short calculation, beginning from Eq.~\eqref{h1}, establishes that this is precisely the stress-energy of a point mass. In other words, a vacuum perturbation $h^{(1)}_{\mu\nu}$ with the local form~\eqref{h1} is {\em identical} to the field sourced by a point mass sitting at $r=0$, a result first shown (to our knowledge) in Ref.~\cite{DEath:75} (see also \cite{Gralla:2008fg,Pound:10a}). Hence, at first  order we can rightly think of the object as a point particle. One can then solve the equation $\delta G_{\mu\nu}[h^{(1)}] = 8\pi T^{(1)}_{\mu\nu}$ directly for $h^{(1)}_{\mu\nu}$, and if necessary, afterward subtract $h^{\S(1)}_{\mu\nu}$ to recover $h^{\R(1)}_{\mu\nu}$. From the point-particle description, one can then also return to the puncture scheme, writing $\delta G_{\mu\nu}[h^{\res (1)}] = 8\pi T^{(1)}_{\mu\nu}-\delta G_{\mu\nu}[h^{\P(1)}]$. With $\delta G_{\mu\nu}[h^{\P(1)}]$ interpreted distributionally, unlike in Eq.~\eqref{effective EFE1}, it cancels the delta function in $8\pi T^{(1)}_{\mu\nu}$, leaving the same nondistributional remainder as if we start with Eq.~\eqref{effective EFE1}.

Section \ref{sec:computation} discusses more technical issues related to both the puncture scheme and the point-particle description. For the moment, we stress the physical picture: we have replaced the object with a singular puncture in spacetime. At linear order, but {\em not} at nonlinear orders, this is precisely equivalent to approximating the object as a classical point particle (i.e., as a delta function material source). The puncture, or the particle, moves on the curve $z^\mu$. We shall next consider its equation of motion.

\subsection{Generalized equivalence principle}\label{equation of motion}

Unlike the derivation of the local field outside the object, the derivation of the equation of motion is essentially different in GR than in Newtonian gravity. In the Newtonian case, the gravitational force is an independent postulate of the theory, separate from the field equation. We can use the field equation to obtain the field as a functional of, say, the body's center of mass $\vec z$ and its multipole moments, but this leaves both $\vec z$ and all the moments freely specifiable. To obtain an equation of motion for $\vec z$, we must then appeal to Newton's law of gravitation. To do this for an asymptotically small body, we must sum the forces on each mass element, from which we find that the center of mass satisfies $m\frac{d^2\vec z}{dt^2} = - \int \rho \vec\nabla\Phi dV$. Due to Newton's third law, all the internal forces cancel one another, leaving us with $m\frac{d^2\vec z}{dt^2} = - \int \rho \vec\nabla \Phi_{\rm ext} dV$. When expanded in powers of $\e$, at leading order this reads $m\frac{d^2 \vec z}{dt^2} = -m \vec\nabla\Phi_0 +O(\e^2)$, recovering the equation of motion of the point particle in Eq.~\eqref{Kepler}.\footnote{ For a more sophisticated treatment of the Newtonian problem, see the review in Ref.~\cite{Harte:2014wya}.} 

In GR, on the other hand, the equation of motion and field equation are inextricably linked. This complicates some derivations, because it means we cannot solve the field equation without simultaneously finding the equation of motion. However, it also enables a different method of deriving the equation of motion. Consider a generic material object with stress-energy $T^{\mu\nu}$. The object's motion is determined by the conservation law $\nabla_\nu T^{\mu\nu} = 0$. By virtue of the contracted Bianchi identity $\nabla_\nu G^{\mu\nu} = 0$, the motion is also equally determined by the Einstein field equation, $G^{\mu\nu} = 8\pi T^{\mu\nu}$. 
In fact, we can go a step further: as first pointed out by Weyl and Einstein~\cite{Weyl:21,Einstein-Grommer:27,Einstein-Infeld-Hoffmann:38,Einstein-Infeld:49}, because the motion is encoded in the field even {\em outside} the object, it can also be determined from the {\em vacuum} Einstein equations in the object's exterior. This means that we can obtain the equation from the same local analysis described in Sec.~\ref{local analysis}, without ever having to consider the object's interior.

Concretely, let us choose some representative worldline $z^\mu$. This can be {\em any} worldline near the object's center of mass, in the sense that it lies in the body zone (or equivalently, in the sense that the mass dipole moment $m_i$ is of order $\e^2$). The linearized Einstein equation~\eqref{vacuum EFE1} applied to the field~\eqref{h1} dictates that at zeroth order, for an arbitrarily structured compact object, the worldline must be a geodesic of the background spacetime $g_{\mu\nu}$~\cite{DEath:75, Gralla:2008fg}:
\beq\label{0th order eqn motion}
\frac{D^2 z^\mu}{d\tau^2} = O(\e).
\eeq
This is the standard equivalence principle, which states that any sufficiently small object moves on a geodesic of the external spacetime (i.e., it is in free fall in that spacetime).

Similarly, Eq.~\eqref{vacuum EFE2} applied to the fields \eqref{h1} and \eqref{h2} determines that at first order, the center of mass (defined by the condition $m_i\equiv0$) satisfies~\cite{Gralla:2008fg,Pound:10a}\footnote{ This result for a spinning object was first derived, in the case of a BH, in Ref.~\cite{Mino-Sasaki-Tanaka:97b} following Thorne and Hartle's approach to matched expansions~\cite{Thorne-Hartle:85}.}
\beq\label{1st order eqn motion}
\frac{D^2 z^\mu}{d\tau^2} = -\frac{1}{2}\e g^{\mu\nu}\left(2h^{\R(1)}_{\nu\rho;\sigma}-h^{\R(1)}_{\rho\sigma;\nu}\right)u^\rho u^\sigma - \frac{\e}{2m}R^\mu{}_{\alpha\beta\gamma}u^\alpha s^{\beta\gamma}+O(\e^2),
\eeq
where the spin tensor $s^{\beta\gamma}$ has components $s^{ij}=\epsilon^{ijk}s_k$ (in the comoving coordinate system introduced above; other components vanish). The first term on the right-hand side is the first-order GSF; the second is the Mathisson-Papapetrou spin force. One can rearrange this equation to equivalently write it in the more compelling form
\beq\label{EOM1 test body}
\frac{\tilde D^2 z^\mu}{d\tilde \tau^2} = -\frac{\e}{2m}\tilde R^\mu{}_{\alpha\beta\gamma}\tilde u^\alpha s^{\beta\gamma} + O(\e^2),
\eeq
where now the proper time $\tilde\tau$, four-velocity $\tilde u^\mu = dz^\mu/d\tilde\tau$, covariant derivative $\tilde D/d\tilde \tau=\tilde u^\mu\tilde\nabla_\mu$, and Riemann tensor are all defined with respect to the effective metric $\tilde g_{\mu\nu} = g_{\mu\nu}+\e h^{\R(1)}_{\mu\nu}+O(\e^2)$. Equation~\eqref{EOM1 test body} is the equation of motion of a (spinning) {\em test body} in the effective metric. This result is directly analogous to the electromagnetic case, in which the particle moved as a test particle in the effective field $\tilde A^\mu$.

In principle, the next-order equation of motion could be obtained from the third-order vacuum Einstein equation, in the same manner as the zeroth- and first-order equations of motion. In practice it has instead been derived~\cite{Pound:2012nt, Gralla:12} by solving the vacuum field equations in the {\em inner} expansion to sufficiently high order and then appropriately transforming to the outer expansion. As of the writing of this review, such a derivation has only been performed in the restricted case of an object with vanishing moments $s_i$, $\delta s_i$, $m_{ij}$, and $s_{ij}$. The equation of motion for such an object, as derived by one of us in Ref.~\cite{Pound:2012nt} (see~\cite{Pound:2017psq} for more details) is found to be
\beq\label{2nd order eqn motion}
\frac{D^2 z^\mu}{d\tau^2} = -\frac{1}{2}\left(g^{\mu\nu}-h_\R^{\mu\nu}\right)\left(2h^\R_{\nu\rho;\sigma}-h^\R_{\rho\sigma;\nu}\right)u^\rho u^\sigma +O(\e^3),
\eeq
where now $h^\R_{\mu\nu}=\e h^{\R(1)}_{\mu\nu}+\e^2 h^{\R(2)}_{\mu\nu}$. This can be rearranged to obtain the geodesic equation in the effective metric $\tilde g_{\mu\nu}=g_{\mu\nu}+h^\R_{\mu\nu}$~\cite{Pound:2012nt,Pound:2015fma},
\beq
\frac{\tilde D^2 z^\mu}{d\tilde \tau^2} = O(\e^3).
\eeq
Once again, we find that the object moves as a test body in the effective metric. This is a generalized equivalence principle: if we neglect finite-size effects (i.e., spin and higher moments), then any object, no matter its internal structure, falls freely with respect to what it perceives as the external metric. In the context of the puncture scheme described in the preceding section, it is the puncture that moves on this worldline, and in Eq.~\eqref{2nd order eqn motion} we replace $h^\R_{\mu\nu}$ with $h^\res_{\mu\nu}$. 

Here we have written all these results in terms of the particular, locally defined regular field introduced in Sec.~\ref{local analysis}. At first glance, it may seem that this regular field is quite remote from the Green's-function-based Detweiler-Whiting field introduced in the electromagnetic case, despite sharing the same key properties. However, the two definitions are intimately related. At linear order, we could have adopted Detweiler-Whiting definitions of $h^{\S(1)}_{\mu\nu}$ and $h^{\R(1)}_{\mu\nu}$ by beginning with the point-particle field equation~\eqref{first-order EFE}, writing the physical solution in terms of a retarded Green's function, and then adopting a split into S and R fields precisely analogous to Eqs.~\eqref{GS curved spacetime} and \eqref{GR curved spacetime}~\cite{Detweiler-Whiting:03}. Despite their seemingly very different definitions, these S and R fields precisely agree with the locally defined ones of Sec.~\ref{local analysis} at least through order $r^2$~\cite{Poisson:2011nh,Pound-Miller:14}. 

Nevertheless, one should keep in mind the warning we gave in Sec.~\ref{local analysis}: the split of the physical field into an R field and an S field is not unique. With the particular definitions used here, taken from Ref.~\cite{Pound:2012nt}, the effective external metric $\tilde g_{\mu\nu}$ at nonlinear orders has all the desirable, ``physical" properties of the Detweiler-Whiting R field: it is a smooth vacuum metric, it is causal when evaluated on the worldline~\cite{Pound:2015fma}, and at least through second order, the object moves as a test body in it. But other choices can be made that preserve all these properties~\cite{Pound-Miller:14}, and one should be wary of ascribing too much physical significance to $\tilde g_{\mu\nu}$. Furthermore, one can make choices that do not possess all the desirable properties of the Detweiler-Whiting split, but which are nonetheless perfectly practical. For example, in Ref.~\cite{Gralla:12} Gralla defines a singular-regular split that can be used in a puncture scheme at second order; but the regular field in his case is not a vacuum solution, and the equation of motion is not a geodesic in his effective metric. Similarly, in Ref.~\cite{Harte:12} Harte defines an effective metric in which the body moves as a test body even in the fully nonlinear regime (in the restricted case of a material body); but his effective metric is not a vacuum metric, and it is not suitable for use in a puncture scheme, as it would actually become singular at the worldline in the point-particle limit.

We also note that, aside from the choice of singular-regular split, there are several other differences between  various equations of motion that have been derived. At first order, the relation between them is thoroughly established, but at second order, few comparisons have been made. The results summarized in this and the preceding section were derived by one of us in Refs.~\cite{Pound:10a, Pound:2012nt, Pound:12b,Pound:2017psq}. Using a very similar method, an alternative form of the second-order equation was derived by Gralla~\cite{Gralla:12}. Though the relation between these results is well understood in broad strokes~\cite{Pound:2015fma,Pound:2017psq}, they have not yet been compared in detail. Section~\ref{sec:evolution} below comments on a principal difference between them. Another result was also derived much earlier, using a more axiomatic approach, by Rosenthal~\cite{Rosenthal:06b}, and Ref.~\cite{Detweiler:12} contains yet another perspective. A main difference between the Pound, Gralla, and Rosenthal results is the {\em gauge} (or class of gauges) in which they are derived. For example, Rosenthal's result was derived in a physically counterintuitive gauge in which the first-order GSF actually vanishes. We turn to this issue of gauge in the next section.

\subsection{Gauge freedom}\label{gauge}



In the context of perturbation theory, GR's gauge freedom corresponds to our ability to make an infinitesimal coordinate transformation, $x^\mu\to x^\mu +\e \xi^\mu+O(\e^2)$. If we expand such a transformation's effect on the metric $g_{\mu\nu}+\e h^{(1)}_{\mu\nu}+O(\e^2)$, we find that it induces a change $h^{(1)}_{\mu\nu}\to h^{(1)}_{\mu\nu}-\nabla_\mu \xi_\nu -\nabla_\nu\xi_\mu$; transformation laws at higher orders in $\e$ are easily derived.


To see how this freedom plays into the discussion in the preceding sections, we once again refer to the Newtonian case. More specifically, we return to the Kepler problem recalled at the beginning of Sec.~\ref{sec:EM self-force}, following an example due to Detweiler and Poisson~\cite{Detweiler:2003ci,Detweiler:2008ft}. Here we can clearly see the singular-regular split: the total gravitational potential of the system, $\Phi(\vec x) = -\frac{m}{|\vec x - \vec z|}-\frac{M}{|\vec x - \vec Z|}$, naturally divides into the S potential sourced by the small mass, 
\beq
\Phi^\S(\vec x) = -\frac{m}{|\vec x - \vec z|},
\eeq
and the regular, external potential sourced by the large mass,
\beq\label{Phi ext}
\Phi_{\rm ext}(\vec x) = - \frac{M}{|\vec x-\vec Z|}.
\eeq
As described by Eq.~\eqref{Kepler}, the motion of $m$ is entirely governed by $\Phi_{\rm ext}$. Nevertheless, even in this scenario we can find what might be called GSF effects---if by that we simply mean a force on $m$ that scales with $m^2$. Let us place the origin of our coordinates at the system's center of mass. This implies $\vec x_{\rm CM} = \frac{M\vec Z + m\vec z}{M+m} = 0$, meaning that the position of $M$ is given by $\vec Z = -\frac{m}{M}\vec z$. If we substitute this into the potential in Eq.~\eqref{Phi ext}, we find
\beq
\Phi_{\rm ext} = -\frac{M}{r} + \frac{m z^i n_i}{r^2} + O\left(\frac{M}{r}\frac{m^2}{M^2}\right)\!,
\eeq
where $r:=|\vec x|$ and $n^i:=x^i/r$. (Note that $r$ and $n^i$ here are not to be confused with the $r$ and $n^i$ of previous sections, which referred to a radial distance and radial unit vector centered on $z^i$.) In this form, the ``external potential'' that $m$ feels has contributions proportional to $m$. In analogy with the definitions in Sec.~\ref{local analysis}, we can idenfity $\Phi_0:=-\frac{M}{r}$ as the ``background'' field, $\Phi^\R_1 := \frac{z^i n_i}{r^2}$ as the ``first-order regular potential'', and 
\beq
F^i_1 = -m\partial^i\Phi^\R_1
\eeq
as the corresponding ``first-order GSF''.  

However, this split of the external field is coordinate dependent. If we had instead chosen the origin of our coordinates to lie at the large mass's position  $\vec Z$, then we would have had $\vec Z=0$. The external potential would then have been simply $\Phi_{\rm ext} = -\frac{M}{r}$. The ``higher-order'' pieces $\Phi^\R_n$ would all have been identically zero, and we would have had $F^i_1=0$. (Though, because the coordinate system in this case is accelerated, pseudo-forces do arise.)\footnote{ The equation of motion in this case is generally written in the effective-one-body form $\mu \frac{d^2z^i}{dt^2} = - \mu (M+m) n^i/r^2$, where $\mu=mM/(M+m)$ is the ``reduced mass" of the ``effective body" in orbit around an effective central mass $M+m$ at $r=0$. In this form, the pseudo-forces have been grouped with the acceleration on the left-hand side.} We can understand this as an example of the gauge freedom described above: if we start in the center-of-mass coordinates, we can transform to $M$-centered coordinates with the small translation $x^i\to x^i +\frac{m}{M} z^i$. This coordinate transformation entirely ``gauges away'' the regular potential $\Phi^\R_1$.

From this example we see that what appears as a perturbative piece of the external field entirely depends on our choice of gauge. The same applies in GR: with an appropriate choice of gauge, the regular field $h^\R_{\mu\nu}$ and its first derivative can be eliminated at all points along the worldline $z^\mu$, and the GSF along with them. Unlike in the Newtonian case, $h^\R_{\mu\nu}$  cannot be {\em entirely} eliminated; it does contain gauge-invariant information. Yet the freedom to alter (or even completely eliminate) the GSF may seem to complicate its interpretation. However, this does not pose a fundamental problem or invalidate the equation of motion~\eqref{2nd order eqn motion}. Consider again the Kepler problem. The Newtonian equation of motion~\eqref{Kepler} is valid in any inertial Cartesian coordinate system; it is invariant under Galilean transformations $x^i\to x'^i = A^i{}_j x^j + v^i t + \zeta^i$, where $A^i{}_j$ is a constant rotation matrix and $v^i$ and $\zeta^i$ are constants. Likewise, if we expand it in powers of $m/M$ as described above, the equation of motion 
\beq\label{Newton's law regular field}
m\frac{d^2z^i}{dt^2} = -m\partial^i \left[\Phi_{0} + \frac{m}{M} \Phi^\R_1 +O\left(\frac{m^2}{M^2}\right)\right]
\eeq
is invariant under infinitesimal Galilean transformations, $x^i\to x^i + (m/M)\xi^i$, where $\xi^i = \e^i{}_{jk}\phi^k x^j + v^i t + \zeta^i$ and $\phi^k$ is the axis of rotation. Under such a transformation, the external field is altered, with $\Phi_{\rm ext}=\Phi_0+(m/M)\Phi^\R_1+O(m^2/M^2)$ becoming $\Phi_0+(m/M)\Phi'^\R_1+O(m^2/M^2)$, where $\Phi'^{\R}_1=\Phi^\R_1 - (m/M)\xi^i\partial_i\Phi_0$ is the new regular field. But the left- and right-hand sides of Eq.~\eqref{Newton's law regular field} both transform in the same way, such that in the new frame, the equation takes the identical form in terms of $\Phi'^{\R}_1$. In the same way, the geodesic equation in the effective external metric, $\frac{\tilde D^2 z^\mu}{d\tilde\tau^2}=0$, is invariant under any smooth coordinate transformation, and its expansion~\eqref{2nd order eqn motion} is correspondingly valid in {\em all} gauges, at least if the transformation between them is smooth. For such transformations, $h^{\S(1)}_{\mu\nu}$ is invariant, while $h^{\R(1)}_{\mu\nu}$ transforms to $h^{\R(1)}_{\mu\nu}-\nabla_\mu \xi_\nu -\nabla_\nu\xi_\mu$, just as if the effective external metric were the complete physical metric. Extensions of these transformation laws to second order are discussed in Ref.~\cite{Pound:2015fma}.

Many of the foundational results in GSF theory were derived in a specific gauge called the Lorenz gauge, in which a ``trace-reversed'' variable $\bar h_{\mu\nu}:=h_{\mu\nu}-\frac{1}{2}g_{\mu\nu}g^{\alpha\beta}h_{\alpha\beta}$ satisfies the gauge condition $\nabla^\mu \bar h_{\mu\nu}=0$, in analogy with $\nabla^\mu A_\mu=0$. In this gauge, the linearized Einstein tensor takes a simple, hyperbolic form, 
\beq\label{dG Lorenz}
\delta G_{\mu\nu}[h] = -\frac{1}{2}\Box \bar h_{\mu\nu} - R_\mu{}^\alpha{}_\nu{}^\beta \bar h_{\alpha\beta}, 
\eeq
where $R_\mu{}^\alpha{}_\nu{}^\beta$ is the Riemann tensor of the background metric. This is the gauge in which the MiSaTaQuWa equation, the first-order Detweiler-Whiting decomposition, the local results in Sec.~\ref{local analysis}, and the second-order equation of motion~\eqref{2nd order eqn motion} were first derived. Among other advantages, it allows one to work with 4D Green's functions in a generic vacuum background spacetime, just as in the electromagnetic case. In terms of the retarded Green's function, the first-order equation of motion~\eqref{1st order eqn motion} (in the absence of spin) takes a form analogous to the electromagnetic equation of motion~\eqref{DeWitt-Brehme}, 
\beq\label{MiSaTaQuWa}
\frac{D^2 z^\mu}{d\tau^2} = -\frac{\e}{2}g^{\mu\nu}(2h^{\rm tail}_{\nu\rho\sigma}-h^{\rm tail}_{\rho\sigma\nu})u^\rho u^\sigma +O(\e^2),
\eeq
where the tail field is given by
\beq
h^{\rm tail}_{\alpha\beta\gamma} = m\int_{-\infty}^{\tau^-}\nabla_\gamma G^+_{\mu\nu\mu'\nu'}u^{\mu'}u^{\nu'}d\tau.
\eeq
This is the original form of the MiSaTaQuWa equation, prior to its reformulation by Detweiler and Whiting.

However, the Lorenz gauge is not always optimal for calculations in particular background spacetimes, such as the black-hole spacetime of an EMRI. Considerable work has gone into understanding the role of gauge in GSF theory~\cite{Barack:2001ph,Gralla:2008fg,Gralla:2011ke,Gralla:2011zr,Pound:2013faa,Pound:2015fma,Pound:2017psq}, and the local field and equation of motion have now been derived in a variety of gauges. Of particular interest are gauges that are either more or less singular than the Lorenz gauge; this takes us outside the class of smoothly related gauges discussed above, demanding fresh consideration of the singular-regular split with each new choice of gauge. Section~\ref{subsec:comp_techniques} below discusses how more singular gauges become useful in practical computations. More recently, Ref.~\cite{Pound:2017psq} has constructed a class of ``highly regular" gauges, which eliminate the most singular, $\sim 1/|x^a-z^a|^2$ term in $h^{(2)}_{\mu\nu}$. We briefly describe the implications of such a gauge choice in the conclusion. 



	
\section{Survey of computational methods}\label{sec:computation}

While the first-order equation of motion (\ref{MiSaTaQuWa}) had been known since 1997, recasting it in a form amenable to numerical computation, concretely for EMRI systems, took some more years of development. In this section we describe the principles of the two main computational frameworks that have been devised and implemented for actual (numerical) computation of the GSF and its effects in EMRI systems: the {\it mode-sum} and {\it puncture} methods. For simplicity we restrict the discussion here to the {\it first-order} GSF; we shall briefly cover recent progress on the second-order problem in the concluding section of this review. Our description here avoids technical detail as much as possible and does not assume familiarity with specific techniques in black-hole perturbation theory. For an expert-level review of GSF computation methods we refer readers to \cite{Barack:2009ux} or the more recent \cite{Wardell:2015kea}.

The starting point for any (first-order) GSF computation is the equation of motion (\ref{1st order eqn motion}) [or the original MiSaTaQuWa form (\ref{MiSaTaQuWa})]. Omitting the spin and $O(\epsilon^2)$ terms in that equation, we write it here in the more compact form 
\beq\label{1st order eqn motion schematic}
m\frac{D^2 z^\alpha}{d\tau^2} = \lim_{x\to z}m \nabla^{\alpha\beta\gamma} h_{\beta\gamma}^R(x) =:F^\alpha(z),
\eeq
where $\nabla^{\alpha\beta\gamma}$ represents the differential operator appearing on the right-hand side of (\ref{1st order eqn motion}), and we have omitted the superscript `$(1)$' in $h_{\beta\gamma}^{R(1)}$ for brevity. We further write $h_{\beta\gamma}^{R}$ as the difference between the full, physical (retarded) metric perturbation sourced by the particle, and the S field:
\begin{equation}\label{GSF_schematic}
F^\alpha(z)=\lim_{x\to z}m \nabla^{\alpha\beta\gamma} h_{\beta\gamma}^R(x)=\lim_{x\to z}m \nabla^{\alpha\beta\gamma} \left[h_{\beta\gamma}(x)-h_{\beta\gamma}^S(x)\right].
\end{equation}
 The argument `$x$' represents a field point in a neighbourhood of the particle's worldline, and `$z$' is the worldline point where the GSF is calculated. We suppose for the moment that the full perturbation $h_{\alpha\beta}$ satisfies the Lorenz-gauge form of the linearized Einstein field equations with a point-particle source, which we write here schematically as
\begin{equation}\label{EFE_schematic}
\tilde\Box h_{\alpha\beta}= S_{\alpha\beta}.
\end{equation}
The explicit form of the wave-like operator $\tilde\Box$ can be found in Eq.\ (\ref{dG Lorenz}) above, and $S_{\alpha\beta}$ is the linearized energy-momentum tensor associated with the particle, with support (a delta function) confined to the particle's worldline. Equation (\ref{EFE_schematic}) has to be solved with physical, ``retarded'' boundary conditions, which are most conveniently imposed at infinity and on the black hole's event horizon. The conditions are that no radiation should be coming in from (past null) infinity, and that no radiation should be coming out from inside the black hole.   

From a computational point of view, solving (\ref{EFE_schematic}) and evaluating (\ref{GSF_schematic}) pose two main technical difficulties. The first difficulty is that the Lorenz-gauge field equations (\ref{EFE_schematic}) constitute a complicated set of coupled partial differential equations (PDEs). Even though these equations are linear and manifestly hyperbolic, solving them numerically is computationally expensive and technically challenging (due, in particular, to the need to resolve the diverging field near the particle with sufficient accuracy, and due also to the occurrence of certain mode instabilities---see below). This problem has often been referred to in GSF literature as the ``gauge problem'': when it was first derived, the correct GSF was only known in the Lorenz gauge, meaning one had to calculate the metric perturbation in a gauge which, despite being convenient for describing the local singularity near the particle, does not sit very well with the global symmetries of the black hole background. The second technical difficulty is the ``subtraction problem'': to implement Eq.\ (\ref{GSF_schematic}) and obtain the GSF, one has to subtract one divergent quantity from another (which is usually given only numerically), before taking the regular limit to the particle. This is obviously problematic in practice. 

In the next two subsections we will describe two methods for tackling the subtraction problem. Then, in subsection \ref{subsec:comp_techniques}, we will see how the gauge problem has been overcome.   
 


\subsection{Mode-sum method}

The mode-sum method \cite{Barack:1999wf,Barack:2001bw} is a general procedure addressing the subtraction problem. The basic idea is very simple.  Instead of directly subtracting the divergent field $\nabla^{\alpha\beta\gamma} h^{S}_{\beta\gamma}$ from the divergent field $\nabla^{\alpha\beta\gamma} h_{\beta\gamma}$, first decompose each of these fields into multipolar-mode components (using a basis of angular harmonics defined on spheres around the {\it large} black hole), and then perform the subtraction at the level of individual modes. Finally, add up all ``regularized'' modal contributions. The benefit of such an approach is twofold: First, due to the particular Coulomb-like form of the singularity in the field near the particle, the individual multipole modes of $\nabla^{\alpha\beta\gamma} h^{S}_{\beta\gamma}$ (and of $\nabla^{\alpha\beta\gamma} h_{\beta\gamma}$) have {\em finite} values even at the location of the particle, so one only ever subtracts {\it finite} quantities in one's calculation. Second, the perturbation field $h_{\beta\gamma}$ is typically solved for mode-by-mode anyway, so the necessary input for the mode-sum procedure is readily available without any extra work. The mode-sum formula simply prescribes the ``correct'' quantity that has to be subtracted from each modal contribution, so that the resulting regularized modal contributions add up to give the correct, physical GSF.  

To describe this more precisely, let us define the fields $F_{\rm full}^{\alpha}(x):=m\nabla^{\alpha\beta\gamma}h_{\beta\gamma}$ and  $F_{\rm S}^{\alpha}(x):=m\nabla^{\alpha\beta\gamma}h^S_{\beta\gamma}$ in the (Kerr) black-hole geometry, and introduce the standard Boyer-Lindquist coordinates $(t,r,\theta,\varphi)$ covering the exterior of the black hole.\footnote{ The Boyer-Lindquist coordinates $(t,r,\theta,\varphi)$ have the standard intuitive meaning of spherical-polar coordinates (+time) in the weak-gravity region far away from the black hole. They coincide with the similarly labelled Schwarzschild coordinates in the limit of a vanishing black-hole spin.} Then let us consider the decomposition of $F_{\rm full}^{\alpha}$ and $F_{\rm S}^{\alpha}$ into spherical-harmonic modes defined on spheres of constant $t$ and $r$ around the black hole: $F_{\rm full}^{\alpha}=\sum_{\ell=0}^{\infty} F_{\rm full}^{\alpha\ell}$, where  $F_{\rm full}^{\alpha\ell}=\sum_{m=-\ell}^{\ell}F_{\ell m}(t,r)Y_{\ell m}(\theta,\varphi)$, with $Y_{\ell m}$ being the usual spherical harmonics, and similarly for $F_{\rm S}^{\alpha}$. Equation (\ref{GSF_schematic}) thus becomes 
\begin{equation}\label{modesum1}
F^\alpha(z)=\lim_{x\to z} \sum_{\ell=0}^{\infty}\left[F_{\rm full}^{\alpha\ell}(x)-F_{\rm S}^{\alpha\ell}(x)\right].
\end{equation}
The individual mode sums of $F_{\rm full}^{\alpha\ell}$ and $F_{\rm S}^{\alpha\ell}$ both diverge at the particle: in the multipolar space, the Coulomb-like particle singularity has turned into a large-$\ell$ (``ultraviolet'') divergence. However, the mode sum of the difference $F_{\rm full}^{\alpha\ell}(x)-F_{\rm S}^{\alpha\ell}$ converges faster than any power of $1/\ell$ everywhere, even at the particle, since $h_{\beta\gamma}^R(x)$ in Eq.\ (\ref{GSF_schematic}) is a smooth field. This implies that $F_{\rm full}^{\alpha\ell}$ and $F_{\rm S}^{\alpha\ell}$ share the same ultraviolet singularity structure. In fact, based on the detailed form of the singular field, one can show \cite{Barack:1999wf,Barack:2001bw} $F_{\rm S}^{\alpha\ell}(z)\sim A^\alpha \ell +B^\alpha + C^\alpha/\ell +\cdots$ at large $\ell$, where $A^\alpha$, $B^{\alpha}$ and $C^{\alpha}$ are $\ell$-independent expansion coefficients that encode the ultraviolet structure. We can thus write 
\begin{eqnarray}\label{modesum2}
F^\alpha(z)&=&\sum_{\ell=0}^{\infty}\left[F_{\rm full}^{\alpha\ell}(z)-A^\alpha \ell +B^\alpha + C^\alpha/\ell\right]
-D^\alpha ,
\nonumber\\
D^\alpha&:=& \sum_{\ell=0}^{\infty}\left[F_S^{\alpha\ell}(z)-A^\alpha \ell +B^\alpha + C^\alpha/\ell\right],
\end{eqnarray}
where the two individual sums are convergent (at least as $\sim\ell^{-1}$). For reasons that will become clear in Sec.~\ref{sec:evolution}, the GSF has historically been calculated by approximating the source orbit as a geodesic, neglecting the acceleration caused by the force. To that end, the explicit values of the so-called ``regularization parameters'',  $A^\alpha$, $B^\alpha$, $C^\alpha$ and $D^\alpha$, have been derived analytically for arbitrary geodesic orbits in Kerr spacetime \cite{Barack:2002mh,Barack:2009ux}.  In particular, it has been shown that when the acceleration is neglected, $C^\alpha$ and $D^\alpha$ always vanish identically (even in a broad class of non-Lorenz gauges~\cite{Gralla:2011zr}). One thus arrives at the final, working form of the mode-sum formula:  
\begin{equation}\label{modesum3}
 F^\alpha(z)=\sum_{\ell=0}^{\infty}\left[F_{\rm full}^{\alpha\ell}(z)-A^\alpha \ell +B^\alpha \right].
\end{equation}
This provides a practical means of evaluating the GSF at any point along a given (geodesic) orbit: First obtain the multipole modes of the physical field $h_{\alpha\beta}$ by solving a suitable version of the linearized Einstein equations mode by mode (this is usually done numerically). From each mode then subtract the analytically given quantity $A^\alpha \ell +B^\alpha$, and finally add up all the modal contributions.

The above schematic description suppresses some important detail. For instance, the values $F_{\rm full}^{\alpha\ell}(z)$ and $F_S^{\alpha\ell}(z)$ usually depend on the direction from which the limit $x\to z$ is taken (and so does the value of the parameter $A^\alpha$), so in the above procedure one is required to specify a particular directional limit. (Of course, the value of the GSF itself is independent of the choice of direction.) Also, the multipole modes into which the field $h_{\alpha\beta}$ is decomposed are not the 
usual spherical-harmonic modes, but (for instance, in the case of a Schwarzschild background) a suitable tensorial generalization thereof. So, obtaining the modal contributions $F_{\rm full}^{\alpha\ell}$ involves the additional step of re-expanding the tensorial modes in a basis of spherical harmonics. There is, in fact, a variant of the mode-sum procedure that employs a fully tensorial mode decomposition (with suitable regularization parameters), avoiding that extra complication \cite{Wardell:2015ada}. There is also a generalization of the mode-sum formula to accelerated (non-geodesic) orbits, allowing one to account for the particle's self-acceleration \cite{Heffernan:2017cad}. Finally, we mention that the mode sum in (\ref{modesum3}) converges only slowly: the summand typically falls off only as $\sim \ell^{-2}$. This means one normally has to compute a large number of $\ell$-modes ($\ell_{\rm max}\sim 50$ is typical), which can become computationally expensive. This problem can be mitigated through the use of ``high-order regularization parameters'', which analytically capture some of the higher-order terms  in the $1/\ell$ expansion of $F_S^{\alpha\ell}(z)$ \cite{Heffernan:2012su,Heffernan:2012vj}; these higher-order terms can be constructed such that they sum to zero, but including them accelerates the rate of convergence.  For more details, refer to \cite{Barack:2009ux,Wardell:2015kea}.

The mode-sum scheme, in its many variants, has been the primary framework for GSF calculations, and it is responsible for many of the results to be presented in later sections. In particular, it was employed in the recent first calculation of the GSF for generic bound geodesic orbits around a Kerr black hole \cite{vandeMeent:2016pee,vandeMeent:2017bcc}.

\subsection{Puncture (or ``effective source'') method}

In Sec.~\ref{subsec:punctures} we introduced the puncture method as a means of circumventing the nonintegrability of the second-order source, and a minimal version of the method was proposed for that purpose at least as far back as Ref.~\cite{Rosenthal:05}. However, the puncture method was first fully developed and implemented~\cite{Barack:2007jh,Barack:2007we,Vega:2007mc} at first order, with a different motivation: tackling problems where a multipole decomposition, and hence mode-sum regularization, is awkward. This is the case with the Lorenz-gauge metric perturbation on a Kerr background, for which no natural basis of harmonics is known. 

As described in Sec.~\ref{subsec:punctures}, the puncture method addresses the subtraction problem differently from mode-sum. Here, the ``regularization'' is performed already at the level of the field equation (\ref{EFE_schematic}). Instead of solving for the physical field $h_{\alpha\beta}$ and then subtracting the singular field, one solves directly for a local approximation to the regular field $h^\R_{\alpha\beta}$. Specifically, we design an analytic function $h_{\alpha\beta}^{\P}(x)$ that approximates the singular field $h_{\alpha\beta}^{\S}(x)$ near the particle sufficiently well that
\begin{equation}
\lim_{x\to z}(h_{\alpha\beta}^{\cal P}-h_{\alpha\beta}^{\S})=0\quad {\rm and\ also}\quad
\lim_{x\to z}(\nabla^{\alpha\beta\gamma} h^{\cal P}_{\beta\gamma}-\nabla^{\alpha\beta\gamma} h^{\S}_{\beta\gamma})=0.
\end{equation}
Then it is perfectly allowable to replace the true singular field in Eq.\ (\ref{GSF_schematic}) with its puncture-field approximant:
 \begin{equation}\label{GSF_res}
F^\alpha(z)=\lim_{x\to z}m \nabla^{\alpha\beta\gamma} h_{\beta\gamma}^{\cal R}(x),
\end{equation}
where we have introduced the residual field $h_{\beta\gamma}^{\res}:=h_{\beta\gamma}-h_{\beta\gamma}^{\P}$. We then make $h_{\beta\gamma}^{\cal R}$ the subject of the field equation (\ref{EFE_schematic}):
\begin{equation}\label{EFE_eff}
\tilde\Box h^{\cal R}_{\alpha\beta}= S_{\alpha\beta}-\tilde\Box h^{\P}_{\alpha\beta}=:S^{\rm eff}_{\alpha\beta}.
\end{equation}
The effective source $S^{\rm eff}_{\alpha\beta}$ contains no delta function on the particle's worldline; its residual non-smoothness there is determined by how well our puncture field approximates the singular field. The field  $h^{\cal R}_{\alpha\beta}$ is at least once differentiable (unlike the physical field $h_{\alpha\beta}$, which is divergent), and it directly yields the GSF, via Eq.\ (\ref{GSF_res}).

In practice, it is convenient to restrict the support of $S^{\rm eff}_{\alpha\beta}$ in Eq.\ (\ref{EFE_eff}) to within a small region around the particle's worldline, so as to avoid having to control the behavior of the puncture field and effective source far from the particle. This can be achieved with a suitable window function, or by introducing a ``worldtube'' around the worldline, such that one solves for $h^{\cal R}_{\alpha\beta}$ inside the worldtube and for the original perturbation $h_{\alpha\beta}$ outside it, with the analytically known value of the puncture field used to communicate between the two variables across the boundary of the tube. This scheme can be implemented numerically without any multipole decomposition, directly evolving the hyperbolic PDE (\ref{EFE_eff}) from initial conditions in 3+1 dimensions (space+time) \cite{Vega:2009qb}.  However, the method has also proved useful when applied in conjunction with a mode decomposition. One can separate the field into azimuthal modes, $\sim e^{im\varphi}$, and evolve each of the $m$-modes separately in 2+1 dimensions \cite{Dolan:2010mt,Dolan:2011dx,Dolan:2012jg,Thornburg:2016msc}; this is possible and useful even on a Kerr background, thanks to its axial symmetry. When a full multipole expansion is possible, such as on a Schwarzschild background, the puncture scheme can be implemented in 1+1 dimensions, multipole mode by multipole mode \cite{Vega:2007mc}. It can even be applied in the frequency domain \cite{Warburton:2013lea,Wardell:2015ada}. 

The utility and significance of the puncture idea becomes fully manifest when coming to solve the second-order field equation (\ref{second-order EFE}). Here, applying the mode-sum method becomes impossible in general. 
Recall that the source term appearing in the second-order equation (\ref{second-order EFE}) is sufficiently singular that the equation does not actually admit a globally valid solution. But even restricting to $x^\mu\neq z^\mu$, the singularity in the second-order solution, described by Eq. (\ref{h2}), is strong enough (a consequence of the distributionally ill-defined source) that its individual $\ell$ modes diverge at the particle.
This means that even if one were given the modes of the retarded field, one could not apply mode-sum regularization to extract the regular field. For these reasons, the puncture idea takes on a more fundamental status in second-order GSF calculations. 
The basic idea of the puncture can also be applied to control the behavior of the second-order solution near the horizon and at large distances, in circumstances where one would otherwise encounter infrared-type divergences~\cite{Pound:2015wva}.

\subsection{Alternative choices of gauge}\label{subsec:comp_techniques}

The application of both the mode-sum and puncture methods involves, in some form, solving linear field equations for the metric perturbation (or for its multipole modes), which must usually be done numerically. In the above discussion we have referred specifically to the Lorenz-gauge form of these equations, (\ref{EFE_schematic}) or (\ref{EFE_eff}). This form is convenient for a number of reasons: First and foremost, it yields the perturbation field in a gauge consistent with that assumed in the original GSF formulation, and so ready to be used directly in calculations. Related to this, the singularity of $h^{\S(1)}_{\mu\nu}$ in the Lorenz gauge has an intuitive, Coulomb-like form. Finally, the field equations themselves are hyperbolic and form part of a mathematically well-posed initial-value problem.  Indeed, much of the initial progress in GSF calculations came from direct Lorenz-gauge implementations \cite{Barack:2005nr,Barack:2007tm,Barack:2010tm}, and the approach is still a popular one, in conjunction with either mode-sum regularization \cite{Osburn:2014hoa} or a puncture \cite{Dolan:2012jg,Wardell:2015ada}.  

However, the direct Lorenz-gauge approach has several serious weaknesses. Prime among these is the fact that the Lorenz-gauge field equations (\ref{EFE_schematic}) cannot be decomposed into individual, decoupled multipole modes on a Kerr background, in any known form. This restricts one to time-domain numerical evolutions in 3+1 or 2+1 dimensions, which are computationally expensive and cumbersome.\footnote{ A Fourier expansion in both $t$ and $\varphi$ would reduce the problem to that of solving an elliptic set in two spatial dimensions, but this approach has not been explored yet.} Second, even on a Schwarzschild background where the equations are separable into (tensorial-type) spherical harmonics, they still constitute a complicated set of 10 equations that couple between the various tensorial components. Third, time-domain evolutions of the Lorenz-gauge equations appear to suffer from linear instabilities associated with certain nonphysical gauge modes \cite{Dolan:2012jg}, whose removal still constitutes an open problem. 

The above complications have motivated the development of methods of calculating the GSF in alternative gauges, facilitated by theoretical work to extend the GSF formalism beyond the Lorenz gauge~\cite{Barack:2001ph,Gralla:2008fg,Gralla:2011ke,Gralla:2011zr,Gralla:12,Pound:2013faa,Pound:2015fma,Pound:2017psq}. These developments have focused on the most traditionally useful gauges in black hole perturbation theory: the Regge-Wheeler-Zerilli gauge~\cite{Regge-Wheeler,Zerilli} (or closely related ``EZ" gauge~\cite{Thompson:2016}) in the Schwarzschild case; and the so-called ``radiation gauge" in the Kerr case~\cite{Chrzanowski:1975wv}. In these gauges, the first-order metric perturbation can be obtained (or ``reconstructed") from one or more scalar quantities that satisfy fully separable field equations, reducing the numerical calculation of $h_{\alpha\beta}$ to solving a set of ordinary differential equations. For example, in the case of the radiation gauge, the scalar quantity is the linear perturbation of one of the Newman-Penrose curvature scalars---either $\Psi_0$ or $\Psi_4$---constructed from the Riemann tensor; the separable field equation is then the well-known Teukolsky equation.

Unfortunately, as alluded to in Sec.~\ref{gauge}, these alternative gauges become poorly behaved in the presence of a point-particle source, introducing pathological singularities into the metric perturbation. These singularities violate the generic form of the local field described in Sec.~\ref{local analysis}. However, by analyzing the local form of the transformation to the Lorenz gauge, Ref.~\cite{Pound:2013faa} showed that nevertheless, the GSF and related quantities can be rigorously calculated from a mode-sum formula in the radiation gauge. Along with subsequent followup work~\cite{Shah:2015nva,Merlin:2016boc,vandeMeent:2017fqk}, this result has effectively resolved the ``gauge problem''.\footnote{ Although no comparable result has been derived in the Regge-Wheeler-Zerilli gauge, current work is expected to soon rectify that~\cite{Thompson}. The usability of this gauge is, anyhow, restricted to the Schwarzschild case.}


Combined with significant development of practical computational methods, these results have enabled vital progress over the past decade, particularly in the Kerr case, where Lorenz-gauge calculations are most cumbersome~\cite{Keidl:2006wk,Keidl:2010pm,Shah:2010bi,Shah:2012gu,Merlin:2014qda,Osburn:2015duj,vandeMeent:2015lxa}. This programme has recently culminated in a calculation of the GSF for generic (eccentric and inclined) bound geodesic orbits, by van de Meent \cite{vandeMeent:2016pee,vandeMeent:2017bcc}. Figures \ref{fig:GSFloops} and \ref{fig:GSFloops2} in Sec.\ \ref{sec:dissipative} below show sample outputs from this calculation, which is an implementation of the radiation-gauge approach.

In both the Regge-Wheeler-Zerilli and radiation gauges, such calculations have been further facilitated by the semi-analytical method of Mano, Suzuki, and Takasugi \cite{Mano:1996vt}, in which solutions to the relevant scalar-like field equations are expressed as infinite sums of hypergeometric functions, involving certain coefficients that are determined numerically. This approach is now enabling very high-precision calculations of the GSF in Kerr spacetime \cite{Shah:2013uya,Johnson-McDaniel:2015vva}. It even makes entirely analytical treatments possible within a post-Newtonian framework, where the equations are solved, and the GSF computed, order by order in a small-frequency expansion. Such analytical calculations have been performed to very high post-Newtonian order~\cite{Bini:2013rfa,Bini:2014nfa,Bini:2015bla,Bini:2015mza,Kavanagh:2015lva,Hopper:2015icj,Kavanagh:2016idg,Bini:2016dvs}.

 

\section{Perturbative approaches to orbital evolution in EMRIs}\label{sec:evolution}

A calculation of the local GSF acting on the small object in an EMRI system is only a first step (crucial as it is) in the programme to model the long-term orbital dynamics and emitted gravitational waves. One must also devise a method that uses the GSF information to construct a sufficiently accurate description of the evolving orbit and emitted radiation. In this section we review the formulation of such a method, based on a systematic perturbative expansion that exploits the adiabatic nature of the inspiral process.

But before we do so, we must review some of the pertinent properties of bound {\it geodesic} orbits in Kerr geometry. The next subsection summarizes the essentials of ``orbital mechanics'' of test particles in Kerr, and in Sec.\ \ref{subsec:evolution} we will move on to discuss the problem of evolution under the effect of the GSF. 

\subsection{Geodesic orbits around a Kerr black hole}\label{sec:orbits}
	
In Keplerian mechanics, the motion of a test mass in a spherically symmetric stationary gravitational potential is very simple. The orbit has a conserved total energy $E$ (kinetic+potential) associated with the time-translation invariance of the background potential, and a conserved orbital angular momentum ${\mathbf L}$ associated with its rotational invariance. The conservation of ${\mathbf L}$ leads to a planar motion, and the particular inverse-distance form of the potential guarantees that bound orbits are exactly periodic. The situation is quite similar in GR when one considers the motion of a test (pointlike, non-spinning) mass in the spherically symmetric and stationary gravitational potential of a Schwarzschild black hole: again, the symmetries of the background potential enable one to identify a conserved total energy $E$\footnote{ In the relativistic treatment, it is customary to include the particle's ``rest mass'', $m c^2$, in $E$, so that $E$ now represents the sum of kinetic, potential, and rest-mass energies.} and a conserved angular momentum ${\mathbf L}$, and the latter's existence means orbits are planar. However, in GR, bound orbits are no longer exactly periodic: the deviation of the potential away from the inverse-distance law produces the well-known relativistic precession effect, whose observational imprint on Mercury's orbit (the ``anomalous'' perihelion advance) was famously used in validating the infant GR theory. Instead, bound geodesic orbits in a Schwarzschild potential exhibit a combination of two periodic motions. Each orbit has an epicyclic period $T_r$, equal to the time between two successive periapsis passages, and a ``rotational'' period $T_\varphi$ associated with the mean azimuthal motion. While in the Keplerian theory $T_r=T_\varphi$, in GR it is always the case that $T_r>T_\varphi$, corresponding to an advance of the periapsis. 

\begin{figure}[htb]
	\begin{center}
        \includegraphics[width=50mm]{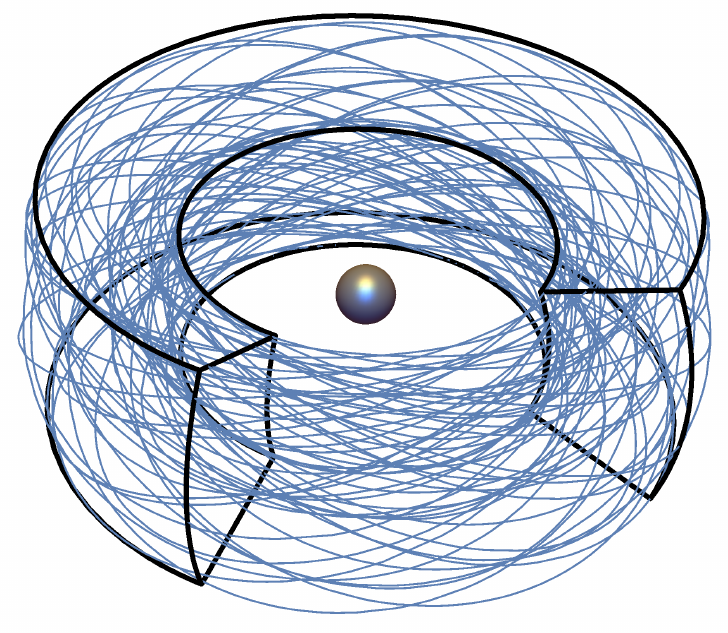}\quad\quad 
        \includegraphics[width=50mm]{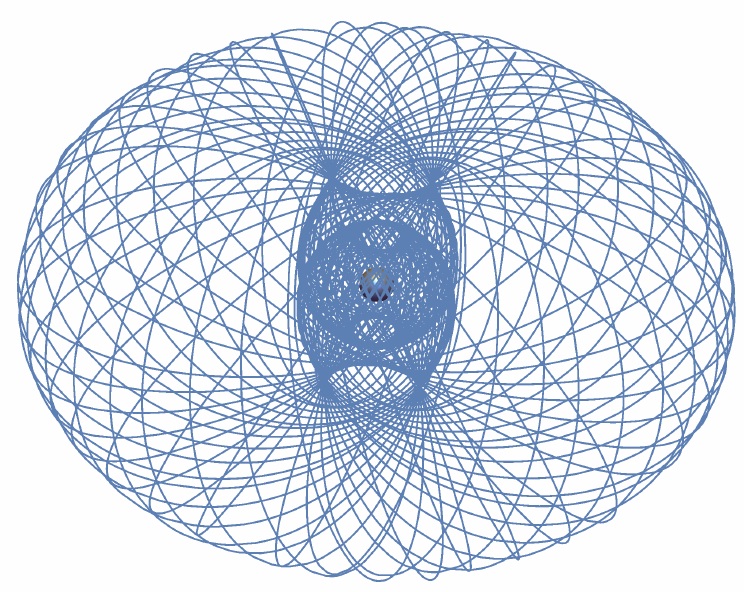}
        \includegraphics[width=20mm]{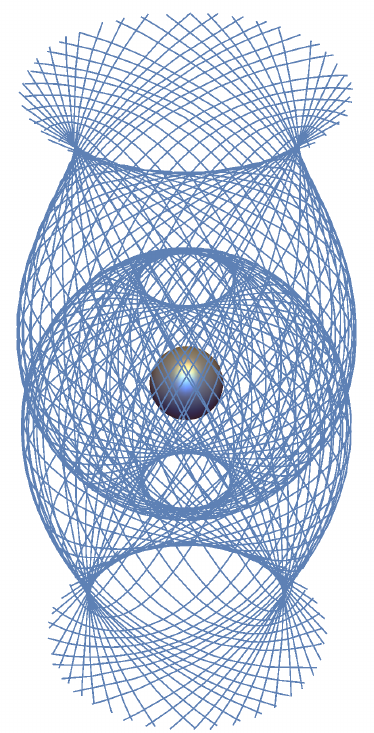}
\caption{{\it Left:} Typical geodesic orbit around a Kerr black hole. The orbit is 3-periodic, and it ergodically fills the interior of the outlined torus-shaped region.  
{\it Right:} The special case of a resonant orbit. Here the radial and longitudinal periods are in a $3\, {:}\, 2$ ratio, and the motion is no longer ergodic. Note how the orbit is instead confined to a certain 2-dimensional surface (topologically, a self-intersecting cylinder).  The rightmost figure expands the central region of the resonant orbit, for clarity.}
        \label{fig:orbit}
	\end{center}
\end{figure}

The situation is more involved when the central black hole is spinning. The gravitational potential of a Kerr black hole is still stationary, so orbital energy is still conserved. However, the potential is no longer spherically symmetric, so the total orbital angular momentum ${\mathbf L}$ is not conserved and orbits are generally non-planar. Nonetheless, the potential remains {\it axially} symmetric, meaning the projection $L_z$ of ${\mathbf L}$ along the symmetry axis (the direction of the black hole's spin) {\it is} conserved. As a result, the orbital plane performs a simple-precession motion about the direction of the black hole's spin. This is an example of the Lense-Thirring effect associated with coupling between spin and orbital angular momentum in GR. Bound geodesic orbits around a Kerr black hole are thus (generically) {\it tri}-periodic: in addition to $T_{r}$ and $T_{\varphi}$, they posses a third, ``longitudinal'' period $T_z$, equal to the time interval between two successive minima of the object's longitudinal angle. The combination of two precessional (libration-type) motions traces out a complicated trajectory, which is generically {\it ergodic} (space-filling): loosely speaking, each point within a certain torus-shaped volume around the black hole eventually gets visited by the particle. Figure \ref{fig:orbit}  displays a sample geodesic orbit, illustrating this behavior (left orbit in the figure; the special behavior of the orbit on the right will be discussed further below).

The conserved quantities $E$ and $L_z$ constitute first integrals of the geodesic equation of motion. The particle's conserved mass, $m$, is a third such integral. Not long after the discovery of the Kerr metric, B. Carter identified a {\it fourth} integral, $Q$, now known as the Carter constant \cite{Carter:1968rr}. $Q$ is associated with a more subtle symmetry of the Kerr geometry. It does not have a simple physical interpretation or a Newtonian analogue, except in the weak-field or Schwarzschild limits, where it roughly corresponds to the square of the ``remaining'' component of angular momentum, $L_x^2+L_y^2$. 
Orbits that are initially equatorial remain equatorial (due to the symmetry of the Kerr geometry under reflection across the ``equatorial'' plane, i.e.~the plane orthogonal to the black hole's spin direction) and have $Q=0$. The trio of constants $\{E,\,L_z,\,Q\}$ completely and uniquely parametrizes all geodesic orbits in Kerr spacetime, up to initial phases.\footnote{ However, curiously, it turns out that the trio of periods $\{T_r,T_\varphi,T_\theta\}$ is {\it not} a good parametrization: there are (infinitely many) pairs of physically distinct orbits that exhibit the same three periods \cite{Warburton:2013yj}.}

The existence of four first integrals of motion---the trio $\{E,L_z,Q\}$, in addition to $m$---allows us to write the geodesic equations of motion in a convenient first-order form. Moreover, as first noted by Carter \cite{Carter:1968rr}, one can choose the ``time'' parameter along the orbit so that the radial and longitudinal libration motions become manifestly decoupled from one another. Let  $x_{\rm p}^{\alpha}(\lambda)=\{t_{\rm p}(\lambda),r_{\rm p}(\lambda),\theta_{\rm p}(\lambda),\varphi_{\rm p}(\lambda)\}$ represent the particle's geodesic trajectory in Boyer-Lindquist coordinates,
where $\lambda$ is Carter's time parameter.\footnote{ $\lambda$ is often referred to as ``Mino time'' in more recent literature, after Y. Mino, who emphasized its utility in the context of EMRI physics \cite{Mino:2003yg}.} The equations of geodesic motion in Kerr geometry take the remarkably simple form
\begin{equation}\label{EOM_BL}
\begin{aligned}
\frac{dr_{\rm p}}{d\lambda}&=\pm\sqrt{R(r_{\rm p})}, \\
\frac{d\theta_{\rm p}}{d\lambda}&=\pm \sqrt{\Theta(\cos\theta_{\rm p})}, \\
\frac{d\varphi_{\rm p}}{d\lambda}&=\Phi_r(r_{\rm p})+\Phi_\theta(\theta_{\rm p}), \\
\frac{d{t}_{\rm p}}{d\lambda}&={\cal T}_r(r_{\rm p})+{\cal T}_\theta(\theta_{\rm p}),
\end{aligned}
\end{equation}
where each of the right-hand-side functions depends only on its indicated argument (as well as on the constant parameters $E/m$, $L/m$ and $Q/m$, and on the black hole's mass and spin); the explicit form of these functions can be found in Carter's paper, Ref.\ \cite{Carter:1968rr}. Given initial conditions [say, $x_{\rm p}^{\alpha}(0)$], the radial and longitudinal motions can be immediately (and independently) determined from the first two of these equations. Then, supplied with $r_{\rm p}(\lambda)$ and $\theta_{\rm p}(\lambda)$, one can solve for the azimuthal motion using the third equation. The fourth of the equations in (\ref{EOM_BL}) relates the parameter $\lambda$ to the standard time coordinate $t$. 

Geodesic motion in Kerr is thus manifestly {\it integrable}, to use the language of dynamical systems. As such, it can be conveniently formulated in terms of action-angle variables. This will prove useful when we later discuss the effect of the self-force. We let $J_{\alpha}=\{E,L,Q,m\}$ be our action variables,\footnote{ Strictly speaking, the action variables are certain integrals over the phase-space torus. But those integrals are (invertible) functions of $\{E,L,Q,m\}$, and so we allow ourselves to refer to $\{E,L,Q,m\}$ themselves as the action variables.} and $q^{\alpha}=\{q^t,q^r,q^\theta,q^\varphi\}$ be our generalized angle variables, associated with the $t$, $r$, $\theta$ and $\varphi$ motions. The equations of geodesic motion in Kerr then take the form 
\begin{equation}\label{EOM_action-angle}
\dot{J}_{\alpha}=0, \quad\quad
\dot{q}^{\alpha}=\omega^{\alpha}(J_{\mu}),
\end{equation}
The four quantities $\omega^{\alpha}$ are generalized ``frequencies'' associated with $q^{\alpha}$; their relationships with the physical (Boyer-Lindquist) frequencies, for bound orbits, were derived by Schmidt \cite{Schmidt:2002qk}. The parameters $J_{\alpha}$ are the orbit's {\it principal} elements. They describe the ``shape'' of the orbit and determine physical attributes such as the orbital eccentricity and semimajor axis. The parameters $q^{\alpha}$ are the orbit's {\it positional} elements. They contain the orbit's phase information and determine physical attributes such as the (time-dependent) direction of the periapsis and orientation of the orbital plane. In Eq.\ (\ref{EOM_action-angle}) we allow the overdots to denote differentiation with respect to any suitable parameter along the orbit---like $\lambda$, $t$, or proper time, with suitable redefinitions of $\omega^{\alpha}$ and $q^\alpha$; we will make a particular choice of parameter later in our discussion.

In general, the radial and longitudinal libration motions are incommensurate, leading to ergodic behavior, as mentioned above. However, there is a special class of geodesic orbits for which $\omega^r/\omega^\theta$ is a rational number. Such ``resonant'' orbits are non-ergodic and precisely periodic: the orbit completes a certain integer number of radial cycles at the same time  it completes a certain integer number of longitudinal cycles. Figure\ \ref{fig:orbit} above shows an example. The unusual periodic nature of resonant orbits manifests itself more profoundly the smaller those integers are; resonances with small integers are sometimes called ``strong''. During the gradual radiative inspiral of an EMRI system, the orbit will become momentarily tangent to  numerous such resonances, including, generically, ones that are strong \cite{vandeMeent:2013sza}. Such resonant crossings have interesting dynamical consequences, mentioned further below.

A distinctive property of orbits around black holes, of no analogue in Keplerian mechanics, is the presence of an {\it innermost stable orbit}. In the Keplerian problem, circular orbits (for example) exist at arbitrarily small radii around the center of attraction: the gravitational pull can always be countered with a sufficiently strong centrifugal force by endowing the particle with a sufficiently large tangential velocity. Moreover, all such orbits are dynamically {\it stable}, in the sense that a small perturbation applied to the orbit  (e.g., one that takes it away from being circular) remains small over time. Contrast this with the situation in GR for, e.g., a Schwarzschild black hole. There, no physically allowable amount of angular momentum can support circular geodesic motion below $r=3M$ (where $M$ is the black hole's mass, with its event horizon at $r=2M$), and the innermost {\em stable} circular orbit (ISCO) is as far out as $r=6M$. In fact, all stable bound geodesic orbits around a Schwarzschild black hole, even highly eccentric ones, reach their perapsis above $r=4M$. For rotating black holes, the ISCO location depends on the amount of spin. A higher spin rate gives rise to a smaller ISCO radius for objects that co-rotate with the black hole, and to a larger ISCO radius for counter-rotation motion.


\subsection{Self-consistent and two-timescale descriptions of the long-term orbital evolution}\label{subsec:evolution}


Though the geodesic dynamics is complex, on the face of it, the long-term inspiral seems fairly simple. We know that the GSF has a dissipative effect, causing the small object to spiral into the central black hole. Because the GSF is small, we also know that this process must be slow; over a single radial period, for example, the worldline traced out by the object must be very nearly a geodesic of the background spacetime. Hence, we can say that the inspiral is approximately {\em adiabatic}, slowly evolving through a sequence of background geodesics. Occasionally this evolution will pass through one of the resonances described in the previous section, but for the most part, the evolution is apparently innocuous.

However, despite the simplicity of this physical description, there is surprising difficulty in obtaining an accurate mathematical one. The challenge arises precisely from what makes the physics deceptively simple: the slowness of the inspiral. If we  continue to think of the inspiral as an evolution through a sequence of geodesics, we can describe it as a slow change of the ``constants'' of motion, $E$, $L_z$, and $Q$. They change at the rate ${\dot E}/{E}\propto m/M^2$, introducing the large time scale $\sim M^2/m$ into the system. This time scale is traditionally called the {\em radiation-reaction time}, denoted $t_{\rm rr}$. Just as the object's small size $\sim m\ll M$ led to a failure of ordinary perturbation theory (motivating the use of matched expansions), the presence of the large scale $t_{\rm rr}$ does likewise. To elucidate this, let us write Eq.~\eqref{g expansion} in the more explicit form of an ordinary Taylor series, 
\beq\label{regular expansion}
{\sf g}_{\mu\nu}(x,\eta) = g_{\mu\nu}(x) + \eta h^{(1)}_{\mu\nu}(x) + \eta^2 h^{(2)}_{\mu\nu}(x) + O(\eta^3), 
\eeq
where $x^\mu$ is any suitable set of background coordinates, such as the Boyer-Lindquist coordinates associated with the central black hole, and $\eta:=m/M$ is the small mass ratio.
(Here we have set $\e=1$ and for clarity redefined each of the perturbations $h_{\mu\nu}^{(n)}$ with a factor $\eta^n$ pulled out.)
Now, it should be clear that the metric perturbation produced by the object will depend on the object's worldline $z^\mu$, which may lead us to expect that each of the perturbations $h^{(n)}_{\mu\nu}$ depend on $z^\mu$. But the worldline satisfying the equation of motion~\eqref{2nd order eqn motion} plainly depends on $\eta$, while the  coefficients $h^{(n)}_{\mu\nu}$ in the expansion~\eqref{regular expansion} are independent of $\eta$. It follows that $h^{(n)}_{\mu\nu}$ cannot depend on the whole of $z^\mu$. Instead, we can only utilize an expansion of the form~\eqref{regular expansion} if we also expand $z^\mu$ in the same way: 
\beq\label{z expansion}
z^\mu(\tau,\eta) = z^\mu_0(\tau) + \eta z^\mu_1(\tau) + \eta^2 z^\mu_2(\tau) + O(\eta^3).
\eeq
In this treatment of the worldline, the zeroth-order worldline $z^\mu_0$ is a geodesic of the background spacetime $g_{\mu\nu}$, and the GSF introduces small corrections to that worldline. Instead of equations of the form~\eqref{2nd order eqn motion}, one obtains evolution equations for the individual terms $z^\mu_n$. $h^{(1)}_{\mu\nu}$ depends only on $z^\mu_0$ and creates the GSF that drives $z^\mu_1$; $h^{(2)}_{\mu\nu}$ depends on $z^\mu_1$ in addition to $z^\mu_0$, and it (together with $h^{(1)}_{\mu\nu}$) drives $z^\mu_2$; and so on. This approach, first used systematically by Gralla and Wald~\cite{Gralla:2008fg} and carried to second order by Gralla~\cite{Gralla:12}, is the only consistent way to apply ordinary perturbation theory to the problem. Unfortunately, it is not suitable for treating long-term effects. Suppose the small object initially moves tangentially to a geodesic $z^\mu_0$. As the inspiral progresses, the object moves further and further from $z^\mu_0$, until the expansion~\eqref{z expansion} breaks down.\footnote{ This breakdown generally occurs not on the radiation-reaction time, but on the much shorter {\em dephasing time}, $t_{dph}\sim M/\sqrt{\eta}$. As we can guess from the form of the equation~\eqref{2nd order eqn motion}, the most rapid growth in $z^\mu_1$ is quadratic in time, such that $\eta z^\mu_1$ grows to order $\eta^0$ after $t_{dph}$; this growth occurs specifically in ``phase" variables such as the azimuthal phase $\varphi$ and the angle variables $q^\alpha$. The dephasing time hence corresponds to the time over which the gravitational wave sourced by a geodesic particle grows significantly out of phase with the wave sourced by a self-accelerated one.}

So ordinary perturbation theory fails. To account for the long-term changes in the worldline, we must avoid the expansion~\eqref{z expansion} of $z^\mu$, and hence the expansion~\eqref{regular expansion}. Instead, we naturally seek an asymptotic expansion that allows the metric perturbation to depend on the full, unexpanded $z^\mu$. We may write this as
\beq\label{self-consistent expansion}
{\sf g}_{\mu\nu}(x,\eta) = g_{\mu\nu}(x) + \eta h^{(1)}_{\mu\nu}(x;z) + \eta^2 h^{(2)}_{\mu\nu}(x;z) + O(\eta^3), 
\eeq
with each coefficient containing an implicit functional dependence on $z^\mu$. This is the type of expansion implicitly used in the preceding sections. It leads to the coupled system of equations~\eqref{2nd order eqn motion}, \eqref{effective EFE1}, and \eqref{effective EFE2}. It is called the {\em self-consistent} approximation, a name first applied by Gralla and Wald~\cite{Gralla:2008fg}, indicating that in it, the trajectory $z^\mu$ is obtained by solving the coupled field equations and equation of motion together, self-consistently. This self-consistent formulation was used in the original derivations of the MiSaTaQuWa equation, and it was put on a systematic basis, extendable to any order in perturbation theory, in Ref.~\cite{Pound:10a}, which first explicitly introduced the expansion~\eqref{self-consistent expansion}.

The self-consistent approximation successfully eliminates the growing errors associated with the expansion~\eqref{z expansion}, and it has the advantage of being formulated in a generic (vacuum) background spacetime. However, it is not quite ideal for the EMRI problem. It does not capitalize on the particular, adiabatic character of the orbital inspiral, which is only slowly evolving and hence very nearly triperiodic. Furthermore, it is not designed to accurately incorporate a second type of slow change in the system: the slow evolution of the large black hole. Over time, the black hole absorbs energy and angular momentum in the form of gravitational waves, causing its mass and spin to slowly change. Because the expansion~\eqref{self-consistent expansion} does not have a built-in way to accurately track this long-term effect, a naive implementation of the self-consistent approximation would exhibit growing errors, just as the expansion~\eqref{regular expansion} exhibited growing errors due to the long-term changes in the orbit. While Eq.~\eqref{self-consistent expansion} can be tweaked to eliminate this inaccuracy~\cite{Flanagan-etal:17}, an approximation more specifically tailored to EMRIs is preferable. Such an approximation is offered by a {\em multiscale expansion} (also known as a ``two-timescale'' expansion), a method of singular perturbation theory~\cite{Kevorkian-Cole:96} that, in the case of an EMRI, expresses the evolving worldline as a function of both ``slow time" and ``fast time" variables. By likewise writing the metric perturbation in terms of these variables, we may split the equation of motion and field equations into corresponding slow- and fast-time equations. At fixed values of slow time, the fast-time equations have the same triperiodicity as a background geodesic, providing ``snapshots'' of the inspiral on the orbital time scale. The slow-time equations then govern the smooth evolution from one snapshot to the next.

This approximation scheme is most easily described in terms of the action-angle variables $(J_\alpha,q^\alpha)$ introduced in the previous section. As a slow time, we may use $\tilde t = \eta t$;\footnote{ Refs.~\cite{Mino-Price:08,Pound:2015wva} discuss more refined choices of slow time.} when Boyer-Lindquist time $t$ is comparable to the radiation-reaction time (i.e., $t\sim M/\eta$), the slow time is of order 1 (i.e., $\tilde t\sim M$). As fast-time variables, we can use the angle variables $q^\alpha$. In place of Eq.~\eqref{regular expansion} or~\eqref{self-consistent expansion}, the expansion of the metric then becomes
\beq\label{multiscale expansion}
{\sf g}_{\mu\nu}(t,x^a,\eta) = g_{\mu\nu}(x^a) + \eta h^{(1)}_{\mu\nu}(x^a, \tilde t, q^\alpha) + \eta^2 h^{(2)}_{\mu\nu}(x^a, \tilde t, q^\alpha) + O(\eta^3), 
\eeq
where $x^a$ can be any set of coordinates on spacetime slices of constant $t$, such as the Boyer-Lindquist coordinates $x^a=(r,\theta,\varphi)$.  The coefficients in the expansion are required to be bounded functions of $\tilde t$, such that they do not grow large with time, and to be periodic functions of $q^\alpha$, with Fourier expansions
\beq\label{multiscale Fourier}
h^{(n)}_{\mu\nu} = \sum_{k_\alpha} h^{(n,k_\alpha)}_{\mu\nu}(x^a, \tilde t)e^{-ik_\alpha q^\alpha},
\eeq
where the sum runs over sets of integer constants $k_\alpha$. From the coefficients $h^{(n,k_\alpha)}_{\mu\nu}$, one can obtain the slow evolution of the ``constants'' $J_\alpha$, which instead of Eq.~\eqref{EOM_action-angle} now satisfy an equation of the form  
\begin{align}\label{J evolution}
\frac{dJ_\alpha}{dt}= \sum_{k_A}\left[\eta G^{(1,k_A)}_\alpha(\tilde t) + \eta^2 G^{(2,k_A)}_\alpha(\tilde t)  + O(\eta^3)\right]e^{-i k_A q^A}.
\end{align}
The Fourier coefficients $G^{(n,k_A)}_\alpha$ are constructed from those of the GSF (and thence, from the coefficients $h^{(n,k_\alpha)}_{\mu\nu}$). Here, unlike in Eq.~\eqref{multiscale Fourier}, the sum is over pairs of integers $k_A=(k_r,k_\theta)$; because the background is stationary and axially symmetric, the ``forces" in Eq.~\eqref{J evolution} are independent of $q^t$ and $q^\phi$. Once the evolution of the action variables is determined, one can obtain the evolution of the angle variables simply by integrating \footnote{ In general, the right-hand side of Eq.~\eqref{dqdt} will contain oscillatory dependence on $q^\alpha$, but such oscillations can be removed with a near-identity transformation~\cite{vandeMeent:2013sza,vandeMeent:2018rms}. (We return to this type of transformation in Sec. 6.2 below.)}
\beq\label{dqdt}
\frac{dq^\alpha}{dt}=\omega^\alpha(J_A) = \omega_{(0)}^\alpha(\tilde t) + \eta \omega_{(1)}^\alpha(\tilde t) + O(\eta^2);
\eeq
that is,
\beq\label{phase evolution}
q^\alpha = \frac{1}{\eta}\int \omega^\alpha d\tilde t = \frac{1}{\eta}\left[q_{(0)}^{\alpha}(\tilde t) + \eta q_{(1)}^{\alpha}(\tilde t) + O(\eta^2)\right]. 
\eeq
Finally, with $J_\alpha(t,\eta)$ and $q^\alpha(t,\eta)$ known, one obtains the metric perturbations~\eqref{multiscale Fourier} as ordinary functions of $t$, $x^a$, and $\eta$. 

A formulation of this sort was first put forward by Hinderer and Flanagan~\cite{Hinderer:2008dm}, who developed the mulsticale expansion of the equation of motion. Besides working out the basic formalism, they also used it to establish which ingredients are needed to compute each of the terms in Eqs.~\eqref{J evolution} and \eqref{phase evolution}. This is particularly important in the case of the phase evolution, since accurately tracking the phase of the waveform is the critical requirement for matched filtering. Hinderer and Flanagan found that in order to obtain the leading term in the phase evolution~\eqref{phase evolution}, one requires only the time-averaged dissipative piece of the first-order force; they refer to this as an ``adiabatic approximation". In order to obtain both the leading and subleading term in~\eqref{phase evolution}, one requires the entire first-order force, along with the time-averaged dissipative piece of the second-order force; Hinderer and Flanagan refer to this as a ``post-adiabatic approximation". Because the subleading term in~\eqref{phase evolution} is of order 1, a post-adiabatic approximation is necessary for accurate modelling, and we can conclude that EMRI science requires at least part of the second-order metric perturbation.
This has been the primary motivation for developing second-order GSF theory, and our primary reason for including a description of it in this review. 

Given its clear advantages, the multiscale expansion is the most promising method of tackling long-term evolution, and following Hinderer and Flanagan's expansion of the equation of motion, work has recently begun on the corresponding expansion of the field equations~\cite{Pound:2015wva,Flanagan-etal:17}. Yet the multiscale method has failings of its own. One limitation is that it breaks down on large distances, where length scales become comparable to the radiation-reaction time scale~\cite{Pound:2015wva}. This failure manifests itself as an infrared divergence in the retarded solution. A similar failure can also occur near the large black hole's event horizon, which (in the context of wave propagation) plays a role similar to that of infinity. Overcoming these failures calls for the introduction of additional, complementary expansions near infinity and the horizon, which can be combined with the multiscale expansion by once again appealing to the method of matched asymptotic expansions. 

\subsection{Transient resonances}\label{subsec:resonances}

Even putting aside the effects on large distances, the multiscale expansion encounters an entirely different, equally problematic  phenomenon: transient orbital resonances. Generically, away from a resonance, all the $k_A\neq0$ modes in Eq.~\eqref{J evolution} are oscillatory, averaging out to zero on the radiation-reaction time. The long-term, average evolution is then driven by the approximately constant, $k_A=0$ modes in Eq.~\eqref{J evolution}, giving rise to an equation of the form
\beq\label{Jav evolution}
\frac{d\langle J_\alpha\rangle}{dt}= \eta G^{(1,0)}_\alpha(\tilde t) + \eta^2 G^{(2,0)}_\alpha(\tilde t)+O(\eta^3),
\eeq
where $\langle\cdot\rangle$ denotes the average over the $(q^r,q^\theta)$ torus. However, this situation changes near a resonance, where for some period of time, one of the  $k_A\neq0$ modes becomes approximately stationary. We can see how this occurs, and its dynamical consequences, by examining how the phase $\psi = k_A q^A$ evolves near a resonance. Suppose a resonance occurs at a time $t_{\rm res}$, meaning that the ratio $\omega^r(t_{\rm res})/\omega^\theta(t_{\rm res})$ is rational. For some integers $k_A=k^{\rm res}_A$, then, the combination $k^{\rm res}_A \omega^A(t_{\rm res})=k^{\rm res}_r\omega^r + k^{\rm res}_\theta\omega^\theta$ vanishes. Now consider the {\em resonant phase} $\psi^{\rm res}:=k^{\rm res}_A q^A$ near $t_{\rm res}$. Expanding $q^A(t)$ around $q^A(t_{\rm res})$, recalling $dq^A/dt = \omega^A(\tilde t)$, 
we obtain 
\beq
\psi^{\rm res}(t) = \psi^{\rm res}(\tilde t_{\rm res}) + k^{\rm res}_A\omega^A(\tilde t_{\rm res})(t-t_{\rm res}) + \frac{\eta}{2} k_A^{\rm res}\left.(d\omega^A/d\tilde t)\right|_{\tilde t_{\rm res}}(t-t_{\rm res})^2 + O[(t-t_{\rm res})^3],
\eeq
In usual circumstances, away from a resonance, the second term on the right dominates the evolution, and the phase varies on the orbital timescale $\sim 1/(k_A\omega^A)\sim M$, causing $e^{-ik_A q^A}$ to oscillate and average to zero. But because $k^{\rm res}_A \omega^A(\tilde t_{\rm res})=0$, near the resonance the third term dominates the evolution. The phase then varies slowly, on the long timescale $\sim 1/\sqrt{\eta k_A^{\rm res}(d\omega^A/d\tilde t)}\sim M/\sqrt{\eta}$. Over periods shorter than this, the terms $G^{(n,k^{\rm res}_A)}_Ae^{-i\psi^{\rm res}}$ in Eq.~\eqref{J evolution} become approximately stationary. They then appear as additional terms on the right-hand side of Eq.~\eqref{Jav evolution}, altering the average rate of change. After the time $M/\sqrt{\eta}$, when the orbit has completed its passage through resonance, the additional driving term in Eq.~\eqref{Jav evolution} will have shifted the action variables $J_\alpha$ by an amount $\sim (dJ_\alpha/dt)/\sqrt{\eta}\sim \sqrt{\eta}$. This induces a corresponding order-$\sqrt{\eta}$ shift to the orbital frequencies, inducing an order-$\sqrt\eta$ term in Eq.~\eqref{dqdt}. Since the remainder of the inspiral, after the resonance, lasts a time of order $M/\eta$, the shift in the frequencies leads to a dramatic, $\sim 1/\sqrt{\eta}$ cumulative shift in the orbital phases. 

\begin{figure}[htb]
	\begin{tabular}{cc}
        \includegraphics[height=50mm]{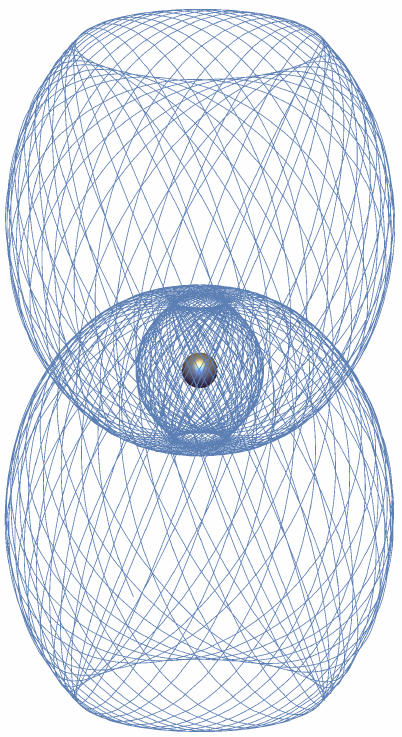} 	\hspace{-3mm}
        \includegraphics[height=25mm]{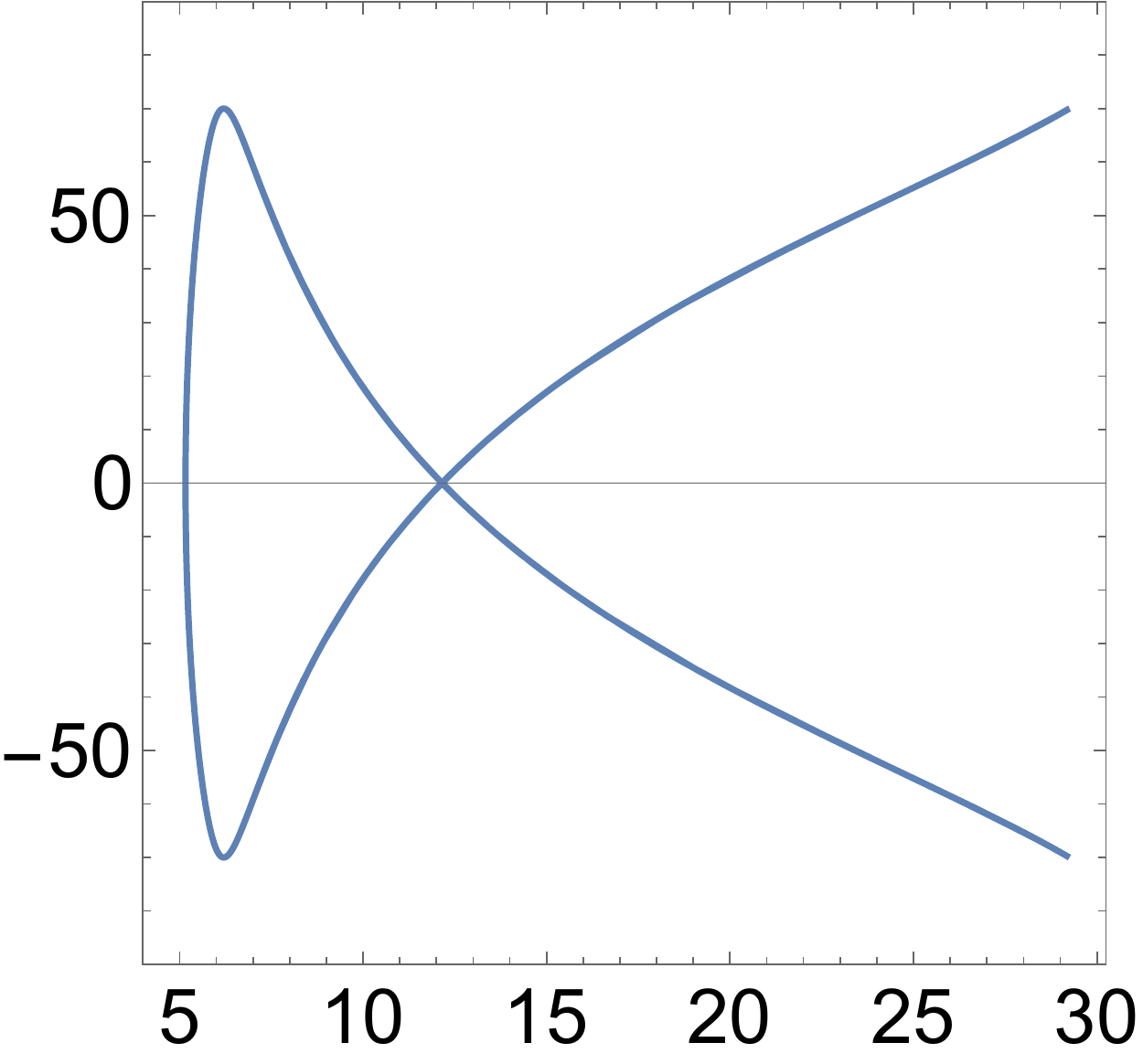}   	 
	    & 
        \includegraphics[height=50mm]{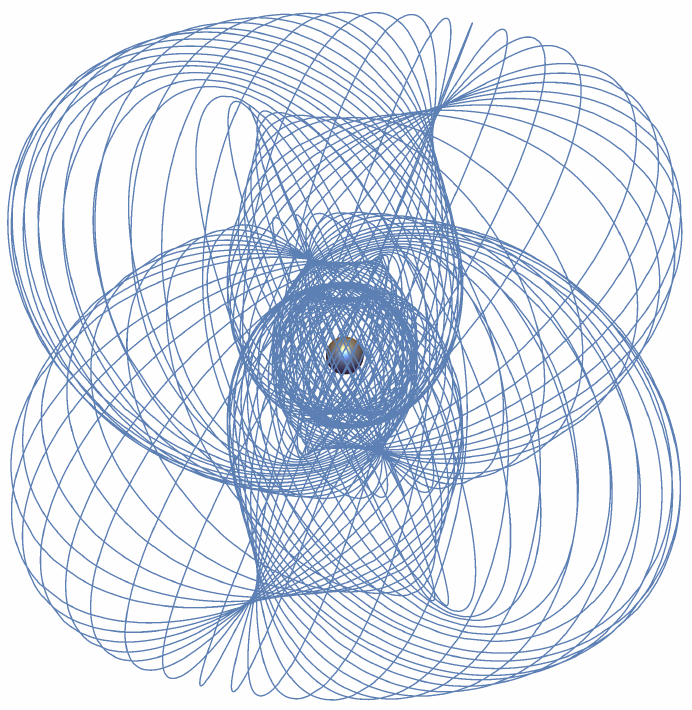}  \hspace{-3mm}    
       	\includegraphics[height=25mm]{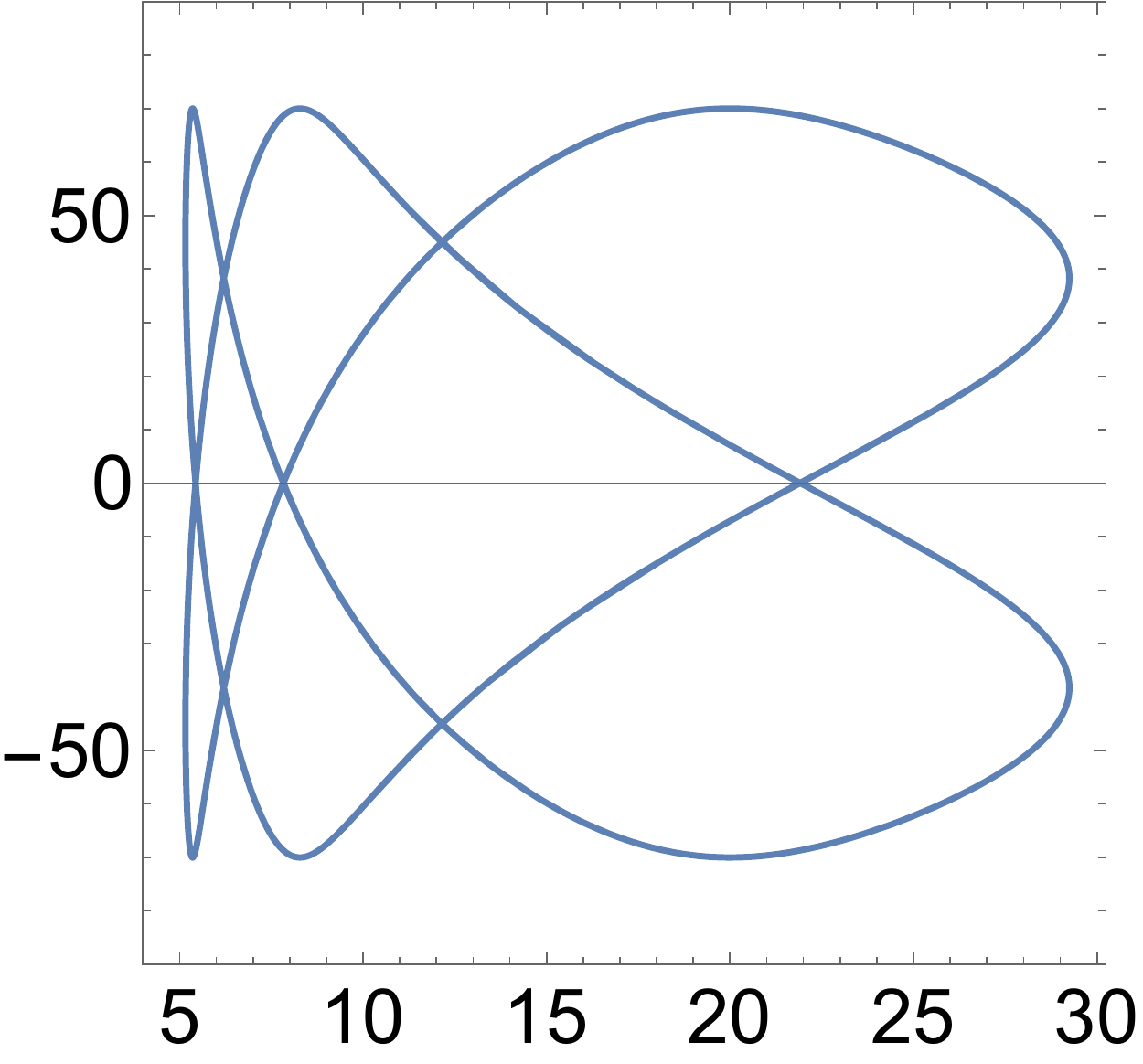}
	\end{tabular}
	\caption{The shape---and radiative dynamics---of a resonant orbit depend sensitively on the resonant phase $\psi^{\rm res}$, here defined as the value of the longitudinal angle $\theta$ at a periapsis of the resonant orbit. The two 3:2 resonant orbits viewed here, and the orbit shown on the right in Fig.\ \ref{fig:orbit}, all have the same parameters $\{E,\,L_z,\,Q\}$ but different resonant phases ($\psi^{\rm res}=20^\circ$ for the orbit in Fig.\ \ref{fig:orbit}, $\psi^{\rm res}=90^\circ$ for the orbit on the left here, and $\psi^{\rm res}=45^\circ$ for the orbit on the right here). The two orbits are viewed from just above the equatorial plane. Next to each orbit we display the corresponding Lissajous curve, showing the orbit in the plane of radius $r$ (horizontal axis, in units of $M$) and inclination angle (vertical axis, measured from the equator in degrees).    }
	\label{fig:resonances}
\end{figure}

The emergence of the dynamical time scale $M/\sqrt\eta$ near resonances violates the assumed form~\eqref{multiscale expansion} of the metric, causing the multiscale approximation to break down. Like the failures at large distances, this can presumably be overcome by introducing another approximation scheme near resonance, one that accurately accommodates the new time scale there. However, even if the passage through resonance is accurately modelled, resonances nevertheless lead to an overall loss of accuracy---a subtle consequence of the sensitive dependence of the dynamics on the resonant phase $\psi^{\rm res}$. As illustrated in Fig.\ \ref{fig:resonances}, the ``shape'' of the resonant orbit is strongly dependent upon the value of $\psi^{\rm res}$. 
Orbits with the same principal parameters (e.g., the same $\{E,L_z,Q\}$) can look very different depending on that  value. Unsurprisingly, this leads to a situation where the details of the radiative dynamics across the resonance depend sensitively on $\psi^{\rm res}$. Indeed, it has been demonstrated with explicit calculations of radiative fluxes from resonant orbits \cite{Flanagan:2012kg} that $\langle{dE/dt}\rangle$, $\langle{dL_z/dt}\rangle$ and (especially) $\langle dQ/dt\rangle$ vary significantly as functions of $\psi^{\rm res}$. Thus, to model the radiative transition across a resonance one must know the exact phase of the orbit as it enters the resonance. 

To appreciate the consequences of such phase dependence, suppose we include the first- and second-order forces prior to the resonance, expecting this to suffice for an accurate model. The error in $\psi^{\rm res}$ upon entering the resonance will then be small, $\sim \e$, but this will generate an order-$\e^2$ error in the driving force, leading to an order-$\e^{3/2}$ error in the frequencies, and from there an order-$\sqrt\e$ error in the phases. While that error remains small, it is larger than the order-$\e$ error we may have expected from including the second-order force. Similarly, if the orbit then goes on to encounter another resonance, this order-$\sqrt\e$ phase error will be boosted to an order-$1$ error. With each successive passage through resonance, we lose half an order of phase accuracy.


However, we should note that a strong resonant effect of this sort only arises when $k^{\rm res}_A$ are {\it small} integers. It is then that the effect of the otherwise-oscillatory forcing terms in Eq.~\eqref{J evolution} can build up coherently over a long time.  In contrast, high-order resonances with large values of $k^{\rm res}_A$ are efficient in ``averaging away'' such coherent buildup, so their resonant effect is suppressed and generally negligible. Studies \cite{Ruangsri:2013hra,Berry:2016bit} show that essentially all astrophysically relevant systems encounter at least one strong resonance while in the LISA band. A recent study \cite{Mihaylov:2017qwn} suggested, however, that LISA-relevant EMRIs are likely to pass through only one such resonance. It estimated  that, as a result, resonances are expected to reduce the number of EMRI detections with LISA by no more than $\sim4\%$.

The basic dynamical features  of transient resonances in EMRIs were first described by Flanagan and Hinderer in \cite{Flanagan:2010cd}, in the framework of the two-timescale expansion. (An alternative, but equivalent, mathematical description was later given by Gair and collaborators \cite{Gair:2011mr,Mihaylov:2017qwn}.) 
Ref.\ \cite{Flanagan:2010cd} demonstrated with an explicit calculation, using a simple post-Newtonian model for the GSF, that a low-order resonant crossing of a typical LISA EMRI can last a few hundred orbital cycles. This was later confirmed in \cite{Ruangsri:2013hra} using a more accurate model of the radiation reaction. 
The impact of resonances on EMRI modelling has been examined in several subsequent works~\cite{Flanagan:2012kg,Isoyama:2013yor,vandeMeent:2013sza,Brink:2013nna,Brink:2015roa,Berry:2016bit}. A complete treatment, including the expansion of the Einstein equations and matching to the multiscale expansion, is still lacking.

\subsection{Spin and finite-size effects}

So far in our discussion of orbital evolution, we have not specified the form of the force. In Sec.~\ref{equation of motion}, we indicated that in addition to the GSF, a small compact object is also subject to forces arising from its finite size. While the GSF arises from a somewhat indirect interaction of a small object with its own field, the finite-size forces arise from direct couplings between the object's multipole moments and the external curvature. This is first seen in the Papapetrou spin term in the first-order equation of motion~\eqref{1st order eqn motion}. At each higher order, one additional pair of multipole moments appear (one mass moment and one current moment). Corrections to the lower-order forces may (potentially) also arise. So, for example, in the second-order equation of motion, we expect the object's quadrupole moments to appear, presumably in precisely the form they do for a test body~\cite{Dixon:74} (though Ref.~\cite{Moxon:2017ozd} has indicated some subtleties in this expectation). We may also expect a correction to the Papapetrou term, coming from coupling of the object's spin to the curvature of $h^{\R(1)}_{\mu\nu}$. Finally, the spin will act as a source in the second-order field equation (or equivalently, appear as a term in the second-order puncture), thereby altering $h^{\R(2)}_{\mu\nu}$ and the GSF that arises from it. 

At present, the precise form of these finite-size effects remains to be confirmed; as mentioned in Sec.~\ref{equation of motion}, all derivations of the second-order equation of motion have specialized to approximately spherical, nonspinning objects. However, based on the two-timescale analysis, we can assess, in advance, which of these terms are required to accurately model an EMRI. Specifically, we require only the time-averaged, dissipative piece of the second-order force. For a test body, the force arising from the quadrupole moments~\cite{Steinhoff:2012rw} seems purely conservative, suggesting that this term is not needed for EMRI modelling. On the other hand, there are tentative hints that the correction to the Papapetrou force will contain dissipative effects~\cite{Dolan:2014pja}. (Note also that spin introduces a new frequency, necessitating the introduction of a new phase variable in the multiscale expansion.) In any case, second-order finite-size forces certainly warrant further investigation.

As part of our discussion of actual EMRI calculations in Sec.\ \ref{sec:dissipative} below, we will review work to incorporate the first-order, Papapetrou force into EMRI models.

\subsection{Gauge dependence}

One crucial issue we have not yet broached is that of the gauge dependence of the orbital evolution. Because the GSF is entirely gauge dependent, the orbital evolution may appear to be pure gauge. In a superficial sense, this is true: as an extreme example, we could choose a gauge in which the GSF identically vanishes, in which case no inspiral would seem to occur at all. However, this choice would lead to secular growth of the metric perturbation, causing perturbation theory to eventually break down. To avoid this and remain within the realm of perturbation theory, we must restrict the class of allowed gauges. For example, we may confine our attention to gauges compatible with the multiscale expansion~\eqref{multiscale expansion}. Physically, this corresponds to adopting a coordinate system that respects the orbit's physical triperiodicity on short time scales. Such restrictions also play an important role in the computation of various physical quantities, as described in Sec.~\ref{sec:conservative}.

\section{Actual calculations for EMRIs: dissipative effects and orbital evolution} \label{sec:dissipative}

We now begin our survey of concrete results from numerical GSF calculations in EMRI systems. We have compiled two sets of such results. The first, to be covered in the present section, explores the radiative effects of the GSF and their impact on the orbital evolution. We focus particularly on two approximation schemes that have been employed to model the evolution; these methods are distinct from the two-timescale expansion described above, but we can assess their accuracy using the two-timescale framework. The second set of results, to be covered in Sec.\ \ref{sec:conservative}, comprises calculations of various specific conservative physical effects of the GSF (along with other quantities constructed from $h^\R_{\mu\nu}$). Section \ref{sec:synergy} then presents tests of GSF results against other approaches to the two-body problem, and describes the productive synergy being pursued between GSF physics and these approaches. 

\subsection{Balance laws and adiabatic evolution}

In the previous section, we noted that to accurately model an EMRI waveform, we must work to at least second order in perturbation theory. However, we also described the inspiral as approximately adiabatic, slowly evolving from one geodesic to the next. For the purposes of simply detecting a sufficiently bright EMRI, without accurately identifying the source orbit's parameters, this leading-order, adiabatic approximation may well suffice. 

In its most primitive form, such an approximation corresponds to simply combining the slowness of the inspiral with the conservation of energy and angular momentum. To motivate this, we return to what is perhaps the most intuitive aspect of GSF physics: back-reaction from emission of gravitational waves. The dissipative piece of the GSF ``does work'' on the particle, dissipating its orbital energy and angular momentum and thereby driving its gradual inspiral deeper and deeper into the potential well of the central black hole. The particle's lost energy and angular momentum is transferred into the gravitational field, which then carries them away to infinity in the form of gravitational waves. In this picture, the loss of orbital energy and angular momentum is ``balanced'' by the emitted radiation. This intuitive picture is basically a valid one (at least at first order in $\eta$), except that---unlike in the analogous flat-space relativistic electrodynamics problem that inspires it---such a balance cannot usually be established in a momentary sense: there is no meaningful way of relating the momentary rate of (say) orbital-energy dissipation to the momentary flux of energy in the gravitational waves. (This impossibility traces back to the fundamental absence of a notion of local energy in GR.)

However, even in GR, in certain circumstances, it is possible to formulate balance relations that hold in a certain ``time-averaged'' sense. For example, it was shown in \cite{Quinn:1999kj} that, for a freely falling point mass whose orbit begins and ends at infinity, the total work done by the first-order GSF exactly balances the total energy carried away in (linearized) gravitational waves. More pertinently for our context, it has been shown for bound geodesic orbits in Kerr that balance relations hold, within an adiabatic approximation (and at first-order in $\eta$), when a suitable orbital averaging is applied~\cite{Sago:2005fn,Ganz:2007rf,Flanagan:2012kg} (see also \cite{Galtsov} for early work to this end). These ``balance laws'' for energy and angular momentum take the simple form 
\begin{equation} \label{balance}
\begin{aligned}
\langle F^{\rm diss}_t/u^t\rangle&=\langle {\dot{\cal E}_{\infty}}\rangle+\langle {\dot{\cal E}_{\cal H}}\rangle,\\
-\langle F^{\rm diss}_\varphi/u^t\rangle&=\langle {\dot{\cal L}_{\infty}}\rangle+\langle {\dot{\cal L}_{\cal H}}\rangle,
\end{aligned}
\end{equation}
where on the right-hand side are time-averaged asymptotic fluxes of energy ($\dot{\cal E}$) and angular momentum ($\dot{\cal L}$) out to infinity (`$\infty$') and down the event horizon of the black hole (`${\cal H}$'), and throughout the rest of this section an overdot denotes $d/dt$. Following the logic of the adiabatic approximation, quantities on both sides are calculated while approximating the orbit as a fixed geodesic: for example, $F^\mu$ is the GSF generated by the field $h^{\R(1)}_{\mu\nu}$ of a point mass moving on a geodesic. We have used $\langle\cdot\rangle$ to denote time averaging (with respect to time $t$); for intrinsically periodic geodesic orbits (any orbit in Schwarzschild, or circular, equatorial or resonant orbits in Kerr) it suffices to average over one period of the motion, but in general (for ergodic geodesics such as the one on the left in Fig.\ \ref{fig:orbit}) one must average over an infinite amount of time (while still treating the orbit as a fixed geodesic), or, equivalently, over the orbital $(q^r,q^\theta)$ torus. On the left-hand side of the balance equations (\ref{balance}) are the averaged $t$ and $\varphi$ components of the local GSF, normalized by the time component of the local four-velocity, $u^t$ (which simply translates from the local proper time used in the definition of the GSF to the usual coordinate time used in defining the asymptotic fluxes). The left-hand sides here are always positive, and so are the fluxes $\langle {\dot{\cal E}_{\infty}}\rangle$ and $\langle {\dot{\cal L}_{\infty}}\rangle$. However, the fluxes $\langle {\dot{\cal E}_{\cal H}}\rangle$ and $\langle {\dot{\cal L}_{\cal H}}\rangle$ of radiation absorbed by the black hole can be either positive or---for certain orbits in Kerr---negative. Negative horizon fluxes mark {\it superradiant} behavior, in which some of the black hole's rotational energy and angular momentum are, in effect, transferred {\it to} the orbit~\cite{Teukolsky:1974yv,Hughes:1999bq,Glampedakis:2002ya}. 

We may also write the balance laws~\eqref{balance} in the more suggestive form
\begin{equation} 
\begin{aligned}
\langle \dot E\rangle&= - \langle {\dot{\cal E}_{\infty}}\rangle - \langle {\dot{\cal E}_{\cal H}}\rangle,\\
\langle\dot L_z\rangle&= - \langle {\dot{\cal L}_{\infty}}\rangle - \langle {\dot{\cal L}_{\cal H}}\rangle.
\end{aligned}
\end{equation}
If we compute the fluxes on the right-hand side for a given geodesic with parameters $E$ and $L_z$, these laws allow us to evolve to a new geodesic with parameters $E+\langle \dot E\rangle\delta t$ and $L+\langle \dot L_z\rangle\delta t$. This gives us a way of deducing the dominant, adiabatic evolution without actually resorting to a calculation of the local GSF. Computational methods for asymptotic fluxes in black-hole perturbation theory have been well developed 
since the early 1970s, and  have been performed many times using frequency-domain methods (e.g, \cite{Teukolsky:1974yv,Poisson:1995vs,Hughes:1999bq,Glampedakis:2002ya,Drasco:2005kz}) as well as time-domain ones  \cite{Martel:2003jj,Poisson:2004cw,Barack:2005nr,Barack:2007tm,Sundararajan:2008zm}. Such calculations are much less computationally expensive than local GSF calculations, and can be done in the convenient framework of Teukolsky's perturbation formalism, working with $\Psi_0$ or $\Psi_4$ instead of the full metric perturbation.  This approach has been taken by Hughes and collaborators \cite{Hughes:2005qb,Hughes:2016xwf} in order to compute the adiabatic evolution of (certain types of) EMRIs, circumventing the need for an explicit GSF calculation.  

Recall, however, that generic geodesic orbits in Kerr are characterized by {\it three}---not two---constants of motion: to get a full description of the adiabatic evolution for a generic orbit one must also
be able to calculate the evolution of the Carter constant $Q$, in addition to the evolution of $E$ and $L_z$.  It is not difficult to write down $\dot Q$ directly in terms of the local GSF \cite{Ori:1997be},
but there is no known way to relate $\langle \dot Q\rangle$ to some asymptotic fluxes of radiation. This problem was solved, at the practical level, thanks to a breakthrough idea by Mino \cite{Mino:2003yg,Mino:2005yw} and follow-up work by Sago and collaborators \cite{Sago:2005fn,Ganz:2007rf}, who derived a simple, practical formula for $\langle \dot Q\rangle$.  The formula involves both quantities that are encoded in the asymptotic radiation (and are readily calculable within the Tuekolsky formalism), and quantities that are locally defined as integrals along the orbit (but are nonetheless also easily calculated).  With this, it is now possible to calculate the evolution of generic EMRI orbits in Kerr, at leading, adiabatic order, without performing an actual GSF calculation. Although originally formulated outside the context of a two-timescale expansion, it can be shown that this evolution scheme is precisely equivalent to solving the leading-order equations in the two-timescale approximation (which Hinderer and Flanagan referred to as ``adiabatic order" for that reason). There is an ongoing programme to numerically implement this approach \cite{Drasco:2005is,Sundararajan:2007jg,Sundararajan:2008zm,Hughes:2016xwf}. 

In addition to enabling adiabatic evolution, the balance law~\eqref{balance} provides an important benchmark for GSF calculations. GSF codes all involve both a calculation of the local GSF and a calculation of the global perturbation field, from which the fluxes $\langle\dot{\cal E}_{\infty/{\cal H}}\rangle$ and $\langle\dot{\cal L}_{\infty/{\cal H}}\rangle$ may readily be extracted. Thus, GSF codes offer the opportunity to test the balance relations (\ref{balance}). [Or, conversely, depending on one's point of view, the balance relations (\ref{balance}) offer the opportunity to test GSF codes.] Explicit numerical calculations demonstrating the validity of (\ref{balance}) have been carried out for a variety of cases, including circular \cite{Barack:2007tm} and eccentric \cite{Barack:2010tm} orbits in Schwarzschild, eccentric orbits in the equatorial plane of a Kerr black hole \cite{vandeMeent:2016pee}, and, most recently, generic (eccentric and inclined) bound orbits in Kerr \cite{vandeMeent:2017bcc}.

Finally, the balance laws have also provided an important test of {\em second}-order calculations. In a recent series of papers \cite{Pound-Miller:14,Pound:2014koa,Pound:2015wva,Miller:2016hjv}, practical methods and computational tools have been developed for calculations of the second-order perturbation produced by a particle on a quasicircular inspiral orbit in Schwarzschild geometry. The monopolar (spherically symmetric) piece of the second-order field equation~\eqref{effective EFE2} contains several dissipative terms, including a term equal to the flux of energy at infinity, a term  equal to the flux of energy through the horizon, and a term describing the rate of orbital-energy dissipation (note that all these quantities are, in fact, second order in $\eta$). It can be analytically shown that this portion of the field equation precisely yields the balance law (\ref{balance}). This result, yet to be published, was derived by one of us (AP) recently. It can be said to be a first concrete physical outcome from a second-order GSF calculation. Because it involves nontrivial properties of the puncture $h^{\P(2)}_{\mu\nu}$ appearing in Eq.~\eqref{effective EFE2},  it stands as a major test of the second-order formalism. 




\subsection{Computation of inspiral orbits with the full (first-order) GSF}

The leading-order, adiabatic, flux-based calculations described above capture the main, average dissipative effect of the (first-order) GSF, but they completely neglect the conservative piece of the force, as well as subleading effects of the dissipative first-order force that average out at leading order in the two-timescale approximation. As discussed in Sec.~\ref{subsec:evolution}, both types of neglected effects of the first-order GSF contribute at the first post-adiabatic order, together with that of the averaged dissipative second-order GSF. These post-adiabatic terms, including the conservative piece of the GSF, do have an important secular effect on the phase evolution in EMRI systems \cite{Pound:2005fs}, and as already stressed, it is important to include them in EMRI models.

The most direct way of incorporating these effects would be to directly integrate the equation of motion in a self-consistent manner, by solving the coupled pair of equations~\eqref{1st order eqn motion} and \eqref{effective EFE1}. This is a computationally challenging task. In a time-domain implementation, one would need to compute the GSF time-step by time-step, and at each step accelerate the orbit that sources the field by a suitable amount. This direct calculation has only been attempted so far using a scalar-field toy model \cite{Diener:2011cc} with short orbital sequences; computing entire EMRI inspirals using this method does not seem to be a realistic prospect without a major improvement in numerical methodology. Frequency-domain implementations seem to be even less suitable for such direct integrations of the equation of motion, as, at least in their current form, they all assume a fixed geodesic source, with a given, unevolving frequency spectrum. 

In Sec.~\ref{subsec:evolution}, we have advocated for the two-timescale approach as an alternative (one which would be amenable to frequency-domain methods). However, to date there has been no direct implementation of that approach. Instead, orbital-evolution calculations so far have been based on the idea of {\it osculating geodesics} (named after the method of osculating orbits in Newtonian celestial mechanics). In this approach, the inspiral orbit is reconstructed as a smooth sequence of geodesics, each lying tangent to (or ``osculating") the true orbit at a particular moment. This amounts to modelling the true orbit as an evolving geodesic with dynamical orbital elements. At each instant $t_0$, one then approximates the GSF by computing it as if, for all $t<t_0$, the particle had been moving on the geodesic that is instantaneously tangential  to the evolving orbit at that instant $t_0$. This GSF is then used to calculate the momentary rate of change in each of the orbital elements---principal as well as, importantly, positional; this is a key difference from the adiabatic approximation, which likewise models the orbit as a smooth sequence of geodesics, but does not account for the GSF's conservative effect on the evolution of the positional elements. Equations governing the forced evolution of all orbital elements in the Schwarzschild case were obtained in Ref.\ \cite{Pound:2007th}, and Ref.\ \cite{Gair:2010iv} later generalized the formalism to Kerr. 

It is important to point out that the osculating-geodesics equations are precisely equivalent to the original equations of motion; they themselves involve no approximation, as long as the forcing term is known exactly for the evolving orbit. The only non-numerical source of error introduced in this procedure is in the use of ``geodesic'' GSF information (i.e., the GSF calculated while treating the orbit as a fixed geodesic rather than as the true, evolving orbit). The resulting error in the momentary self-acceleration formally scales with the square of the small mass ratio, and is {\it a priori} comparable to the error introduced by neglecting the second-order GSF. Assessment of the actual magnitude of error from using the ``geodesic'' GSF would require comparison with calculations based on a direct time-domain evolution. 

To implement the above idea in practice, rather than calculating the ``geodesic" GSF on the fly, one prepares a large database of GSF values covering the region of interest in the parameter space of geodesic orbits (this is a 2D space in the case of a nonspinning particle in Schwarzschild, and a 3D space if either the particle or central black hole is spinning). Each entry in that database (labelled by, e.g., $\{E,L_z,Q\}$) contains numerical GSF values calculated as a function of the phase variables (e.g., $\{q^r,q^\theta\}$) along a fixed geodesic orbit. This can be done relatively quickly using frequency-domain GSF codes. The data is then interpolated across the space, so that one now has analytic interpolation formulas for the GSF components at each point in the (4D or 6D) phase space of the problem. The preparation of this database is computationally heavy, but done once and for all. One can then use the interpolated GSF information to evolve any orbit, with any given initial conditions.  

\begin{figure}[htb]
	\begin{center}
        \includegraphics[width=90mm]{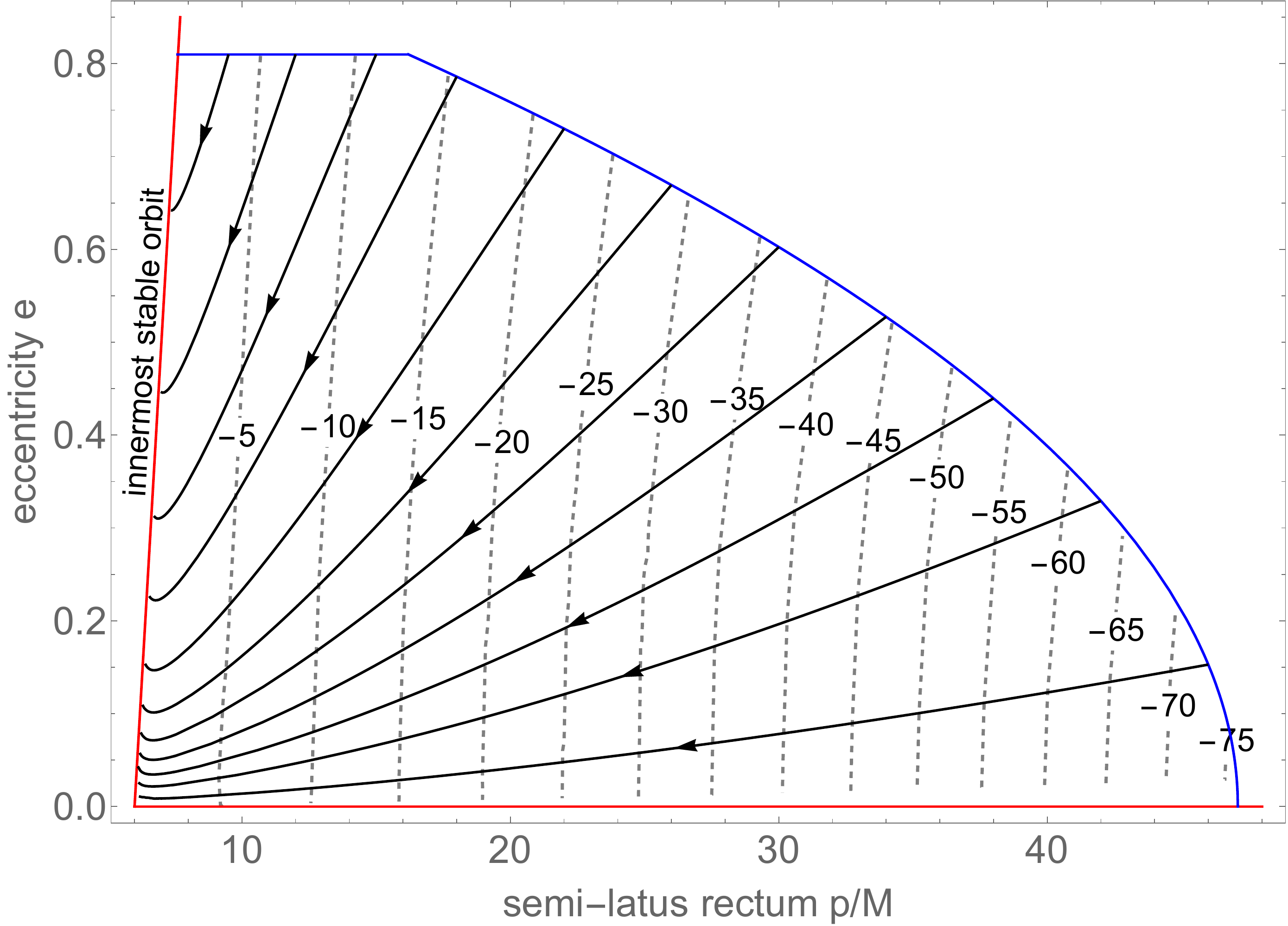}
\caption{Sample inspiral orbits around a Schwarzschild black hole, as calculated by Osburn {\it et al.}~in \cite{Osburn:2015duj} for a mass ratio $\eta= 10^{-5}$. Each solid black curve tracks the evolution of a particular EMRI orbit in the $e$--$p$ plane (eccentricity vs.\ semilatus rectum), from entering the LISA band (blue curve, assuming a black-hole mass of $M=10^6 M_{\odot}$), until reaching the innermost stable orbit (red line on the left). During the evolution, the conservative piece of the GSF acts to decrease the rate of periastron advance; the dashed contour lines (with associated numerical values) indicate the total amount of periastron phase (in radians) accumulated due to this effect, from a given moment until plunge.}
        \label{fig:evolution}
	\end{center}
\end{figure}

The first implementations of the osculating-geodesics method, with the full first-order GSF in Schwarzschild and using the ``geodesic'' GSF approximation, were presented by Warburton and collaborators in \cite{Warburton:2011fk,Osburn:2015duj}. Figure \ref{fig:evolution}, reproduced from Ref.\ \cite{Osburn:2015duj}, shows sample results from this work. Here, the eccentricity, $e$, and semilatus rectum, $p$, are used as convenient, geometrically intuitive parameters for bound geodesic orbits in Schwarzschild. The plot shows the computed evolution of several EMRI orbits with mass ratios $\eta=10^{-5}$ in the plane of $e$ and $p$, for a range of initial conditions; each solid (black) curve represents a single EMRI orbit, starting at the blue curve on the right, and evolving leftward (i.e., inward) until the last stable orbit, represented by the near-vertical (red) curve on the left. We can see how radiation reaction drives the orbit to be more and more circular (until very near the final plunge where the eccentricity can briefly increase). Important information is provided in this plot by the dashed curves: these are contours that mark the number of radians by which the periastron position will rotate under the effect of the conservative GSF, from a given point until plunge. Stated differently, they indicate the total phase error that one would be making by neglecting the effect of the conservative GSF. (The negative values of the indicated phase implies that the conservative GSF works against the usual periastron advance.)  These results illustrate the importance of accounting for conservative effects in EMRI models.  

The full inspiral trajectories of the osculating-geodesics method are still rather slow to evaluate in practice (Ref.\ \cite{Osburn:2015duj} reports a computation time of minutes to hours for each such trajectory, depending on mass ratio). That is because the forcing terms in the evolution equations depend explicitly on the orbital phases, so one must resolve the inspiral trajectory over the short orbital timescale. Van de Meent and Warburton \cite{vandeMeent:2018rms} showed how this can be avoided using a simple trick, by means of what is known in the theory of dynamical systems as a {\it near-identity transformation}.  The idea is to apply a ``small'' transformation to the phase-space variables of the problem, such that the resulting forcing terms in the evolution equations no longer depend on the orbital phase, while the solution to the modified problem remains uniformly close to the original solution. In effect, this procedure ``pushes'' the phase dependence one order higher in the small parameter of the problem (mass ratio, in our case), effectively averaging away all orbital-scale oscillations at leading order. Ref.\ \cite{vandeMeent:2018rms} applied this method to inspiral orbits in Schwarzschild spacetime, demonstrating the substantial gain in computational efficiency that it offers. 

While the osculating-geodesics evolution model of Refs.\ \cite{Warburton:2011fk,Osburn:2015duj,vandeMeent:2018rms} represents significant progress (compared to the leading-order adiabatic model), it still misses several important pieces, all formally second order in the GSF approximation, but all expected to contribute at the first post-adiabatic order just like the first-order conservative GSF. Missing is the dissipative piece of the second-order GSF, and missing too are the aforementioned corrections due to the use of the ``geodesic'' GSF approximation. If the small particle is spinning, the Papapetrou force must also be included, along with second-order dissipative forcing terms associated with the spin. The first step in this direction was taken in recent work \cite{Warburton:2017sxk}, which explores the impact of the Papapetrou term on the orbital evolution. Ref.~\cite{Pound:2015tma} has also suggested how the osculating geodesics method might be extended to second order, potentially providing an alternative to the two-timescale expansion.

\begin{figure}[htb]
     \begin{center}
         \includegraphics[width=50mm]{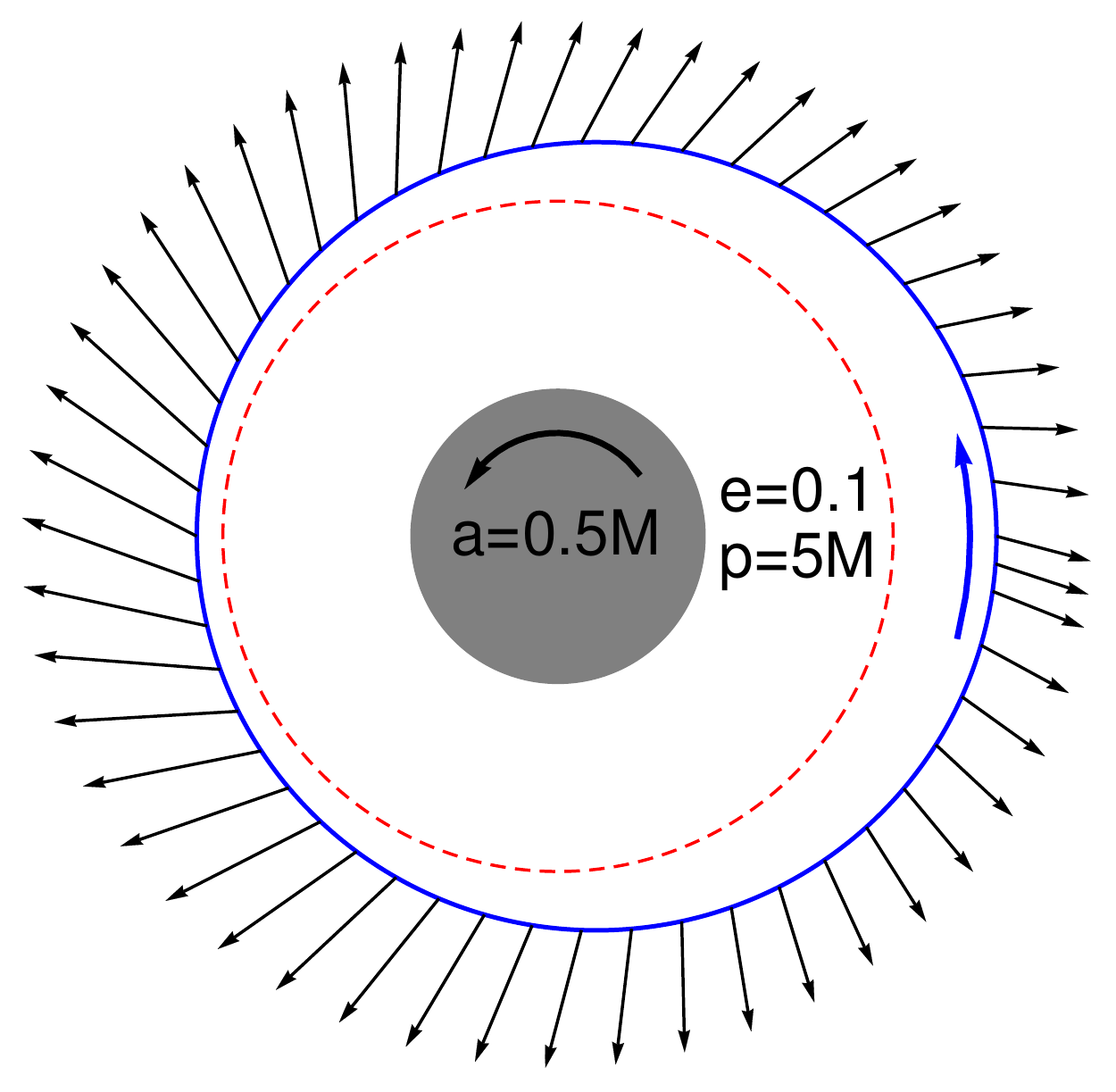}
         \includegraphics[width=68mm]{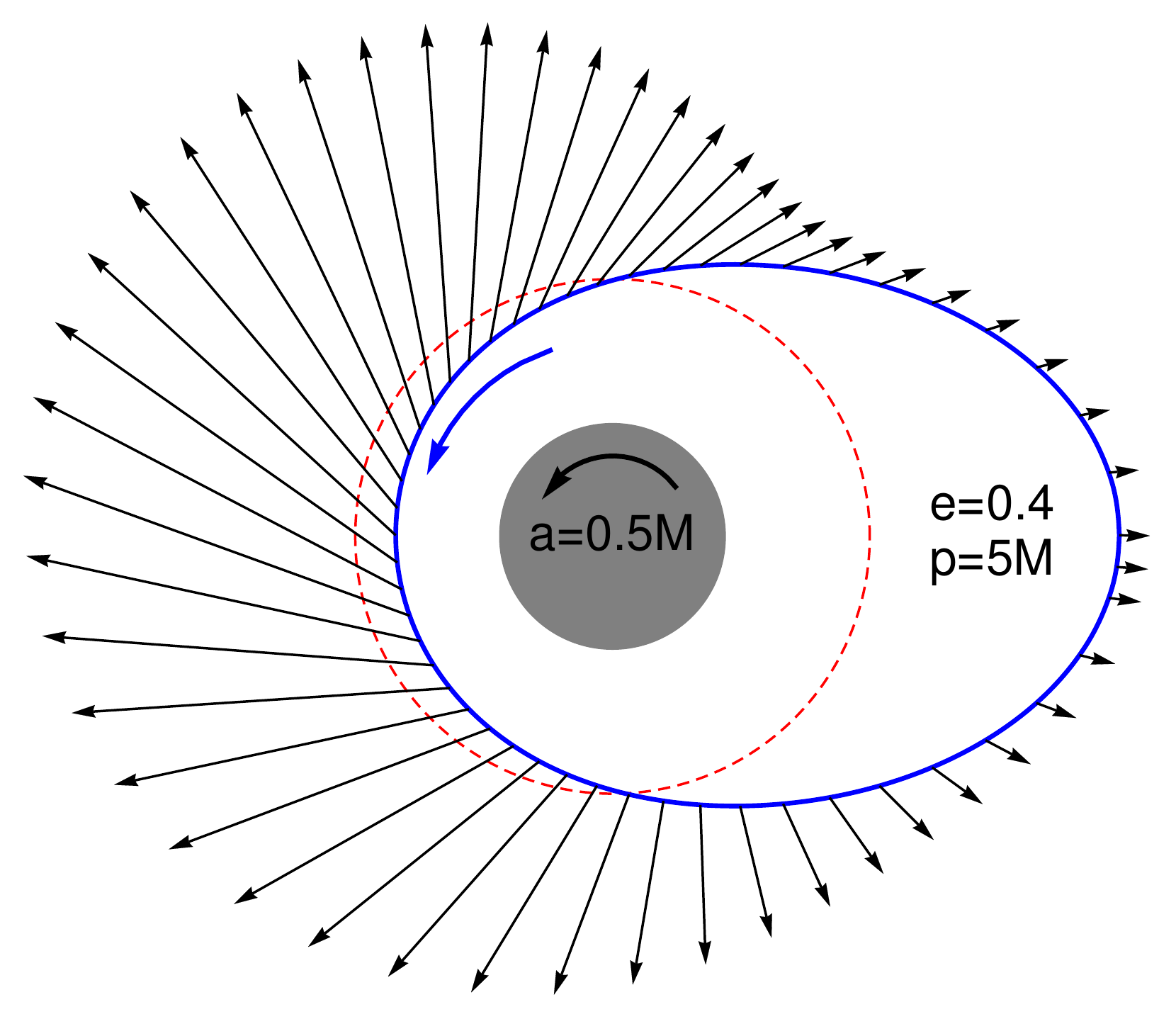}
\caption{Sample output from van de Meent's code 
\cite{vandeMeent:2016pee}, showing the first-order GSF 
along eccentric geodesic orbits in the equatorial plane of a Kerr black 
hole. We show here two orbits, of the same semilatus rectum but 
differing eccentricities: $e=0.1$ on the left and $e=0.4$ on the right. 
In both cases the orbital angular velocity is aligned with the black 
hole's spin, which is set at half the maximal allowable rate. For 
clarity, the orbits are plotted in the $r$--$q^r$ plane, where $q^r$ is 
the phase of the radial epicyclic motion (not the usual azimuthal phase 
$\varphi$; this is why the actual precessional motion of the orbit is 
not visible here). The two orbits are shown on the same scale, and the 
central holes are also to scale. Dashed (red) circles mark the innermost 
circular stable orbit (ISCO), for reference. Along each orbit we indicate with arrows
the spatial projection of the GSF onto the 
equatorial plane. The normalization of the GSF vector projection was 
chosen arbitrarily, but it is fixed along each orbit and between the two 
orbits; the actual magnitude of the GSF projection at the pericenter of 
the orbit on the left is $\sim 0.0237\eta^2$.
The code presented in \cite{vandeMeent:2016pee} is a frequency-domain 
implementation of the radiation-gauge approach with mode-sum 
regularization, and in its latest version \cite{vandeMeent:2017bcc} it 
is capable of computing the GSF along (essentially) any bound geodesic 
orbit around any Kerr black hole.
}       \label{fig:GSFloops}
     \end{center}
\end{figure}

\begin{figure}[htb]
     \begin{center}
         \includegraphics[width=64mm]{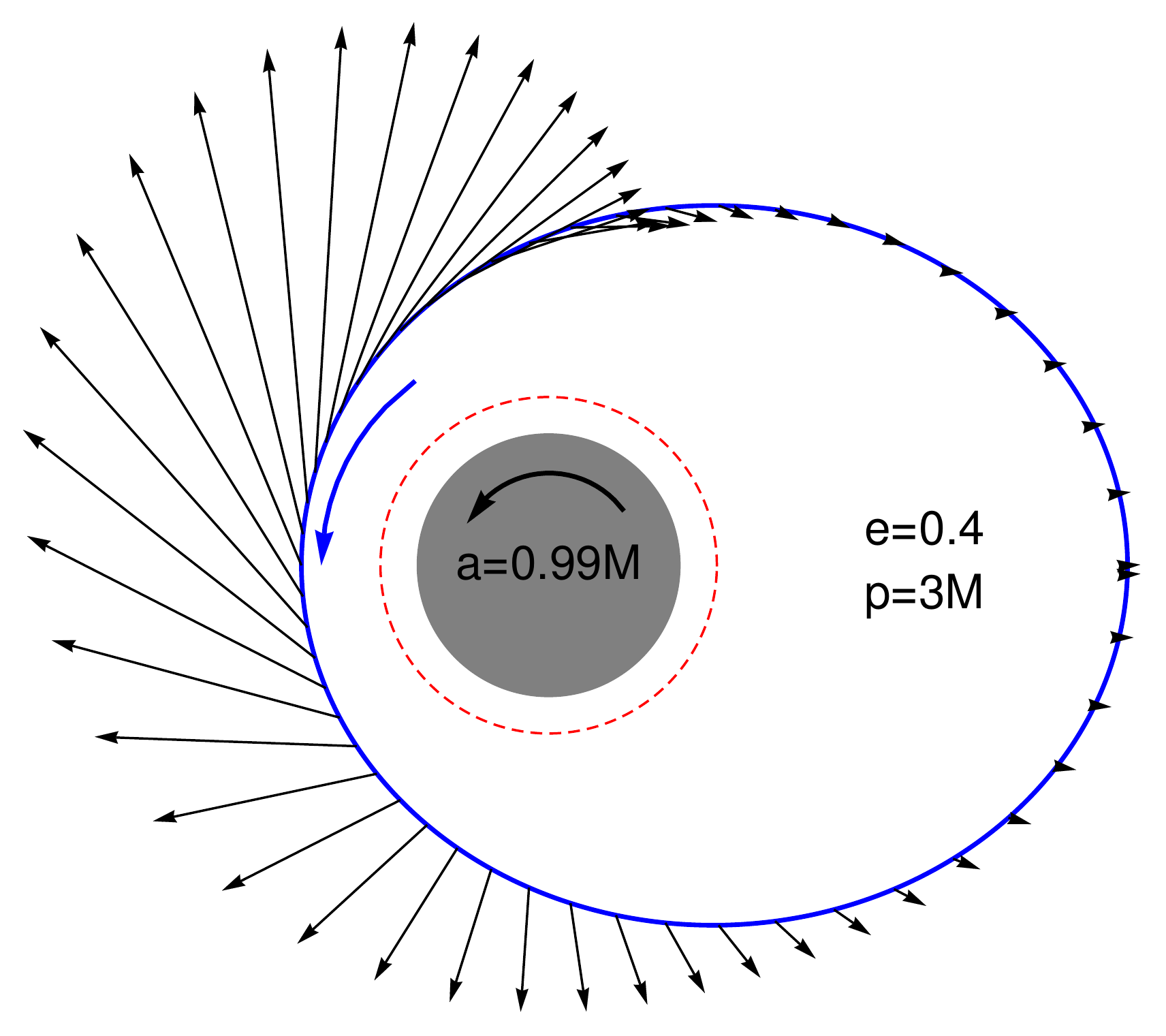}
         \includegraphics[width=64mm]{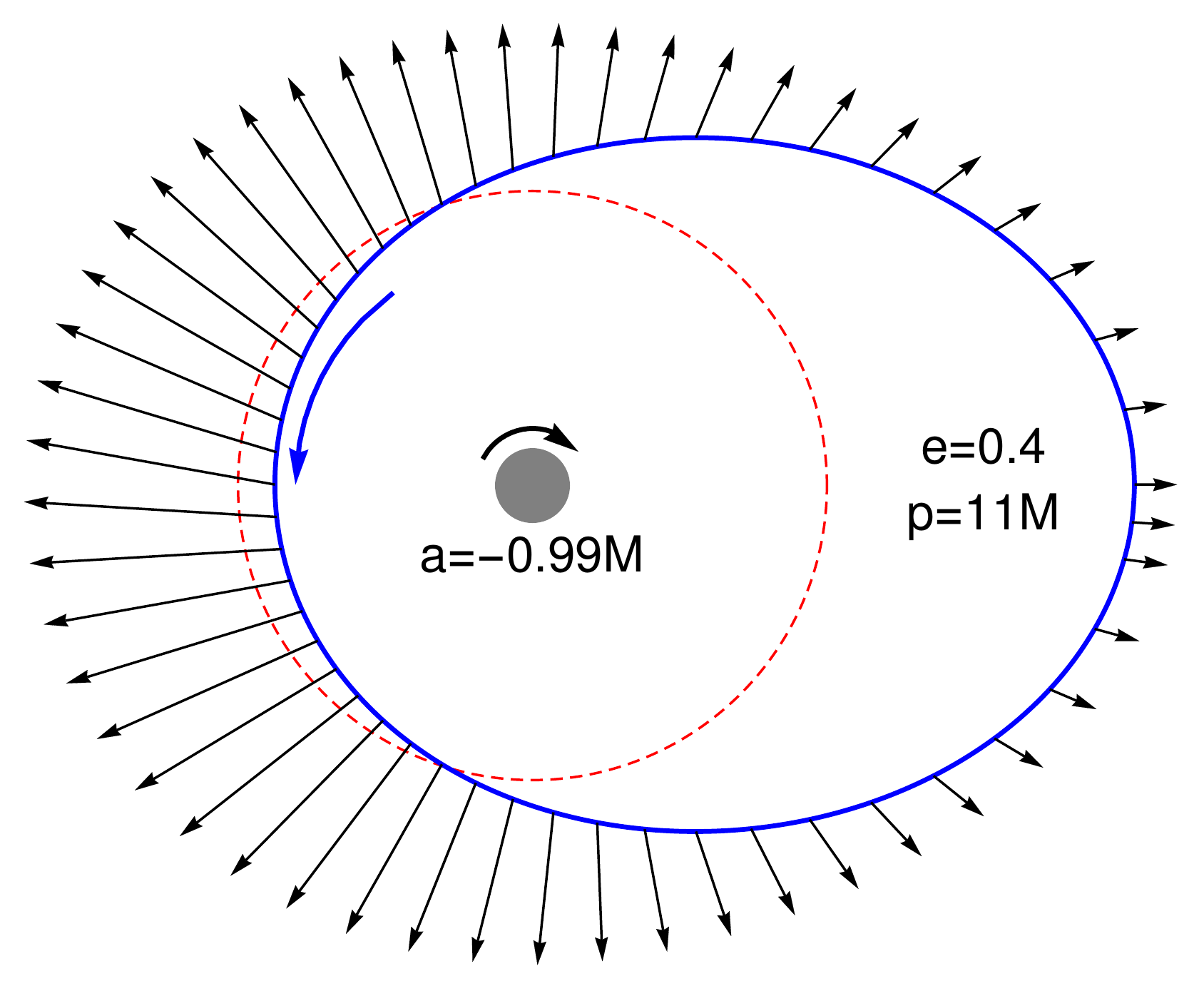}
\caption{Same as in Fig.\ \ref{fig:GSFloops}, this time with a central 
hole spinning at 99\% the maximal allowable rate. The orbit on the left 
co-rotates with the black hole, while that on the right is 
counter-rotating.  (Counter-rotating orbits encounter their ISCOs 
relatively far out, so their adiabatic radiative evolution terminates 
while still relatively far from the event horizon.) Note the two orbits 
are {\it not} shown on the same scale, and use a different normalization 
of the GSF vector. The actual magnitude of the GSF vector projection at 
pericenter is $\sim 0.3856\eta^2$ for the left orbit and (a mere) $\sim 
0.00768\eta^2$ for the orbit on the right.
}
         \label{fig:GSFloops2}
     \end{center}
\end{figure}

Work on orbital evolution has so far concentrated on the Schwarzschild problem. However, ``geodesic" GSF calculations are now available for generic bound geodesic orbits in Kerr~\cite{vandeMeent:2016pee,vandeMeent:2017bcc}---Figs.~\ref{fig:GSFloops} and \ref{fig:GSFloops2} display sample outputs of those calculations. With these results, we expect the osculating-geodesics programme to now extend its reach to the Kerr case. The main hurdle here is computational: one must populate the (now 3D) space of Kerr geodesics sufficiently densely with GSF data, and devise an efficient interpolation method over that space. One would need to extend the application of the near-identity-transformation method \cite{vandeMeent:2018rms} to the Kerr case. There is, additionally, ongoing work to develop more efficient ways to compute the orbital evolution in Kerr based on a two-timescale expansion with a judicious choice of orbital elements and gauge \cite{Fujita:2016igj}. Inspirals in Kerr also generically undergo resonant crossings, which, as discussed before, demand a special treatment.

\subsection{Sustained resonances?}

An intriguing question, explored in another work by van de Meent \cite{vandeMeent:2013sza}, is that of the  possibility of {\it sustained} resonances in EMRIs. In the general theory of resonant dynamical systems, a sustained resonance is one in which the driving force does not take the system across the resonance and away from it (as in the usual, transient-type resonances discussed in Sec.\ \ref{subsec:resonances}), but rather it keeps the system oscillating around the resonance for a prolonged amount of time---typically $\Delta t\propto\eta^{-1}$ rather than $\eta^{-1/2}$, in the EMRI case. The conditions for a sustained resonance can be deduced from an effective-potential equation satisfied by the resonant phase, of the form $\dot{\psi}^2=K-V(\psi)$, where $K:=V(\psi^{\rm res})$, and in the EMRI case, the potential $V$ depends on the self-forcing terms \cite{vandeMeent:2013sza}. A sustained resonance requires both that $V(\psi)$ has a minimum in which $\psi$ can be ``trapped'', and importantly, that such trapping can actually occur during the normal evolution of the EMRI. The latter condition requires that the system enters the resonance with an appropriate value of $K$. Ref.\ \cite{vandeMeent:2013sza} established that the ``window'' of suitable K values, as expressed, e.g., in terms of an interval of orbital phase, is extremely narrow, being proportional to $\eta^{1/2}$. Thus, even if sustained resonances are dynamically possible in EMRIs (i.e., there exist configurations for which $V$ has a minimum), which is not yet clear, the proportion of systems likely to be captured into one will be very small. It appears unlikely that LISA will observe a sustained resonance in an EMRI system.

\subsection{Resonant kicks}

There is also a different type of resonance that occurs in EMRI systems. Eccentric orbits, even around a Schwarzschild black hole, can experience resonances between the epicyclic $r$ motion and the rotational $\varphi$ motion: instances when $n_r T_r=n_\phi T_\phi$ with $n_r,n_\varphi$ some (small) integers. In such instances, the precession of the periastron momentarily halts. The intrinsic dynamics is unaffected, but the usual azimuthal symmetry (on average) of the emitted gravitational waves is broken, and the effect of asymmetric emission can build up coherently over a large number of orbits---the episode lasts an amount of time $\propto\eta^{-1/2}$, just as an $r$--$\theta$ resonance. During such an $r$--$\varphi$ resonance, gravitational waves carry to infinity a net amount of linear momentum in a particular direction, resulting in a net ``kick'' to the system's center of mass in the opposite direction. The magnitude of the kick velocity is expected to scale as $\eta^{3/2}$---a factor $\eta^2$ for the momentary magnitude of linear momentum in the gravitational waves, times a factor $\eta^{-1/2}$ for the buildup duration. 

\begin{figure}[htb]
\center
     \begin{tabular}{cc}
\includegraphics[width=110mm]{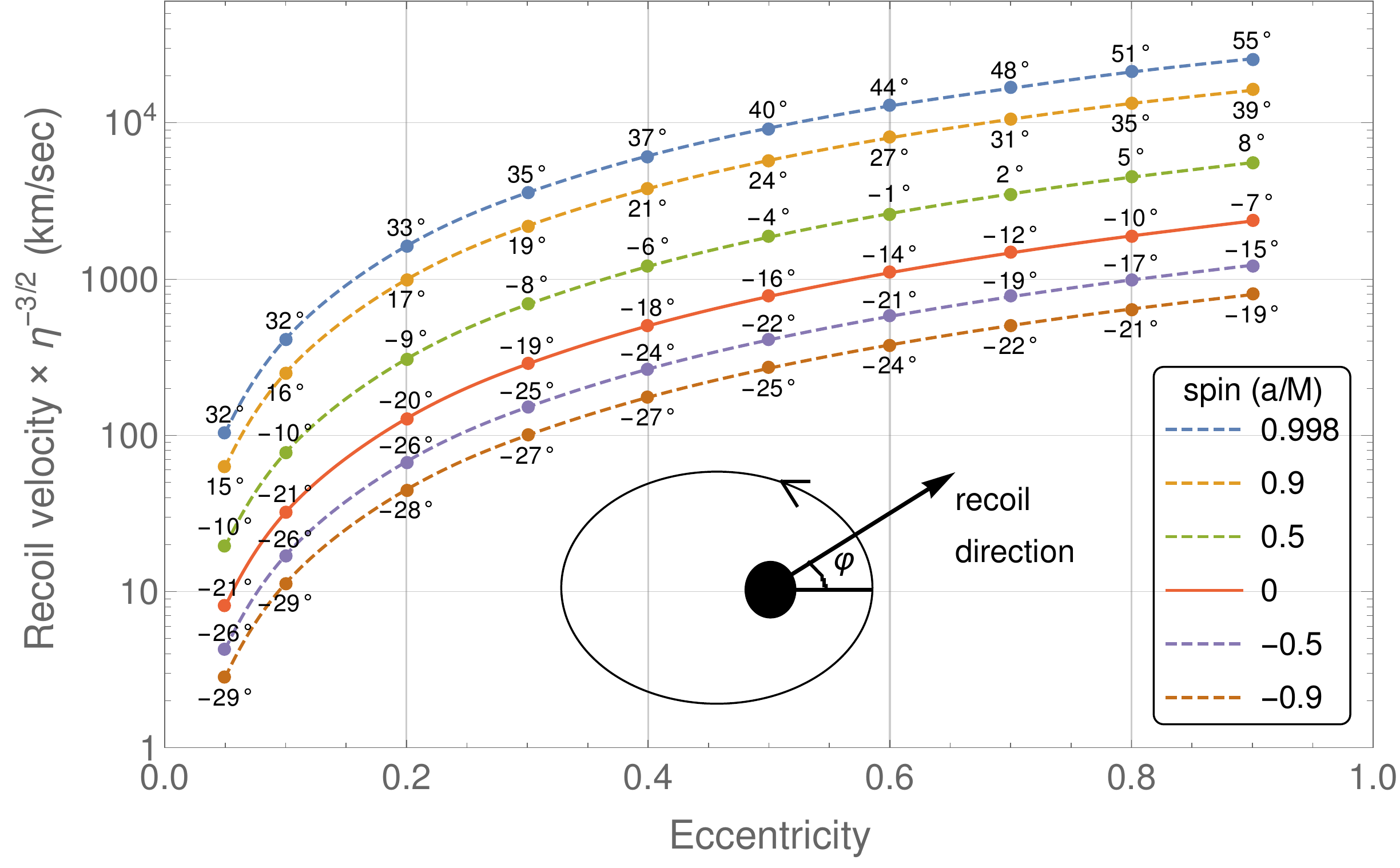}
     \end{tabular}
     \caption{Resonant kicks imparted to eccentric equatorial orbits 
around a Kerr black hole as they encounter a 1:2 $r$--$\varphi$ 
resonance (based on data from Ref.\ \cite{vandeMeent:2014raa}). Plotted is 
the magnitude of recoil velocity (divided by $\eta^{3/2}$), as a 
function of orbital eccentricity, for a range of black-hole spins [from 
top to bottom: $a/M=0.998,0.9,0.5,0,-0.5$ and $-0.9$, with positive 
(negative) values corresponding to prograde (retrograde) orbits]. 
Indicated next to each data point is the direction of recoil in the 
equatorial plane, measured as an azimuthal angle $\varphi$ relative to 
the on-resonance direction of the periastron (with $\varphi$ increasing 
in the direction of motion).}
     \label{fig:kicks}
\end{figure}

In Ref.\ \cite{vandeMeent:2014raa}, van de Meent performed a detailed analysis of $r$--$\varphi$ resonances for equatorial EMRIs in Kerr, numerically computing the magnitudes and directions of the resulting kicks. (This followed a similar analysis by Hirata for $\theta$--$\varphi$ circular-orbit resonances \cite{Hirata:2010xn}, where, however, the effect appeared to be much less pronounced.) His main results are summarized in Fig.\ \ref{fig:kicks}. An intriguing feature is the misalignment of the kick direction with the direction of periastron: this is due to the ``light-bending'' effect of the strong gravitational field around the black hole. The maximal kick velocities attainable are around $v\sim 30,000\times\eta^{3/2}$ km/sec, for a 1:2 $r$--$\varphi$ resonance at a very high eccentricity outside a very rapidly spinning Kerr black hole. Even in such extreme (and astrophysically less likely) configurations, the effect can become observationally important only with $\eta$ in the not-so-extreme regime. For example, $\eta=10^{-2}$ yields $v=30$ km/sec, only just comparable to the escape velocity of a typical globular cluster (the likely nesting place for putative intermediate-mass black holes). It appears very unlikely that resonant kicks will have observational implications for LISA-type EMRIs.

\section{Actual calculations for EMRIs: Conservative effects} \label{sec:conservative}

In our discussion so far we have encountered several times the fact that the GSF in EMRIs has, in addition to the obvious dissipative effect responsible for the emission of gravitational waves and for the orbital decay, also a range of non-dissipative, or {\it conservative} effects. This was manifest, for example, in Fig.\ \ref{fig:evolution}, which illustrated how the GSF acts to decrease the rate of periastron
advance in what leads to a significant cumulative effect on the long-term phase evolution. This effect comes from a local order-$\eta$ correction to a {\it positional} element of the orbit (i.e., a phase), which (at first order) has no impact on the rate of energy or angular momentum dissipation. Heuristically, the presence of the small mass ``deforms'' the background gravitational potential in a way that also affects the details of the dynamics of positional elements. 

At least at first order, this conservative effect can be cleanly disentangled from that of dissipation by writing the GSF as a sum of a time-symmetric piece $F_\alpha^{\rm cons}$ (``conservative force'') and a time-antisymmetric piece $F_\alpha^{\rm diss}$ (``dissipative force''),\footnote{ The precise definition of $F_\alpha^{\rm cons}$ and $F_\alpha^{\rm diss}$ alludes to the retarded and advanced metric perturbations (the gravitational analogues of the potentials $A_-^{\mu}$ and $A_+^{\mu}$ of Sec.~\ref{sec:EM self-force}):  $F_\alpha^{\rm cons}$ is the force exerted by the perturbation $\frac{1}{2}(h_{\alpha\beta+}+h_{\alpha\beta-})-h_{\alpha\beta}^{\rm S}$, while  $F_\alpha^{\rm diss}$ is exerted by $\frac{1}{2}(h_{\alpha\beta+}-h_{\alpha\beta-})-h_{\alpha\beta}^{\rm S}$.} and then considering the equation of motion with the full GSF replaced with either  $F_\alpha^{\rm cons}$ or $F_\alpha^{\rm diss}$. What we may call the ``conservative dynamics'' is thus described by the solution of the equation of motion  
\beq\label{1st order eqn motion cons}
m\frac{D^2 z^\mu}{d\tau^2} = F_{\rm cons}^\alpha(z),
\eeq
obtained from (\ref{1st order eqn motion schematic}) by setting $F_{\rm diss}^\alpha$ to zero. 

As a simple example, consider the circular orbit of a non-spinning particle around a Schwarzschild black hole. Working in Schwarzschild coordinates $(t,r,\theta,\varphi)$ and setting (without loss of generality) $\theta=\pi/2$ and $\dot{\theta}=0$ at some initial moment, the full (first-order) equations of motion (\ref{1st order eqn motion schematic}) explicitly read, for {\em any} orbit in Schwarzschild,
\begin{eqnarray}
\dot{E}&=&- F_t, \label{EoMt}
\\
m\,\ddot{r}&=& -\frac{m}{2}\frac{dV(r;L_z)}{dr} + F^r ,  \label{EoMr}
\\
m\,\ddot{\theta}&=& F^{\theta} ,\label{EoMtheta}
\\
\dot{L}_z&=& F_\varphi,  \label{EoMphi}
\end{eqnarray}
where $E=m\left(1-\frac{2M}{r}\right)\dot{t}$, $L_z=mr^2 \dot\varphi$, an overdot denotes (here and throughout this section) differentiation with respect to proper time, and $V(r;L_z)$ is a certain effective potential [see Eq.~\eqref{effectivepotential} and Fig.~\ref{fig:isco} below]. Since, from symmetry, we have $F^\theta=0$ on the equatorial plane, Eq.\ (\ref{EoMtheta}) tells us that the orbit remains confined to that plane, as expected. In the absence of a GSF, Eqs.\ (\ref{EoMt}) and (\ref{EoMphi}) turn into conservation equations for the geodesic energy $E$ and angular momentum $L_z$. Now, in the specific case of a {\em circular} orbit, the particular symmetries of the setup imply $F_\alpha^{\rm diss}=\left\{F_t,0,0,F_\varphi\right\}$ and $F_\alpha^{\rm cons}=\left\{0,F_r,0,0\right\}$, so that the $t$ and $\varphi$ component of the GSF are purely dissipative while its $r$ component is purely conservative. To consider the purely conservative dynamics in this example thus amounts to setting $F_t=0=F_\varphi$, turning Eqs.\ (\ref{EoMt}) and (\ref{EoMphi}) into conservation equations. However, a conservative component of the GSF, $F_r$, still features in Eq.\ (\ref{EoMr}).   Let us explore the effect of this conservative forcing term, highlighting the role of gauge freedom in the problem.

To begin, we notice that, in the absence of the GSF, Eq.\ (\ref{EoMr}) describes motion in a one-dimensional effective potential. Circular orbits are defined via the conditions $\dot{r}=0=\ddot{r}$, which determine the values of $E$ and $L_z$ as functions of the orbital radius, $r_0$. They also determine the value of the orbital frequency\footnote{ To be precise, $\Omega$ is not a frequency but an angular velocity (i.e., frequency times $2\pi$). In calling it frequency we succumb here to convention.} $\Omega:=\frac{d\varphi}{dt}=\sqrt{M/r_0^3}$ (the same frequency as a Keplerian orbit of the same radius). The conservative forcing term $F^r$ in Eq.\ (\ref{EoMr}) modifies slightly the coordinate location of the circular-orbit equilibrium, and with it also the functional relation between $\Omega$ and $r_0$: a short calculation gives
\begin{equation}\label{Omega}
\Omega=\sqrt{\frac{M}{r_0^3}}\left(1-\frac{r_0^2(r_0-3M)}{2mM(r_0-2M)}\, F^r\right).
\end{equation}
We see that the conservative force $F^r$ appears to slightly ``shift'' (by an amount of order $\eta$) the value of the orbital frequency at a given coordinate radius $r_0$. But there is an important subtlety here. We recall that both the radius and the GSF are gauge dependent. Expressing the perturbed metric in different gauges changes the identification of the value $r_0$ with a particular physical radius, and it also changes the value of $F^r$. In fact, a short calculation \cite{Barack:2007tm} reveals that a small coordinate transformation $r\to r-\xi^r$ (where $\xi^r$ is order-$\eta$) shifts the radius $r_0$ and the value of $F^r$ in precisely the right way so as to cancel each other's order-$\eta$ contribution on the right-hand side of Eq.\ (\ref{Omega}), leaving $\Omega$ {\it invariant} under any such transformation. The orbital frequency, unlike the coordinate radius, is a genuine, physical, ``gauge-invariant'' attribute of the orbit.\footnote{ The sense in which $\Omega$ is said to be gauge invariant will be clarified further below, in Sec.\ \ref{quasi-invariants}. $\Omega$ is not actually invariant under arbitrary gauge transformations.} In this example, the conservative GSF does not have a clear physical (``observable'') significance. Rather, it simply tells us how the invariant frequency is related to radius in any given gauge.


Our first example serves to highlight the kind of subtleties one faces in trying to interpret the physical effect of the conservative GSF, given the gauge ambiguity of the underlying perturbation theory. However, one should by no means conclude that the conservative effect of the GSF is a pure gauge artifact. Indeed, in the rest of this section we will explore a handful of genuine physical effects of the conservative force, which influence the conservative sector of the post-geodesic dynamics in an (essentially) gauge-invariant way. We will also describe invariant quantities constructed directly from the effective metric~$\tilde g_{\mu\nu}$, which have no direct reference to the GSF. Over the past decade a significant proportion of work in the field has been dedicated to obtaining a quantitative description of these effects in Schwarzschild or Kerr spacetimes. The main motivation for such work is to enable comparisons and synergies with other approaches to the two-body problem, as we shall describe later, in Sec.\ \ref{sec:synergy}.

Like in the preceding section, we focus here on first-order effects, and we omit the superscript ``(1)'' from quantities such as $h^{\R(1)}_{\mu\nu}$.




	\subsection{Frequency of the innermost stable circular orbit}\label{sec:ISCOshift}
	
\begin{figure}[htb]
\center
	\begin{tabular}{cc}
        \includegraphics[width=95mm]{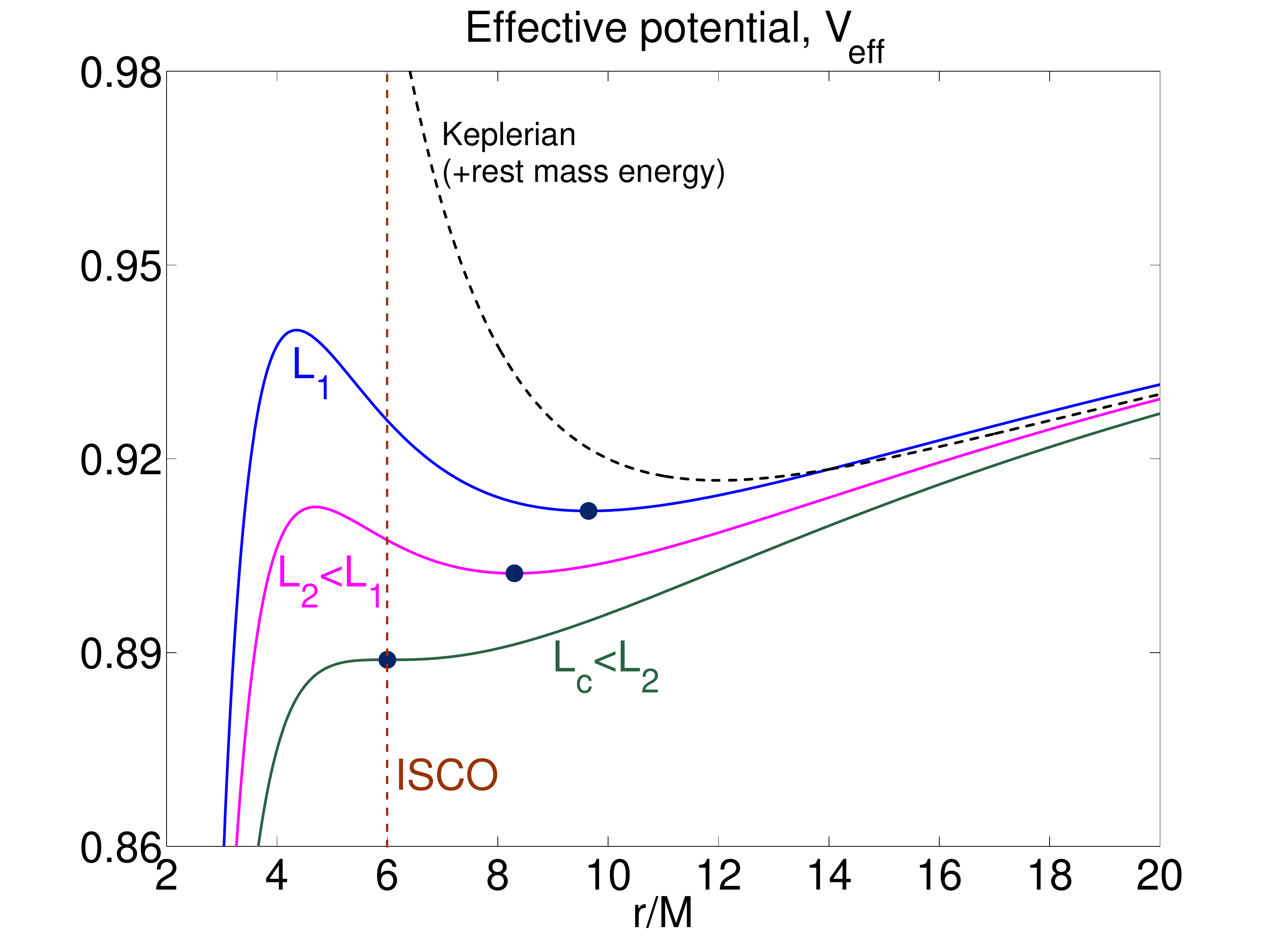}
	\end{tabular}
	\caption{Effective radial potential for geodesic orbits around a Schwarzschild black hole. The radial velocity is governed by $\dot{r}^2=(E/m)^2-V(r;L_z)$, with the effective potential $V(r;L_z)$ [Eq.\ (\ref{effectivepotential})] shown here for three values of the angular momentum $L_z$. For comparison we also show (dashed line) the familiar Keplerian effective potential (with rest mass energy $m$ added, to match the convention for $V$ in the relativistic case). Stable circular orbits correspond to minima of the effective potential, and no such orbits exist for $L<L_c$.   }
	\label{fig:isco}
\end{figure}

In Sec.\ \ref{sec:orbits} we mentioned the existence of innermost stable orbits as a distinct characteristic of the strong-field dynamics around black holes in GR. Here we will consider the deformation of that innermost orbit due to the conservative piece of the GSF. To this end, we start with a slightly more detailed discussion of the situation in the geodesic case, restricting first to circular orbits around a Schwarzschild black hole. In this case, the existence and location of an ISCO can be readily inferred from the radial equation of motion (\ref{EoMr}), where for now we omit the GSF term. In that equation, the effective potential is given by 
\begin{equation}\label{effectivepotential}
V(r;L_z)= \left(1-\frac{2M}{r}\right)\left(\frac{L_z^2}{m^2r^2}+1\right).
\end{equation}
Figure \ref{fig:isco} shows $V(r)$ for a number of $L_z$ values. It starts off resembling the familiar Keplerian potential at large distances, and like it (for sufficiently large $L_z$) it reaches a minimum at some radius depending on $L_z$. But whereas the Keplerian potential goes off to diverge at small separations, the Schwarzschild potential further develops a peak before dropping (to zero) towards the horizon. The minimum of the effective potential marks the location of a circular orbit, where the effective radial force vanishes; it is a {\it stable} circular orbit since the effective radial force acts as a restoring force in the vicinity of the minimum [note the sign of $\ddot{r}$ in Eq.\ (\ref{EoMr})]. The radius of the stable circular orbit decreases with decreasing angular momentum $L_z$. In the Keplerian case one can support a stable circular orbit at arbitrarily small separations by selecting a suitable value of the angular momentum,\footnote{ In the limit $r\to 0$ one would have to take $L_z$ to zero as $\sim r^{1/2}$, corresponding to a diverging tangential velocity $v\sim r^{-1/2}$.} but not so in the Schwarzschild case: here, as $L_z$ decreases, the maximum of $V(r)$ migrates outward to larger radii, until, at some critical value $L_z=L_c$, it merges with the minimum, at which point both extrema disappear. No minimum, or maximum, occur for $L_z<L_c$. The point at which the maximum and minimum merge, and the effective radial force vanishes, is the ISCO.  

The effective-potential picture can also be invoked to describe the situation when the orbit evolves under the effect of the full first-order GSF (including dissipation). Again, for simplicity, consider the adiabatic inspiral of (quasi-)circular orbits in a Schwarzschild geometry. As angular momentum is radiated away, the effective potential $V(r;L_z)$ evolves through a sequence similar to that shown in Fig.\ \ref{fig:isco}, with the minimum migrating leftward. It has been shown \cite{Kennefick:1995za} that, in such a situation, orbital energy dissipates at just the right rate so as to keep an initially circular orbit circular.  Thus, the orbit evolves through a sequence of circular geodesics, sitting (roughly) at the bottom of the potential well as it gradually shifts leftward in Fig.\ \ref{fig:isco}. This adiabatic process proceeds so long as $L_z$ is sufficiently above $L_c$. As the angular momentum approaches its critical value, the effective radial restoring force weakens, and at some point can no longer counter the object's effective radial inertia inwards. At that point, the gradual inspiral transitions into a direct {\it plunge} into the black hole, which proceeds approximately along a fixed geodesic, with (almost) fixed $E$ and $L_z\sim L_c$. The details of the transition from adiabatic inspiral to plunge across the ISCO were first worked out (for a Kerr black hole) in Ref.\ \cite{Ori:2000zn} (see also \cite{Buonanno:2000ef}), which, in particular, concluded that the radial thickness of the transition regime scales as $\Delta r\propto \eta^{2/5}$. Thus, for a true inspiral process under the effect of the full GSF, the notion of a well localized ISCO is replaced with that of a ``smeared out'' transition regime. 

However, a precisely localized notion of an ISCO is recovered when considering the conservative dynamics alone, ``switching off'' the dissipative piece of the GSF in the equations of motion (\ref{EoMt})--(\ref{EoMphi}). The remaining, conservative piece of the GSF perturbs the effective restoring force and thereby perturbs the value of $r$ at which it vanishes (i.e., the ISCO's coordinate location). The coordinate radius $r_{\rm isco}$ of the GSF-shifted ISCO is, of course, gauge dependent, but the order-$\eta$ frequency shift $\Delta\Omega_{\rm isco}:=\Omega(r_{\rm isco})-\Omega_0(6M)$ [where $\Omega(r_0)$ is given in Eq.\ (\ref{Omega}) and $\Omega_0(r_0)=\sqrt{M/r_0^3}$] provides a genuine, physically meaningful measure of the conservative GSF effect. To compute $\Delta\Omega_{\rm isco}$ in practice one has to calculate the contribution of the conservative GSF to the effective restoring force acting on slightly eccentric geodesic orbits, in the limit of vanishing eccentricity, near the ISCO. This calculation was first performed, numerically, for a Schwarzschild black hole, in Ref.\ \cite{Barack:2009ey}.\footnote{ A similar calculation, of the ISCO shift in the toy problem of the scalar-field self-force acting on a scalar charge, was performed earlier in Ref.\ \cite{DiazRivera:2004ik}.} The result is
\begin{equation}\label{ISCOshift}
\frac{\Delta\Omega_{\rm isco}}{\Omega_0(6M)}= 0.25101546\eta ,
\end{equation}
with an error bar $\pm 5\cdot 10^{-8}\eta$.\footnote{ We quote here the more numerically accurate value obtained later in \cite{Akcay:2012ea}. Ref.\ \cite{Barack:2009ey} used a definition of frequency based on the Lorenz-gauge time coordinate, which suffers from a certain pathology. The value given here corrects for that pathology, following \cite{Damour:2009sm}. (More on that in Sec.\ \ref{quasi-invariants} below.)} This order-$\eta$ frequency shift is hardly an ``observable'' effect: after all, for true EMRIs it is typically much smaller than the radial extent of the transition regime, which is proportional to $\eta^{2/5}$. Nonetheless, the value provides an invaluable handle on the strong-field conservative dynamics and a benchmark against which calculations based on other approaches can be tested. In Sec.\ \ref{sec:synergy} we will describe the important role that the GSF ISCO-shift result (\ref{ISCOshift}) has played in advancing studies of the two-body problem in GR.  

So far we have identified the ISCO via the condition of a vanishing effective restoring force. It is natural to ask whether one could instead locate the ISCO by minimizing a suitable conserved energy. Indeed, for circular {\it geodesic} orbits, there is a well-defined conserved energy, given as a function of radius by $E=m(1-2M/r_0)(1-3M/r_0)^{-1/2}$, whose minimization correctly gives $r_0=6M$. One does not {\it a priori} expect there to exist a similar notion of conversed energy in the {\it perturbed} EMRI spacetime, which is lacking a time-translation symmetry. However, in 2011 Le Tiec and collaborators \cite{LeTiec:2011dp} have proposed precisely such an energy function, defined on the conservative sector of the perturbed geometry for circular orbits, whose existence and form came as direct consequences of the so-called ``first law of binary black-hole mechanics'', which had been formulated at around the same time \cite{LeTiec:2011ab,Blanchet:2012at}. Le Tiec {\it et al.}\ showed, as an important test, that minimization of their energy recovers the direct GSF result (\ref{ISCOshift}), providing a first strong vindication of the first law in the strong-field regime. The same notion of energy, generalized to the Kerr geometry, came out later from a systematic Hamiltonian formulation of the conservative dynamics \cite{Fujita:2016igj}, and in Ref.\ \cite{Isoyama:2014mja} it was used to calculate the ISCO frequency shift, through order $\eta$, for orbits in the equatorial plane of a Kerr black hole (as a function of the black hole's spin). This calculation was more recently confirmed with a direct GSF restoring-force analysis in Ref.~\cite{vandeMeent:2016pee}.

In Sec.\ \ref{redshift} we shall return to the first law of binary black holes in our discussion of redshift and its relation to binding energy.

\subsection{Periastron advance of slightly eccentric orbits} \label{sec:precession} 

Consider a circular geodesic orbit around a Schwarzschild black hole, with radius $r_0>6M$, like the one represented by the black dot sitting at the minimum of the upper curve in Fig.\ \ref{fig:isco}. Now consider a small perturbation of that orbit away from circularity (e.g., by endowing it with a small additional amount of energy at a fixed angular momentum). At first order, the radial motion will then exhibit simple harmonic oscillations around the base circular orbit: expanding the right-hand side of Eq.\ (\ref{EoMr}) (without the GSF term) about $r=r_0$, recalling $dV/dr=0$ at $r=r_0$, gives $\ddot r\simeq -\omega^2(r-r_0)^2$, with $\omega^2=\frac{M(r_0-6M)}{r_0^3(r_0-3M)}$. This frequency corresponds to proper time $\tau$. To convert it to a frequency with respect to the usual coordinate time $t$---call it $\Omega_r$---we write $\Omega_r=(d\tau/dt)\omega$, where the ``redshift'' factor is $d\tau/dt=(1-3M/r_0)$. Thus we have 
\begin{equation}\label{Omegar}
\Omega_r = \Omega\left(1-\frac{6M}{r_0}\right)^{1/2},
\end{equation}
where $\Omega=(M/r_0^3)^{1/2}$, recall, is the azimuthal frequency of the base circular geodesic. Note how $\Omega_r$ vanishes at the ISCO, as expected, and how this radial oscillation frequency is always smaller than $\Omega$: it takes the object longer to complete one radial oscillation than it does to complete one $\varphi$ rotation. This is the origin of the celebrated effect of relativistic periastron advance. The frequency of periastron precession is $\Omega-\Omega_r$, and over a full radial period, $\Delta t=2\pi/\Omega_r$, it gains a $\varphi$-phase equal to $(\Omega-\Omega_r)\Delta t = 2\pi(\Omega/\Omega_r-1)$. Thus, recalling Eq.\ (\ref{Omegar}), we obtain that the amount of periastron advance per radial cycle is given by
\begin{equation}\label{delta}
\delta=2\pi\left[\left(1-\frac{6M}{r_0}\right)^{-1/2}-1\right].
\end{equation}

This is a textbook result, and the reason we have elaborated on its derivation here is that it should now be straightforward to understand how it is modified by the action of the conservative GSF. The GSF modifies (at order $\eta$) the effective restoring force near the minimum of the potential, and with it it modifies (at order $\eta$) the value of $\Omega_r$ for a given circular orbit. At this point we must be mindful of gauge ambiguity, and we therefore adopt the frequency $\Omega$, and not the coordinate radius $r_0$, as a label of the base circular orbits. We then say that the conservative GSF modifies the {\it functional relation} between the two invariant frequencies $\Omega_r$ and $\Omega$. We express this in the form 
\begin{equation}\label{OmegarGSF}
\Omega_r(\Omega)=\Omega_r^{(0)}(\Omega)+\Delta\Omega_r(\Omega),
\end{equation}
where $\Omega_r^{(0)}(\Omega)$ is the background geodesic relation, and $\Delta\Omega_r(\Omega)(\sim\eta)$ is the GSF correction to it. The background relation itself is defined by 
\begin{equation}\label{Omegar0}
\Omega_r^{(0)}(\Omega) := \Omega\left(1-\frac{6M}{r_{\Omega}(\Omega)}\right)^{1/2},
\end{equation}
obtained from (\ref{Omegar}) by replacing $r_0$ with its geodesic expression in terms of $\Omega$,
$r_{\Omega}(\Omega)=(M/\Omega^2)^{1/3}$---but note this replacement is not unique, and one sometimes finds it more natural to define $r_{\Omega}(\Omega)=[(M+m)/\Omega^2]^{1/3}$ (this recalls the actual relation in the analogue Kepler problem). Note that how we choose to define the background function $\Omega_r^{(0)}(\Omega)$ affects what we mean by the ``GSF correction'' $\Delta\Omega_r$, and one should be aware of this ambiguity; what remains unambiguous is the {\em sum} of the two terms on the right-hand side of Eq.\ (\ref{OmegarGSF}). Finally, now that we have $\Omega_r(\Omega)$ through order $\eta$, the derivation of the periastron advance $\delta(\Omega)$ proceeds just as in the geodesic case, leading to an expression of the form 
\begin{equation}\label{deltaGSF}
\delta(\Omega)=\delta^{(0)}(\Omega)+\Delta\delta(\Omega),
\end{equation}
where the background relation $\delta^{(0)}(\Omega)$ is the one obtained from (\ref{delta}) by  replacing $r_0\to r_\Omega(\Omega)$.

A calculation of $\Delta\delta(\Omega)$ requires knowledge of the conservative piece of the GSF on slightly eccentric orbits, in the limit of vanishing eccentricity. This calculation was first performed (numerically) in Ref.\ \cite{Barack:2010ny} for a Schwarzschild black hole, and later for equatorial orbits in Kerr in Ref.\ \cite{vandeMeent:2016hel}.  The GSF results were shown to agree nicely with post-Newtonian calculations in the weak-field (large $r_0$) regime. 

The function $\delta(\Omega)$ provides a powerful diagnostic of the conservative dynamics in EMRIs, and it has been utilized in that role on many occasions. The comparison with PN calculations, besides providing much reassurance, also tells us about the convergence properties of the PN expansion (at least through order $\eta$) and benchmarks its performance in the strong field. As we shall describe in Sec.\ \ref{sec:synergy}, it was even used to decide between two initially conflicting recent results for the fourth-order PN equation of motion. Also in Sec.\ \ref{sec:synergy} we will describe how $\delta(\Omega)$ has been used to facilitate rewarding 3-way comparisons between GSF, PN and full NR calculations, and to help calibrate the potentials of EOB theory.


	\subsection{Spin precession and self-torque}
	
Another familiar effect of GR dynamics is the one referred to (somewhat nondescriptively) as the ``geodetic effect'', or ``geodesic precession'': the behavior associated with the failure of a spin vector to return to itself after being parallelly transported along a closed curve in curved spacetime.	This is also known as the {\it de Sitter precession}, after W.\ de Sitter, who, in as early as 1916, predicted that the rotation axis of the Earth--Moon system as it moves around the Sun should experience a precession of $\sim 1.9$ arcseconds per century due to GR corrections to the Newtonian dynamics \cite{1916MNRAS..76..699D}  (a prediction since confirmed using Lunar laser ranging experiments). A much larger de Sitter effect of $\sim 6.6$ arcseconds per {\it year} was measured using a gyroscope in a polar Earth orbit as part of the Gravity Probe B experiment \cite{PhysRevLett.106.221101}. Larger still is the effect observed in the (only known) double-pulsar system, PSR J0737-3039, where the spin of one of the pulsars appears to be precessing at a rate of $\sim 4.8$ {\it degrees} per year \cite{Breton:2008xy}. Given the binary's 2.45-hour orbital period, this translates to $\sim 3.7\times 10^{-6}$ radians of precession angle per radian of orbital revolution---still a rather meagre effect when measured this way.

A much more extreme manifestation of spin precession can be found in the late inspiral of compact-object binaries, including EMRIs, where its rate can reach $O(1)$ radians per radian of orbital revolution. We can quantify this precisely in the case of a spinning test particle in a circular orbit around a Schwarzschild black hole, in the small-spin limit where the particle can be assumed to follow a geodesic of the background geometry. Then the angle of spin precession per radian of orbital revolution is given in exact form by the simple expression
\begin{equation}\label{psi}
\psi=1-\left(1-\frac{3M}{r_0}\right)^{1/2},
\end{equation}
and we see that $\psi$ can be as large as $\sim 0.3$ (at the ISCO, $r_0=6M$) in this scenario: that is some $105^\circ$ rotation of the spin axis over a single orbital period! This primary geodesic effect is perturbed when the particle's finite (nonzero) spin magnitude and/or mass are considered. A finite spin magnitude affects the motion in two ways: by giving rise to a Mathisson-Papapetrou force term in the equation of motion [recall Eq.\ (\ref{1st order eqn motion})], and by perturbing the metric of spacetime. Ignoring both these effects (by taking the limit of vanishing spin magnitude), we can ask how the finite mass of the particle modifies the precession rate $\psi$ in the conservative dynamics (i.e., ignoring the dissipative piece of the GSF as elsewhere in this section).  

At linear order in $m$, the answer can be easily found thanks to a result by Harte \cite{Harte:12}, who showed that, as the particle moves on a geodesic of the effective metric $\tilde g_{\alpha\beta}$, its spin vector gets parallelly transported in that same effective metric. Just as geodesic motion in $\tilde g_{\alpha\beta}$ corresponds to self-acceleration in $g_{\alpha\beta}$, parallel transport in $\tilde g_{\alpha\beta}$ corresponds to a {\em self-torque} in $g_{\alpha\beta}$: an effective torque exerted on the spinning particle by its own perturbation, which, in particular, modifies the rate of precession $\psi$ away from its geodesic value given in (\ref{psi}). Again using the orbital frequency $\Omega$ as an ``invariant'' label of the circular orbit, we write
\begin{equation}\label{psiGSF}
\psi(\Omega)=\psi^{(0)}(\Omega)+\Delta\psi(\Omega),
\end{equation}
where the background relation $\psi^{(0)}(\Omega)$ is the one obtained from (\ref{psi}) by  replacing $r_0\to r_{\Omega}(\Omega)$. An explicit formula for the $O(\eta)$ self-torque correction $\Delta\psi(\Omega)$, which involves $h_{\alpha\beta}^\R$ and its first derivatives on the particle, was derived in Ref.\ \cite{Dolan:2013roa}. This work also performed a first numerical calculation of $\Delta\psi(\Omega)$, showing agreement with the predictions of PN theory at large $r_0$ (small $\Omega$) and quantifying its gradual breakdown in the strong field. As we describe in the next section, follow-up work extended this comparison to very high order in the PN expansion, and also used the GSF results for $\Delta\psi(\Omega)$ to provide further calibration of the EOB model. In this fashion, the self-torque effect has been utilized as an additional diagnostic of the strong-field dynamics, complimentary to that provided by $\delta(\Omega)$.

In Ref.\ \cite{Akcay:2016dku}, Akcay {\it et al.}~introduced a generalized notion of $\psi$ applicable to any bound (eccentric) geodesic orbit in Schwarzschild geometry. Here, the self-torque effect is computed over a full radial cycle, and expressed in terms of the two frequencies $\Omega$ and $\Omega_r$. Again, a successful comparison was made with equivalent expressions calculated analytically in a PN framework. This programme of comparison was further pursued in \cite{Kavanagh:2017wot}. Similar calculations are underway in Kerr \cite{Akcay:2017azq,vdMeent:pc}.

	\subsection{Self-tides}
	
Spin-precession information, just like the GSF, is encoded in the {\it first} derivatives of the effective metric $\tilde g_{\alpha\beta}$ (evaluated on the orbit). Other invariant quantities of interest may be constructed from higher-order derivatives, as first suggested by Bini and Damour in \cite{Bini:2014ica}.  In Ref.\ \cite{Dolan:2014pja}, Dolan {\it et al.}~considered a set of {\it tidal} invariants constructed from second derivatives of $\tilde g_{\alpha\beta}$. To motivate these quantities, we refer back to the Newtonian scenario of a small body immersed in an external gravitational potential $\Phi_{\rm ext}$. The body's center of mass $z^i$ is subject to an acceleration $a_i=-\partial_i \Phi_{\rm ext}(z)$. Relative to the center of mass, a mass element inside the body, at $z^i+\delta x^i$, experiences a {\em tidal} acceleration $a_i^{\rm tidal}=-[\partial_i\Phi_{\rm ext}(z+\delta x)-\partial_i\Phi_{\rm ext}(z)] = -\Phi_{,ij}^{\rm ext}(z)\delta x^j+O[(\delta x)^2]$. The quantity ${\cal E}_{ij}=\Phi^{\rm ext}_{,ij}(z)$, referred to as the quadrupole tidal moment, characterizes the (leading-order) tidal forces on the body.

In GR, tidal forces are described by the Riemann curvature tensor. Thinking of $\tilde g_{\alpha\beta}$ as the external metric, we can construct a relativistic tidal moment ${\cal E}_{\alpha\beta}:=\tilde R_{\alpha\beta\gamma\delta}\tilde u^{\beta}\tilde u^{\delta}$, where $\tilde R_{\mu\nu\beta\delta}$ is the Riemann tensor associated with $\tilde g_{\alpha\beta}$, and $\tilde u^\alpha$ is the particle's four-velocity normalized in $\tilde g_{\alpha\beta}$. Unlike in Newtonian gravity, in GR there is an additional tidal moment, ${\cal B}_{\alpha\beta}:=\frac{1}{2}{\epsilon_{\alpha\gamma}}^{\mu\nu}\tilde R_{\mu\nu\beta\delta}\tilde u^{\gamma}\tilde u^{\delta}$, where $\epsilon_{\alpha\beta\mu\nu}$ is the fully antisymmetric Levi-Civita tensor. While ${\cal E}_{\alpha\beta}$ characterizes stretching, ${\cal B}_{\alpha\beta}$ characterizes twisting; they are sometimes referred to as electric- and magnetic-type tidal moments, respectively, or as the tidal field and the ``frame-drag" field. Focusing on circular orbits in the equatorial plane of a Kerr black hole, Dolan {\it et al.} showed that the eigenvalues of these tensors are invariant quantities. 
They identified four nontrivial such eigenvalues, three associated with ${\cal E}_{\alpha\beta}$ and one with ${\cal B}_{\alpha\beta}$. Letting $\lambda\in\{\lambda_i^E,\lambda^B\}\ (i=1,2,3)$, they then split each $\lambda$ into ``background'' and ``particle'' contributions in the usual way,
\begin{equation}\label{psiGSF}
\lambda(\Omega)=\lambda^{(0)}(\Omega)+\Delta\lambda(\Omega),
\end{equation}
giving explicit expressions for the ``self-tides'' contribution $\Delta\lambda(\Omega)$ in terms of $h_{\alpha\beta}^{\R}$ and its first and second derivatives on the particle. Ref.\ \cite{Dolan:2014pja} then performed a numerical calculation of $\Delta\lambda(\Omega)$ in the case of a Schwarzschild background, and demonstrated how the results agree at small $\Omega$ with the analytical predictions of an independent post-Newtonian calculation. The latter was truly independent in that it had no reference to a point particle or self-force: it was based on \cite{Taylor:2008xy} and \cite{JohnsonMcDaniel:2009dq}, where the approximate form of the metric near a small black hole immersed in a weakly-curved tidal environment was derived. The agreement is thus nontrivial and highly informative. 
     
However, there is good reason to question whether $\lambda$ genuinely describes the physical tidal environment of the small object. After all, $\tilde g_{\alpha\beta}$ is {\it not} the true metric; it is only an effective one. We know that the particle follows a geodesic in this effective metric and that its spin gets parallelly transported in it. But what is the physical significance of the tidal field associated with it? Given the degree of ambiguity involved in the very definition of the field $h_{\alpha\beta}^\R$ (recall our discussion in Sec.~\ref{local analysis}), should we expect it to have a clear physical meaning at all? The agreement with the post-Newtonian analysis suggests that, in fact, it has. Additional light was shed on this question in Ref.~\cite{Pound:2017psq} using a systematic matched-asymptotic-expansions analysis. This work demonstrated explicitly that ${\cal E}_{\alpha\beta}$ and ${\cal B}_{\alpha\beta}$ are {\em not} equal to at least one particular pair of tidal moments that have been used to describe the local spacetime geometry near a tidally perturbed small object. But it also showed that due to an inherent ambiguity in the definition of tidal moments in GR, one {\em can} write the metric of the local spacetime in such a way as to interpret ${\cal E}_{\alpha\beta}$ and ${\cal B}_{\alpha\beta}$ as the physical tidal moments. Hence, while there is some danger of over-interpreting $h^\R_{\alpha\beta}$, the tidal moments constructed from it {\em can} be thought of as one {\em particular} characterization of the object's tidal environment. 

In \cite{Nolan:2015vpa}, Nolan {\it et al.}~have advanced this programme one step further by defining and calculating {\it octupolar} tidal invariants associated with $\tilde g_{\alpha\beta}$, which involve {\it third} derivatives of the perturbation $h_{\alpha\beta}^{\R}$. Here too, focusing on circular orbits in Schwarzschild, they were able to obtain accurate numerical results and demonstrate their consistency with analytical PN calculations.

	\subsection{Redshift and binding energy}  \label{redshift}

We have kept for last a conservative GSF effect that, in fact, had been the first to be identified and analyzed (by Detweiler in 2008), the first to enable a direct comparison between GSF and PN results (again by Detweiler), and one whose fundamental role in perturbation theory has since been gradually revealed through the work of many. If we have kept it for last it is only because its physical interpretation is somewhat more subtle. In \cite{Detweiler:2008ft},  Detweiler introduced the {\it redshift variable} $z(\Omega)$ as a diagnostic of the conservative dynamics  of circular orbits in Schwarzschild geometry.\footnote{ To be pedantic, we note that Ref.\ \cite{Detweiler:2008ft} actually considered the inverse redshift, $z^{-1}=\tilde u^t$. The use of $z$ became prevalent in more recent literature.} Now often referred to as ``Detweiler's redshift'', it is defined as follows. Consider the particle's four-velocity $\tilde u^\alpha$ ,
as normalized in the effective metric $\tilde g_{\alpha\beta}$ [i.e., $\tilde g_{\alpha\beta}\tilde u^\alpha \tilde u^\beta=-1$  through $O(\eta)$]. Detweiler's redshift is then
\begin{equation}\label{zdef}
z:=(\tilde u^t)^{-1}= \frac{d\tilde\tau}{dt},
\end{equation}
where, like in Sec.~\ref{equation of motion}, $\tilde\tau$ is proper time as measured in $\tilde g_{\mu\nu}$. Labelling the circular orbit by its frequency $\Omega$, as usual, we then extract the $O(\eta)$ effect by writing 
\begin{equation}\label{z}
z(\Omega)= z^{(0)}(\Omega)+\Delta z(\Omega).
\end{equation}
The value of $z$ on the zeroth-order geodesic in the Schwarzschild background is $z^{(0)}(\Omega)=1-\frac{3M}{r_\Omega(\Omega)}$. Detweiler showed that $\Delta z(\Omega)$ can be extracted from the value of the field $h^\R_{\alpha\beta}$ on the particle alone (the derivatives of $h^\R_{\alpha\beta}$ are not needed), and that $z$ depends on $h^\R_{\alpha\beta}$ only through the simple scalar contraction $h^\R_{\alpha\beta}u^\alpha u^\beta$. He also showed that $\Delta z(\Omega)$ is an invariant quantity, in a certain useful sense (to be clarified in the next subsection).

What is the physical interpretation of $z$? It is tempting to suggest that $z$ measures gravitational redshift: the ratio between proper-time lapse along the orbit and proper-time lapse of inertial observers far away. However, we must recall $z$ is calculated in the effective geometry $g_{\alpha\beta}+h^\R_{\alpha\beta}$, not in the true, physical geometry (the actual redshift in the latter is predominantly due to the singular piece of the physical metric perturbation, which diverges on the particle). In the case that the small object is a black hole, a more strictly physical interpretation has emerged in recent years: $z$ is (up to a proportionality factor) equal to the {\it surface gravity}, $\kappa$, of the small black-hole's horizon. This relationship was first surmised based on the fact that the two quantities play analogous roles in the first law of binary black hole mechanics (discussed further below)~\cite{LeTiec:2011ab}. Using a matched-asymptotic-expansions analysis, one of us~\cite{Pound:unpublished} later showed that this relationship is {\em exact} through linear order in the mass ratio. The surface-gravity interpretation was impressively confirmed by Zimmerman {\it et al.} even in fully nonlinear GR, using full NR simulations of black-hole binaries \cite{Zimmerman:2016ajr}.\footnote{ More precisely stated, the surface-gravity interpretation of $z$ was confirmed in Ref.\ \cite{Zimmerman:2016ajr} assuming the validity of the first law. Conversely, if one assumes that interpretation, the results of \cite{Zimmerman:2016ajr} can be considered a test of the first law.} 
Thus, $\Delta z$ describes the $O(\eta)$ contribution to the surface gravity of the small hole's horizon due to back-reaction from its own field.  

The redshift variable $z(\Omega)$ has played a central role in the development of both GSF and PN methods over the past decade. Soon after its introduction, it was used as a benchmark in first comparisons of numerical GSF results obtained in different coordinate gauges \cite{Sago:2008id,Shah:2010bi}, in facilitating a first contact with PN theory \cite{Detweiler:2008ft}, and in  benchmarking first results in Kerr \cite{Shah:2012gu}. It later enabled high-precision comparisons with PN theory and the accurate calibration of the EOB potentials, both activities to be described in Sec.\ \ref{sec:synergy} below. Because $z(\Omega)$ requires only the perturbation field and not its derivatives, it is relatively easy to extract from numerical solutions of the field equations, and thus provides a first testing point in developing new GSF codes. 

But besides possessing practical value, $z$ also turned out to be of some fundamental significance. In 2011, Le Tiec and collaborators \cite{LeTiec:2011ab} formulated the ``first law of binary mechanics'': a conjectured\footnote{ The validity of the first law was rigorously established in a PN context and for a range of certain circumstances---cf.\ \cite{Tiec:2015cxa,Blanchet:2017rcn}.} variational formula, which, remarkably, when applied to the circular-orbit EMRI setup, relates the global properties of the system---binding energy and total angular momentum---to the local redshift $z$ as measured on the particle.   Using this formula it is possible to write down explicit expressions for the binary's binding energy and angular momentum through $O(\eta^2)$ (i.e., including the leading-order self-gravity term) in terms of $\Delta z(\Omega)$ alone \cite{LeTiec:2011dp}. This is remarkable, because a similar calculation of the global energy and angular momentum using systematic perturbation theory alone would require knowledge of the second-order metric perturbation. The first law thus appears to provide a strikingly ``easy'' access to second-order information! At the practical level, one then gains access to an additional, important gauge-invariant property of the binary, namely its binding energy, with an ability to compute the GSF correction to it without having to go to second order. This, indeed, has been exploited in many synergistic studies with PN theory and EOB. Knowledge of the binding energy also allowed, as we have seen before, an ``easy'' route to the determination of the ISCO frequency shift in the Kerr case \cite{Isoyama:2014mja} (through minimization of that energy---a calculation later confirmed with a direct GSF calculation). The relation between Detweiler's redshift and binding energy manifests itself clearly in the recently developed Hamiltonian formulation of EMRI dynamics \cite{Fujita:2016igj}, in which the two-body interaction piece of the Hamiltonian is expressed as a simple function of $\Delta z$. 

The original (Detweiler's) definition of $z$ as an invariant variable relies crucially on the existence of helical symmetry in the problem, as in the circular-orbit setup (ignoring dissipation).  Later on, in \cite{Barack:2011ed}, a generalized notion of redshift was introduced, applicable to any bound periodic orbit in Schwarzschild. The Barack--Sago generalized redshift is given by 
\begin{equation}
z(\Omega,\Omega_r) = \langle \tilde u^t \rangle^{-1} = \frac{\text{$\tilde\tau$ period}}{\text{$t$ period}},
\end{equation}
where $\langle \cdot \rangle$ denotes an average with respect to proper time over a complete radial period of the motion, and on the right-hand side we have, one finds, simply the ratio between the radial period measured in proper time $\tilde\tau$ and the radial period measured in $t$. This, it was shown, is invariant within a class of gauges consistent with the periodicity of the setup (i.e., ones in which the metric perturbation has a bi-periodic spectrum with fundamental frequencies $\{\Omega,\Omega_r\}$). These two frequencies are also used as an invariant label of the bound geodesics,\footnote{ As mentioned in an earlier footnote, the two frequencies are not actually good global parameters, due to the isofrequency phenomena studied in \cite{Warburton:2013yj}. However, one can still use them as parameters with due caution in local regions of the parameter space. } and in splitting $z$ into ``background'' and ``self-force'' contributions in the usual way. Calculations of the generalized $\Delta z(\Omega,\Omega_r)$, with detailed comparisons to PN predictions, were carried out in, e.g., \cite{Akcay:2015pza,Bini:2016qtx} for Schwarzschild and \cite{vandeMeent:2015lxa,Bini:2016dvs} for Kerr. The introduction of a generalized redshift has motivated the formulation of a similarly generalized first-law of binary mechanics---by Le Tiec in 2015 \cite{Tiec:2015cxa}---which divorced itself from its own reliance on helical symmetry.   This, in turn, provided an important link between $\Delta z(\Omega,\Omega_r)$ and some of the EOB potential terms associated with radial momentum, giving a new handle on the latter \cite{Akcay:2015pjz,Bini:2015bfb,Bini:2016dvs}.

	\subsection{Comment on invariants and quasi-invariants}  \label{quasi-invariants}
	
In closing this section, it is appropriate that we make more precise the notion of gauge invariance implied in our discussion. We have announced that the frequency $\Omega$ was a gauge-invariant parametrization of circular orbits, and that $\Delta\delta(\Omega)$, $\Delta\phi(\Omega)$, $\Delta\lambda(\Omega)$ and $\Delta z(\Omega)$ were gauge-invariant functional relations. 
But in truth, none of these quantities are gauge invariant in the usual mathematical sense. In perturbation theory, a linearized quantity (e.g., a field) is said to be gauge invariant if its value is unaffected by the action of any linear diffeomorphism of the form $x^{\alpha}\to x^\alpha-\xi^\alpha$, where $\xi^\alpha$ is a small gauge displacement (of order $\eta$, in our case). Examples are the components of the Riemann tensor linearized about flat spacetime, or the components of the Ricci tensor linearized about any vacuum spacetime.\footnote{ For a Kerr background, Aksteiner and B\"{a}ckdahl \cite{Aksteiner:2018vze} have identified a set of 18 independent gauge-invariant linearized perturbations (including the 10 components of the Ricci perturbation), defined from linear combinations of second and third derivatives of the metric perturbation, which constitutes a minimal ``generating'' set in the sense that all other geometrical invariants can be derived from it by taking derivatives and linear combinations.} But consider, for example, the frequency $\Omega=d\varphi/dt$ of a circular orbit. It is easy to find gauge displacements $\xi^\alpha$ that {\it do} change $\Omega$: take, for instance, $\xi^\alpha=\varpi\, t\, \delta^\alpha_\varphi$ (where $\varpi$ is some order-$\eta$ constant), under which $\varphi\to \varphi-\varpi t$ and thus $\Omega\to \Omega-\varpi$. This gauge transformation, which is perfectly legitimate mathematically, has clearly modified the ``gauge-invariant'' frequency. What happened here, of course, is that $\xi^\alpha$ generated rotation with respect to the original system. It is not surprising to find that the frequency defined with respect to the rotating system differs from that defined in the original system. 

In light of this example, we must refine our language. When we say $\Omega$ is ``gauge invariant'', what we really mean is that it is invariant under a certain, suitably restricted class of ``physically reasonable'' gauge transformations, which must exclude, in our case, ones that artificially generate rotation. But our example points to a true practical dilemma often encountered in GSF calculations. 
In perturbation theory gauge conditions are rarely given as explicit coordinate conditions and more typically formulated in terms of conditions on the metric perturbation itself. Given a metric perturbation in a particular gauge, how do we tell if the gauge is ``physically reasonable''? Or, equivalently, given two perturbations that, we are told, are related via a ``nonphysical'' gauge transformation (as in the above example), how can we tell which of the two gauges (if any) is the ``good'' one?

These are important questions, because they impact on our ability to assign a proper physical interpretation to our GSF results and, as a result, on our ability to compare our results with those obtained in other gauges or using other methods. To address these questions, we must accompany the definition of each of our ``invariant'' quantities [$\Omega$, $\Delta z(\Omega)$, etc.] with a precise statement about the class of ``physically reasonable'' gauges within which it remains invariant. That statement should be in the form of conditions on the metric perturbation, and these conditions must refer to some genuine invariant attributes of the spacetime in question.  In the case of circular orbits, one such attribute is {\it helical symmetry}, and another is {\it asymptotic flatness}. One demands that the metric perturbation respects the helical symmetry, meaning $(\partial_t+\Omega\partial_\varphi)h_{\alpha\beta}=0$, and that it has the right fall-off at infinity. (In our above example of a ``rotating'' gauge it would be the form of the metric perturbation at infinity that would identify it as a ``noninertial'' one and thus exclude it from consideration.) Alternative conditions must be imposed in setups that are not helically symmetric, such as eccentric-orbit configurations or inclined orbits in Kerr. These can rely on an ``averaged'' notion of symmetry or on the periodic properties of the problem, in combination with asymptotic flatness \cite{Barack:2011ed,vandeMeent:2016pee}. The formulation of conditions to define a class of suitable gauges is a subtle matter, which depends on the invariant quantity being calculated and on what the calculation aims to achieve. For example, when the goal is to enable comparison with a PN calculation, one must ensure that both calculations are done within the same class of gauges.  

In practice, however, one often finds in GSF calculations that the gauges most convenient to work with (e.g., the Lorenz or radiation gauges) fall outside the class of gauges deemed suitable for a particular calculation. One then has to introduce, {\it a posteriori}, a correction to the quantity being calculated, which accounts for the effect of a gauge transformation from the working gauge (say, Lorenz or radiation) onto the class of physically suitable gauges.  As an example, consider the calculation of the ISCO frequency shift described above, in Sec.\ \ref{sec:ISCOshift}. The Lorenz-gauge metric perturbation used for that calculation in \cite{Barack:2009ey} turns out to suffer from a minor gauge pathology: the monopole part of the perturbation does not decay at infinity but rather approaches a constant value. This is easily cured with a simple gauge transformation (a small shift of the time coordinate) away from Lorenz---but that gauge transformation changes the value of $\Delta\Omega_{\rm isco}$. The value quoted in Eq.\ (\ref{ISCOshift}) is the gauge-corrected one, and it is that value one can meaningfully compare to, say, PN results. Similar gauge corrections were necessary, and have been introduced, in other calculations of ``invariant'' quantities, such as in the radiation-gauge calculation of the periastron advance in Kerr \cite{vandeMeent:2016hel}.

In summary, the various physical quantities whose GSF corrections we have considered in this section---$\Omega_{\rm isco}$, $\delta$, $\psi$, etc.---are not gauge invariant in the strict mathematical sense, but we may refer to them as {\it quasi-invariants}. They are invariant within a certain class of gauges that must be defined {\it ad hoc}, depending on the purpose of our calculation. The process of identifying a suitable class of gauges, and of performing any necessary gauge adjustments, is an important prerequisite in the programme of comparison and synergy with other methods. We turn to discuss this programme next.

\section{Comparisons and synergies}\label{sec:synergy}

As we have seen, comparison of results from GSF calculations with the analytical predictions of the PN theory provides a powerful overall check of both the numerical calculation itself and the GSF theory underpinning it.  In fact, this is a mutual test of both GSF and PN methods. Each of these frameworks relies on subtle regularization procedures and involves a chain of subtle computational manipulations. Thus, finding an agreement on the ``final value'' of a concrete physical quantity is greatly nontrivial and highly gratifying. 

But such comparisons can achieve more than just mutual tests. The GSF and PN methods are both systematic approximation approaches to the two-body problem in GR, each based on a perturbative expansion in a different limiting domain of the problem. The PN method expands about the limit of large separation (weak, Newtonian gravity), keeping the mass ratio $\eta$ arbitrary, while the GSF method expands about the limit $\eta\to 0$ (geodesic motion), keeping the separation arbitrary. Combining these two complementary approximation methods can be useful in at least three different ways.
First, using GSF results as a benchmark, one can directly study the convergence properties of the PN expansion in the strong field, at least through $O(\eta)$. Similarly, analysis of how PN terms depend on $\eta$ may provide hints about the convergence properties of the GSF expansion. 
Second, one can use GSF information to determine high-order terms in the PN expansion that are hard to obtain with a direct PN calculation (hence pushing the validity of the PN expansion further into the strong-field regime). Similarly, one can use PN information to ``predict'' certain high-order GSF terms. 
Finally, one can hope to inform a model of the two-body problem in an intermediate domain that is perhaps not accessible to either of the approximation methods separately. 

\begin{figure}[htb]
\center
        \includegraphics[width=85mm]{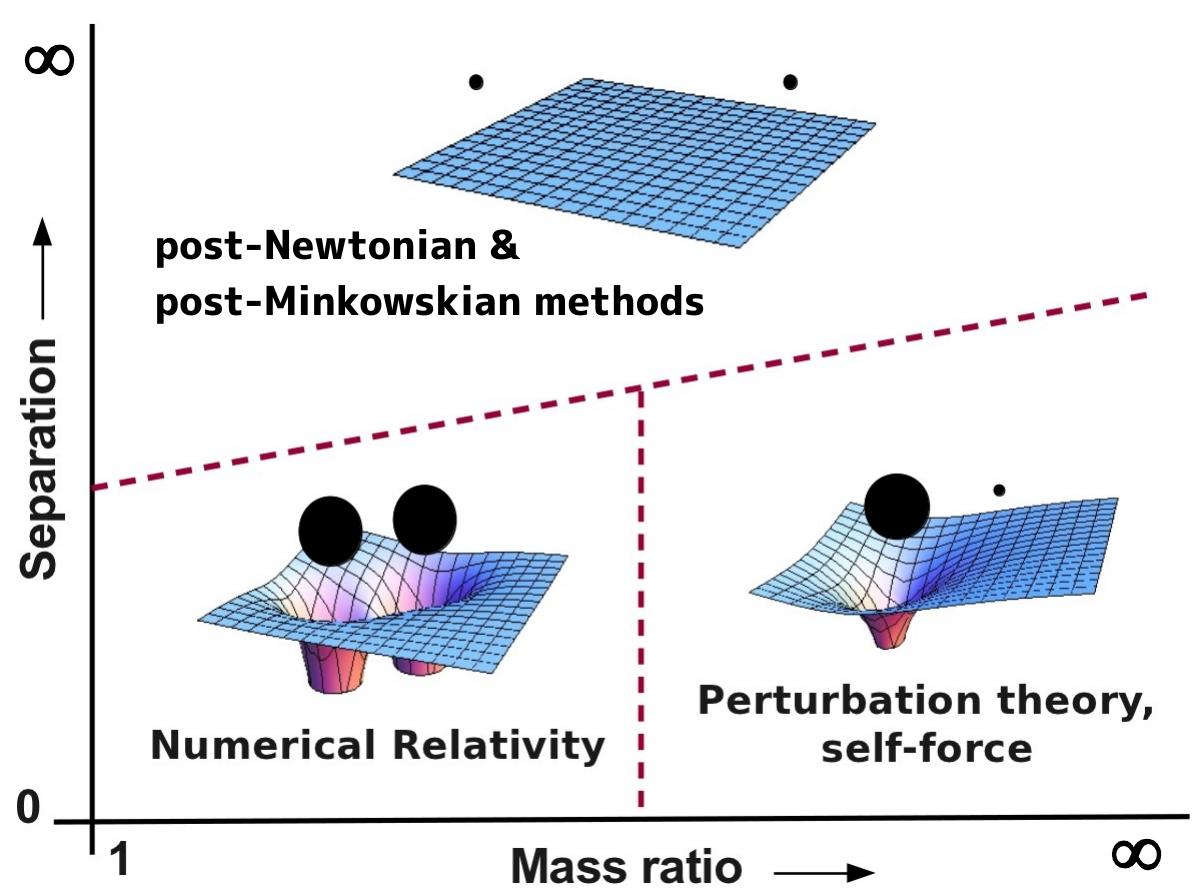}
	\caption{Domains of the two-body problem in GR. The dashed lines schematically delineate the applicability domains of the three main approximation methods. Cartoons illustrate the principle behind each approach: PN expands about flat space, GSF expands about the exact fixed geometry of a central black hole, and NR tackles the full nonlinear dynamics. Synergistic work seeks to interface between results from the three types of calculations. }
	\label{fig:twobody}
\end{figure}

This synergistic programme can be expanded to include NR, which directly solves the fully nonlinear Einstein field equations that describe the inspiral and merger dynamics in black-hole binaries. The idea of such a three-fold synergy is illustrated in Fig.\ \ref{fig:twobody}, showing the respective applicability domain of each of the three methods, NR, PN and GSF, in the essential two-parameter space of the two-body problem: mass ratio vs.\ separation (or strength of gravitational interaction). NR simulations can model the two-body dynamics of two comparable-mass black holes close to their merger, but become computationally prohibitive when the separation is large or the mass ratio too small ($\lesssim 1\!\!:\!\!10$, in practice). In each of these situations, a ``small parameter'' presents itself in the problem, making it amenable to a perturbative description: PN for large separations, and GSF for small mass ratios. The synergistic studies to be reviewed in the rest of this section seek to explore the interfaces between the three approaches and reach across them. This is often done in the framework of the EOB approach, which aims to provide a universal, semi-analytical model of the two-body dynamics across its entire parameter space; this, too, will be reviewed below.

For a fuller review of work on the overlap between the various approaches to the binary black-hole problem, see \cite{Tiec:2014lba}.

	\subsection{Interface with post-Newtonian theory}

As mentioned in Sec.\ \ref{sec:conservative}, a first direct contact with PN theory was made by Detweiler in a 2008 paper \cite{Detweiler:2008ft}, utilizing the relation $z(\Omega)$ for circular orbits in Schwarzschild as a benchmark for the comparison. Using large-$r_0$ fits to increasingly more accurate numerical GSF results as they became available, later calculations have extracted the $O(\eta)$ piece of $z(\Omega)$ to a very high PN order \cite{Blanchet:2009sd,Blanchet:2010zd,Shah:2013uya,Blanchet:2014bza,Johnson-McDaniel:2015vva}. Such high-order PN terms of $z(\Omega)$ have also been obtained via a direct analytical calculation \cite{Bini:2015bla,Kavanagh:2015lva}, using the semi-analytical approach based on the Mano--Suzuki--Takasugi method, which we have mentioned in Sec.\ \ref{subsec:comp_techniques}. 
Conversely, $O(\eta^2)$ terms of $z(\Omega)$ have been used to ``predict'' certain components of the second-order GSF, which are yet to be directly calculated \cite{Bini:2016cje}. There is a large body of work on the PN expansion of other physical GSF quantities---see, e.g., \cite{Hopper:2015icj,Bini:2016dvs,Kavanagh:2017wot,Bini:2017wfr} and references therein. 

The ISCO-shift calculation described in Sec.\ \ref{sec:conservative} typifies the way in which a GSF result can serve as a strong-field benchmark against which the performance of the PN approximation may be assessed. Long before the advent of GSF methods, the value of the ISCO shift had been providing a comparison point for various resummation techniques that have been suggested over the years for improving the convergence of the PN expansion in the strong field (see, e.g., \cite{Damour:2000we,Blanchet:2002mb}). The GSF calculation of the ISCO shift in 2009 supplied a much-sought-for ``exact'' value that could reliably, for the first time, discriminate between the various PN models (and between the various ways of usefully defining the ISCO within a PN framework). Work by Favata \cite{Favata:2010yd} examined the performance of about two dozen PN schemes against the GSF value of the ISCO shift, ruling most of them out. 

Another opportunity for GSF calculations to inform the development of PN theory was provided more recently, with the 
direct derivation of the fourth-PN equation of motion, independently (and using different methods) by Damour {\it et al.}~\cite{Damour:2014jta,Damour:2015isa} and by Bernard {\it et al.}~\cite{Bernard:2015njp}. Both derivations used GSF results to determine the values of certain parameters associated with ambiguities in their respective regularization methods. Furthermore, accurate GSF results for the periastron precession \cite{vandeMeent:2016hel} were later utilized to resolve an initial discrepancy between the results of the two derivations \cite{Damour:2016abl,Bernard:2016wrg}. (A complete, ambiguity-free, fully PN derivation has very recently been performed, arriving at the same result as the earlier, GSF-informed derivations~\cite{Marchand:2017pir}.)

\subsection{Interface with Numerical Relativity}

The regime of the binary black-hole problem where the two masses are comparable and strongly interacting can only be tackled reliably using NR tools. This is the regime most relevant to LIGO's black-hole merger sources, and indeed, the analysis of signals from these sources has been relying on theoretical waveforms produced using NR methods. It is desirable to extend the reach of NR methods to systems of more disparate masses: this is important for future LIGO observations \cite{Abbott:2016wiq}, and crucial for potential detections of intermediate-mass-ratio inspirals (IMRIs) with space-based or third-generation ground-based detectors \cite{Veitch:2015ela}. The problem is that NR simulations become fundamentally intractable for masses that are too disparate, because of the extra computational burden that comes with the need to resolve the two lengthscales and to track the evolution over a larger number of orbital cycles. This motivates a synergistic approach that meshes together results from NR and GSF calculations. The ultimate goal is to construct a model that is valid across all mass ratios, but the synergy can pay significant dividends along the way, as illustrated by work already done, which we now review. 
	
The potential of a GSF-NR synergy was first demonstrated in Ref.\ \cite{LeTiec:2011bk}. That work exploited two then-recent developments: on one hand, the calculation of a concrete physical effect of the GSF---its contribution to the periastron advance of orbits around a Schwarzschild black hole \cite{Barack:2010tm,Barack:2010ny}; and, on the other hand, the development of a highly accurate spectral NR code, enabling, for the first time, the extraction of local orbital phase information \cite{Boyle:2007ft}.  In Ref.\ \cite{LeTiec:2011bk}, the periastron advance $\delta(\Omega)$ for a range of mass ratios between $\eta=1\!:\!1$ and $\eta=1\!:\!8$ was calculated in NR and compared with the ``predictions'' of GSF and PN calculations (as well as with those of the EOB model; see below). The analysis allowed, for the first time, researchers to quantitatively assess the performance of the first-order GSF approximation in the comparable-masses regime, using the NR result as a benchmark. It nicely demonstrated how that approximation improves with decreasing $\eta$, and gave a first handle on the magnitude of second-order (and higher) GSF effects. Most remarkably, it illustrated how the first-order GSF approximation can prove applicable far beyond its natural $\eta\ll 1$ domain: with the simple replacement of $\eta$ with the symmetric mass ratio $mM/(M+m)^2$ [$=\eta+O(\eta^2)$], it was shown that the first-order GSF reproduces the NR result for $\delta(\Omega)$ with a good accuracy for {\it any} mass ratio---even for $\eta=1$! This curious result has since been demonstrated also with physical quantities other than $\delta(\Omega)$, but it is yet to be explained on theoretical grounds.

In Ref.\ \cite{Tiec:2013twa}, Le Tiec and collaborators used NR results for spinning black holes to predict (in effect, ``calculate'') the GSF correction to $\delta(\Omega)$ for orbits around a Kerr black hole, before direct GSF calculations in Kerr were available. Their results were confirmed (and improved upon) only very recently, with the direct GSF calculation by van de Meent \cite{vandeMeent:2016hel}. Further work (mentioned above), by Zimmerman and collaborators \cite{Zimmerman:2016ajr}, has shown how the redshift function $z(\Omega)$, including GSF corrections, can be extracted from full NR simulations. This programme has motivated the development of precision tools for extraction of orbital frequencies from NR simulations \cite{Lewis:2016lgx}. Here, comparison with existing GSF results provides a benchmark for testing the extraction procedure. Conversely, NR results for quantities for which direct GSF results are not yet available, such as the orbital-plane precession, may in the future provide a benchmark for testing the eventual GSF results. Ref.\ \cite{Lewis:2016lgx} also studied the impact of resonance transitions in NR evolutions. No clear effect has been observed for the mass ratios studied ($\eta\geq 1\!:\!7$), but future work with smaller $\eta$ may allow for an interesting comparison with corresponding GSF calculations.

The GSF-NR synergy has a large potential scope that is only now beginning to be realized. Besides testing each other, GSF and NR techniques may be directly combined to improve the performance of either. GSF codes can greatly benefit from the incorporation of advanced numerical methods that are used routinely in NR (like mesh refinement, or the use of highly accurate spectral methods for solving PDEs). On the other hand, GSF information may be directly incorporated in NR codes in order to ease the computational burden when the mass ratio is small. According to one such proposal \cite{Schutz}, one could ``excise'' a region around the smaller black holes out of the computational domain, using GSF information to set boundary conditions on the excision boundary. It is hoped that such, or similar ideas could enable 
one to tackle the IMRI regime of the two-body problem, which remains seemingly inaccessible to any one of the individual approaches on its own.

\subsection{Calibration of the Effective-One-Body model }

The EOB formalism  \cite{Buonanno:1998gg,Buonanno:2000ef,Damour:2001tu}  (see \cite{Damour:2011xga} for a review) is an analytical framework that aims to provide a universal description of the dynamics of coalescing binaries of black holes and other compact objects, valid for any mass ratio and over the entire merger process. To achieve this, the model seeks to generalize the effective-one-body reduction of the Kepler problem: just as the Newtonian two-body problem can be reduced to the motion of an effective body in a central potential, the EOB model posits that the two-body problem in GR (at least in its conservative sector) can be reduced to geodesic motion of an effective body in a deformed Schwarzschild geometry. By design, the model automatically reproduces the true dynamics in the test-particle ($\eta\to 0$) limit and to all known PN orders. Dissipative terms and spin effects are incorporated into the model in a modular way, based on resummed PN expressions.  Beyond that, the analytical EOB model involves certain {\it a priori} unspecified ``potentials'', and the idea is that these EOB potentials may then be ``calibrated'' using whatever information becomes available from other, more systematic calculations of the two-body dynamics. There is an ongoing, intensive programme to calibrate the EOB potentials using numerical data from NR simulations; see \cite{Buonanno:2009qa,Taracchini:2013rva,Bohe:2016gbl} and references therein. This activity has proven hugely successful, and NR-calibrated EOB waveform templates now play a central role in LIGO data analysis.  The EOB idea has also been applied to tidally interacting neutron-star systems (e.g.,  \cite{Damour:2009wj,Bernuzzi:2014owa}) and to the problem of hyperbolic scattering of black holes 
 \cite{Damour:2014afa,Bini:2017wfr}.

In a 2010 paper \cite{Damour:2009sm}, Damour first highlighted the potential of (the then-just-emerging) GSF results to provide a new, powerful type of calibration for the EOB model. Information from GSF calculations is highly numerically accurate, conveniently comes split into conservative and dissipative effects, and most importantly, gives a handle on the small-mass-ratio portion of the problem, which is inaccessible to NR. Damour demonstrated in \cite{Damour:2009sm} how then-available GSF results usefully constrain the $O(\eta)$ piece of the EOB potentials, and proposed a list of physical GSF quantities whose transcription into EOB language could provide further calibration. 

Thus began a fruitful exchange between the EOB and GSF programmes. The initial transcription between the GSF and EOB languages was done using the periastron advance $\delta(\Omega)$ \cite{Damour:2009sm,Barack:2010ny} as a physically meaningful invariant quantity. Later work utilized Detweiler's redshift for that purpose, for circular orbits in Schwarzschild \cite{Akcay:2012ea,Bini:2013rfa,Bini:2014nfa,Bini:2015bla} and Kerr \cite{Bini:2015xua}, then for eccentric orbits in Schwarzschild \cite{Bini:2015bfb,Bini:2016qtx}, and finally for eccentric orbits in Kerr \cite{Bini:2016dvs}. Further work utilized the self-torque on circular \cite{Bini:2014ica} or eccentric \cite{Kavanagh:2017wot} orbits, and also the self-tides \cite{Bini:2014zxa}. Ref.\ \cite{Bini:2016cje} explored the prospect of calibrating the EOB potentials at $O(\eta^2)$ using a future GSF calculation of the redshift through second order. The calibration of the dissipative sector of the EOB model has also been improved, in the EMRI domain, using perturbative energy-flux calculations \cite{Nagar:2011aa,Nagar:2016ayt,Harms:2016ctx,Lukes-Gerakopoulos:2017vkj}.

There are many prospects for future GSF calculations to further inform the EOB model (and the implications to EOB of some recent GSF results, such as the ISCO shift in Kerr \cite{Isoyama:2014mja}, have yet to be considered). Referring back to Fig.\ \ref{fig:twobody}, it can be said that the EOB approach provides a convenient framework for fusing together results from the very distinct methods of NR, PN and GSF, with the aim of covering the entire plane shown in the figure. GSF information is particularly valuable, as it gives a handle on a ``remote'', hitherto inaccessible corner of the parameter space. The hope is that such an EOB-facilitated universal model could one day provide an accurate, reliable, and computationally efficient description of all binary sources relevant to gravitational-wave astronomy, from LIGO's merging comparable-mass black holes and neutron stars to LISA's EMRIs, and everything in between.


\section{Frontiers}\label{sec:frontiers}

The self-force programme provides a fine example of experiment-driven progress in theoretical physics. The renewed interest in this old problem of classical relativity, the volume of activity over the past two decades, and the significant progress that has been achieved, all no doubt owe themselves to the exciting prospects of observing gravitational waves from EMRIs in the not-so-distant future. It remains a central goal of the programme to provide an accurate theoretical model of astrophysical EMRIs to enable the identification and interpretation of EMRI signals in the datastream of LISA. There is still much to be done to achieve that goal, and in this concluding section we will review the main outstanding issues, both foundational and computational. 


\subsection{Open problems and prospects: foundational issues}

The foundations of self-force theory are by now fairly well developed. Given an arbitrarily structured, small, compact object, there is a concrete algorithm, involving matched asymptotic expansions, for obtaining the metric (up to the smooth vacuum perturbation $h^\R_{\mu\nu}$) in the object's local neighbourhood to any order in perturbation theory. This local metric can be converted into a singular puncture in the spacetime, effectively reducing the object to a point-particle-like structure characterized by the object's multipole moments. The position of the puncture then acts as a representation of the ``position'' of the object, and the equation of motion governing it follows from two things: the vacuum Einstein field equations in the puncture's local neighbourhood, and a condition that identifies it with the object's center of mass. Once the puncture and its equation of motion are in hand, the physical field everywhere outside the object, an effective field inside it, and its mean motion can all be computed (through second perturbative order) from the coupled set of equations~\eqref{effective EFE1}, \eqref{effective EFE2}, and \eqref{2nd order eqn motion}. At first perturbative order, this description can be reduced to that of a point particle interacting with the Detweiler-Whiting free radiation field $h^{\R(1)}_{\mu\nu}$. At both first and second order, the physical picture that emerges is that of a pointlike object that moves as a test body in an effective external spacetime.

However, work remains to be done even at this general level. As stressed in Sec.~\ref{subsec:evolution}, for optimal EMRI signal analysis we require the equation of motion through second perturbative order. Although it has been derived explicitly through that order, the derivations have so far been restricted to the special case of a nonspinning, spherical object. For a generic object, the second-order equation of motion will involve the object's spin and quadrupole moments. At least hypothetically, those forces can be extracted from the fully nonlinear results of Harte~\cite{Harte:2014wya}. However, those results are, at least in principle, not applicable to black holes. One would also need to be sure of how Harte's multipole moments (defined by integration over a material stress-energy tensor) relate to those that would appear in the metric outside the body. Hence, a first-principles derivation based on matched asymptotic expansions would be preferable. We can expect two new subtleties to arise in such a derivation. First, there should be multiple possible ``good'' definitions of center of mass, corresponding to the multiple ``spin supplementary'' conditions studied in PN, test-body, and fully nonlinear contexts~\cite{Costa:2014nta}. Second, the evolution of the object's quadrupole moments cannot be determined from the vacuum EFE alone; instead, it will require knowledge of the object's internal composition. If the object is a black hole, then the no-hair theorem dictates that at leading order its quadrupole moments are those of the Kerr metric, uniquely determined by its mass and spin. But if it is a star, the quadrupole moments must be determined from the star's equation of state.
 
Beyond this extension to nonspherical, spinning objects, there are several other foundational issues to explore. For example, the second-order punctures in Refs.~\cite{Pound:12b} and~\cite{Gralla:12}, which are in different gauges, should be directly compared. The class of ``highly regular gauges'' in Ref.~\cite{Pound:2017psq} should also be examined; this is a class of gauges in which the singularity in the second-order Einstein tensor $\delta^2 G_{\mu\nu}$ is significantly ameliorated. Counter to the general discussion in Sec.~\ref{perturbation theory}, in this special class one {\em can} define a distributional stress-energy $T^{(2)}_{\mu\nu}$ and a distributionally well-defined field equation of the form~\eqref{second-order EFE}. With a well-defined distributional source, one could then define useful quasilocal, Green's-function-based definitions of the second-order singular and regular fields, analogous to the Detweiler-Whiting definitions at first order. This would complement the purely local definitions described in Sec.~\ref{local analysis}. It would also potentially enable an alternative path to calculations of second-order effects~\cite{Wardell:2014kea}. However, $T^{(2)}_{\mu\nu}$ itself has not yet been derived, and many practical details of numerical schemes in a highly regular gauge have yet to be worked out.

Another area that should be explored is the effective-field-theory approach to self-force theory~\cite{Galley:2008ih}. This approach, based on a presumed point-particle effective action together with some regularization prescription, has only been applied at first order. But if extended to second order, it would provide an alternative path to a Green's-function formulation at second order~\cite{Galley:2010xn,Galley:2011te}. Given that effective-field-theory calculations are generally quite efficient, it may also enable simple derivations of finite-size effects and even of the third-order self-force~\cite{Galley:2010xn,Galley:2011te}.

There are also more abstruse open problems. Self-force theory could be extended to the case of a photon, for example, perhaps building on work on the ultrarelativistic limit~\cite{Galley-Porto:13}. And although there have been several studies in the case of an object that is both massive and charged~\cite{Zimmerman-Poisson:14,Linz-Friedman-Wiseman:14,Zimmerman:15a}, and of objects in alternative theories of gravity~\cite{Zimmerman:15b}, there is considerable room to develop these cases from first principles. Extensions to higher dimensions could also be pursued~\cite{Harte:2018iim}.

But the greatest remaining challenges do not lie in purely foundational issues. Instead they sit at the level of practical formulations in the specific case of an EMRI. We enumerate some of those challenges below. 


\subsection{Open problems and prospects: computational issues}

There are three main (interrelated) open problems facing concrete calculations of EMRI spacetimes: improving upon and extending the reach of current first-order GSF calculations; developing and implementing an evolution scheme that remains accurate over a complete inspiral; and finding practical ways of calculating second-order GSF effects.

At first order, as we have seen, state-of-the-art numerical codes are now capable of outputting the value of the first-order GSF along (essentially) any given, fixed, bound geodesic orbit around a Kerr black hole. The most numerically efficient approach is based on a semi-analytical treatment of the frequency-domain perturbation equations in the Teukolsky formalism, followed by metric reconstruction and mode-sum regularization in the radiation gauge~\cite{vandeMeent:2016pee}. But even this method remains, in general, very computationally expensive, and suffers from several weaknesses. In particular, it does not perform well at large orbital eccentricities---not only because of the slow convergence of the frequency-mode sum, but also because of what appears to be delicate cancellations between different mode contributions~\cite{vandeMeent:2016pee}. 
Continuous-spectrum methods could help tackle high-eccentricity scenarios, as well as unbound (parabolic- or hyperbolic-type) orbits; preliminary work by Hopper and collaborators \cite{Hopper:2017qus,Hopper:2017iyq} nicely illustrates the applicability of such an approach. 

However, the problems of highly eccentric and unbound orbits are perhaps more naturally tackled using a time-domain approach. Time-domain methods and working codes have greatly improved over the past decade, but many challenges remain. In the most direct time-domain approach, based on the Lorenz-gauge formulation, there remains an unsolved problem of linearly growing gauge modes that contaminate numerical evolutions \cite{Dolan:2012jg}. And there remains a problem of slowly damped junk radiation due to the imprecise determination of (the unknown) initial conditions for the evolution.
There has been much recent work to improve the performance of time-domain codes, with a variety of techniques being introduced. These included the use of adaptive mesh refinement \cite{Thornburg:2009mw,Thornburg:2010tq,Thornburg:2016msc}, hyperboloidal and compactified coordinates \cite{Harms:2014dqa}, gauge-damping drivers \cite{Dolan:2012jg}, and pseudospectral methods \cite{Canizares:2010yx}. There is an ongoing programme of development, and one should expect to see a gradual improvement in code performance over time. We expect to see a step-function improvement from moving to work with curvature scalars (instead of metric components) as numerical variables, using the method proposed in Ref.\ \cite{Barack:2017oir}. This approach, which also automatically resolves the problem of linearly growing gauge modes, requires further development; in particular, the nontrivial extension from Schwarzschild to Kerr must be carried out. Further work is also required on eliminating initial junk radiation, perhaps through the formulation of more precise initial data for numerical evolutions. 

All of the above applies to calculations for fixed geodesics. There are significantly greater challenges in the way of accurately simulating complete inspirals. The multiscale treatment of the system appears to be the optimal method for tackling this problem, as it naturally accounts for the system's two time scales and would enable efficient frequency-domain calculations. But it has not yet been fully worked out. In Sec.~\ref{subsec:evolution}, we noted that a naive implementation of the multiscale expansion leads to infrared divergences, which must be resolved by again using the method of matched asymptotic expansions, matching the two-timescale expansion to alternative expansions in the far field and near the horizon. Furthermore, the multiscale approach is complicated by the presence of transient resonances. These resonances introduce a third, transient time scale into the system, leading to a breakdown of the two-timescale expansion. The obvious way of handling this time scale is to utilize yet another expansion, this one tailored to the near-resonance regime. A complete description of the system would then be formed from the combination of the two-timescale, near-resonance, near-horizon, and near-infinity expansions.

Several alternatives to the two-timescale expansion could be pursued. Most obviously, one could directly tackle the coupled system of equations~\eqref{effective EFE1}, \eqref{effective EFE2}, and \eqref{2nd order eqn motion}, self-consistently evolving the gravitational field and the orbit that sources it. This may have substantial advantages in the less adiabatic regions of the parameter space where a two-timescale expansion would struggle (e.g., the transition from inspiral to plunge, highly eccentric orbits, and perhaps even the transition across resonances). But such an approach faces numerous obstacles. As described in Sec.~\ref{subsec:evolution}, a naive implementation would incur secularly growing errors due to the long-term changes in the central black hole. Although this can likely be overcome~\cite{Flanagan-etal:17}, self-consistent evolution would prohibit many of the methods that have succeeded for fixed geodesic sources. Because the orbital frequencies change with time, discrete-spectrum frequency-domain methods would not apply. This suggests a time-domain approach. Currently, there is yet to be developed a time-domain code that computes the self-consistent evolution of the orbit under the GSF effect---such a calculation has only been performed so far for a scalar-charge toy model \cite{Diener:2011cc}. A code of that sort could come with a prohibitive computational cost: time-domain implementations of the perturbation equations are extremely computationally expensive, and with current technology it cannot be hoped that one would be able to self-consistently evolve an EMRI orbit for more than $O(10)$ orbits. However, as described above, time-domain methods are making advances, and they could ultimately be capable of performing a full self-consistent evolution.

Some more radical ideas for self-consistent evolutions include (1) the use of a fully spectral algorithm akin to SpEC \cite{spec} for solving the self-consistent perturbation equations; or (2) the use of a fully nonlinear NR code to solve for the full, nonperturbative metric of the EMRI system, with an excision boundary in an intermediate-scale buffer zone around the black hole \cite{Schutz}. A third idea would be to attempt self-consistent evolution without resorting to the time domain. The time-evolving frequency spectrum may be addressed using a {\it wavelet} decomposition instead of the standard Fourier decompositions: unlike Fourier harmonics, wavelets are localized in time (as well as in frequency), and in principle, a basis of such functions could naturally describe the slowly evolving perturbation in the EMRI problem. The details of all three proposals are yet to be worked out and their applicability and advantage assessed.

Another obvious, less radical alternative to the two-timescale method would be an extension of the method of osculating geodesics. To date, this is the only method that has been used to evolve orbits in practice (aside from leading-order, adiabatic evolutions), but currently it can only accommodate the first-order GSF. Accurate EMRI modelling will require some second-order effects, and the method will have to be extended to account for them. Reference~\cite{Pound:2015tma} provided an initial sketch of how that might be done.

Making that leap to second order, regardless of which evolution scheme is used, is now the most significant challenge within the programme to obtain accurate EMRI waveforms for LISA. So far, work to that end has been restricted to the simple test case of quasicircular orbits in Schwarzschild geometry, and even in that case only partial initial results have been obtained. We expect this to remain at the forefront of activity in the field over the coming years. The challenges here are significant. Even in the Schwarzschild case, extending the current frequency-domain (two-timescale), Lorenz-gauge-based method \cite{Wardell:2015ada,Miller:2016hjv} to noncircular orbits is quite nontrivial. The principal technical issue here is how to construct the second-order perturbation from a sum over inhomogeneous frequency modes while avoiding complications associated with the non-smoothness of the field at the particle. We envisage using a variant of the method of ``extended particular solutions'' introduced in \cite{Hopper:2012ty} (itself a generalization of the method of ``extended homogeneous solutions'' \cite{Barack:2008ms}), which elegantly circumvents the problem. But the details are yet to be worked out. 

A yet more challenging task is the extension of second-order calculations to Kerr: here, the existing frequency-domain Lorenz-gauge-based approach is invalid in its current form, due to the lack of separability of the field equations in Kerr. One could possibly work with coupled field equations, but we suspect the resulting scheme would be  cumbersome to the point of intractability. It might be possible to tackle the second-order Kerr problem in the time domain, using self-consistent evolution in the Lorenz gauge; but in addition to the problems described above, this would involve an evolution of the field equations with a non-compact source. That presents its own potential difficulties, and it is yet to be attempted. It would be highly desirable to instead extend the radiation-gauge method that has facilitated calculations at first order. This would require a formalism of second-order metric reconstruction from curvature scalars satisfying separable equations on Kerr. The prospects for obtaining such a formalism in the near future remain unclear. For the purpose of EMRI modelling, it may also be possible to work entirely with curvature scalars. Sufficiently accurate EMRI models do not require {\em all} the information from second order, but only certain time-averaged dissipative effects. At first order, these effects can be computed entirely from curvature scalars; if the same is true at second order, one could bypass the need for a second-order metric-reconstruction formalism.

Besides ultimately providing accurate waveform templates for analysis of LISA data, self-force calculations can, on a shorter timescale, inform the gradual development of approximate EMRI waveforms that can be used in data-analysis studies. Such approximate models---known as ``kludges''---may even prove useful as crude search templates in a first stage of a hierarchical, multi-step search algorithm. The idea of kludge templates has a long history in EMRI studies. The first kludge models \cite{Barack:2003fp,Babak:2006uv} incorporated back-reaction effects in an approximate manner, using post-Newtonian formulas, but there is a long-running programme \cite{Huerta:2008gb, Huerta:2014wka, Chua:2015mua, Chua:2017ujo} aimed to improve the accuracy of the models using numerical data from actual self-force calculations. Efforts so far have concentrated on the simple Schwarzschild problem, but as self-force calculations in Kerr are now becoming available there is much scope for further development of kludges of increasing precision across the full parameter space of astrophysical EMRIs. We envisage using a simple, perhaps semi-analytical kludge model as an adjustable platform, with parameters that can be calibrated by fitting to numerical self-force data (in much the same way that the EOB model of merging black hole binaries is calibrated using NR data). It is not unreasonable to expect that some form of an advanced self-force-calibrated kludge model could ultimately suffice for both detection and parameter extraction of LISA EMRIs. 

\ack 
We are grateful to Sam Dolan for many discussions that helped shape the structure and style of this work. We thank Niels Warburton for help in producing Fig.\ \ref{fig:evolution}, and Maarten van de Meent for supplying numerical data for the production of Figs.\ \ref{fig:GSFloops}, \ref{fig:GSFloops2} and \ref{fig:kicks}. We also thank Anna Heffernan for pointing out several errors in a previous version of this text. L.B. acknowledges support from STFC through Grant No. ST/R00045X/1. A.P. acknowledges support from a Royal Society University Research Fellowship.

\raggedright

\vspace{5mm}

\bibliography{RoPP}

\end{document}